\documentclass[a4paper,11pt]{article}
\usepackage{jheppub} 
\usepackage{amsfonts,amssymb,epsfig,amsmath,mathtools}
\usepackage{color}
\usepackage{comment}
\usepackage{url}
\usepackage{tikz}
\usepackage[compat=1.1.0]{tikz-feynman}
\allowdisplaybreaks
\usepackage{slashed,cancel}
\usepackage{mathrsfs}
\usepackage{graphicx,mdframed,overpic,subfig}
\usepackage{hyperref}
\definecolor{ForestGreen}{RGB}{34,139,34}

\usepackage{changepage}

\newcommand{\be}{\begin{eqnarray}}
\newcommand{\ee}{\end{eqnarray}}

\newcommand{\bn}{\begin{enumerate}}
	\newcommand{\en}{\end{enumerate}}

\def\centerarc[#1](#2)(#3:#4:#5)
{ \draw[#1] ($(#2)+({#5*cos(#3)},{#5*sin(#3)})$) arc (#3:#4:#5); }

\DeclareMathOperator{\Tr}{Tr}

\renewcommand{\Im}[0]{\operatorname{Im}}
\renewcommand{\Re}[0]{\operatorname{Re}}
\newcommand\e{\epsilon}
\newcommand\x{{\vec x}}
\newcommand\kv{{\vec k}}

\newcommand\p{{\vec p}}

\newcommand\q{{\vec q}}

\renewcommand{\d}{\partial}
\newcommand\vv{{\vec v}}
\newcommand\g{\mathfrak{g}}

\newcommand\h{\mathfrak{h}}

\newcommand\n{{\hat n}}
\newcommand\s{{\hat s}}

\title{\boldmath 
	\vspace*{1.5cm}
    Symmetry and causality constraints on Fermi liquids
	\vspace{-0.2cm}
}

\author[a,1]{Luca V. Delacr\'etaz,\note{lvd@uchicago.edu}}
\author[a,b,2]{Subham Dutta Chowdhury,\note{sdutta\_c@ictp.it}}
\author[c,3]{Umang Mehta,\note{umang.mehta@colorado.edu}}
\affiliation[a]{Kadanoff Center for Theoretical Physics \& James Franck Institute,\\
	University of Chicago, Illinois 60615, USA}
\affiliation[b]{The Abdus Salam ICTP,\\ Strada Costiera 11, 34151, Trieste, Italy}
\affiliation[c]{Department of Physics and Center for Theory of Quantum Matter,\\
	University of Colorado, Boulder CO 80309, USA
	\vspace{-0.3cm}}

\abstract{ 
We investigate symmetry and causality constraints on interacting Fermi liquids. Whereas Galilean or Lorentz boost symmetry leads to a well-known constraint on the first Landau parameter $F_1$, we show that scale invariance similarly constrains $F_0$. In the case of conformal Fermi liquids, we show that causality constraints on the particle-hole continuum and on zero sound strongly restrict the available parameter space for interacting Fermi liquids. 
We also consider nonlinear response, which we show is sensitive to additional ``generalized Landau parameters'' even at lowest orders in the long wavelength limit. We impose Galilean, Lorentz and scale symmetry on these generalized Landau parameters, obtaining further nonlinear constraints. We test our constraints in several microscopic models that enter a Fermi liquid phase.}

\makeatletter
\gdef\@fpheader{}
\makeatother

\begin{document}
\maketitle
\flushbottom

\section{Introduction and Results}

\begin{figure}
\centering
	\hfill
	\subfloat[]{
	\begin{overpic}[page=1,width=0.23\linewidth,tics=10]{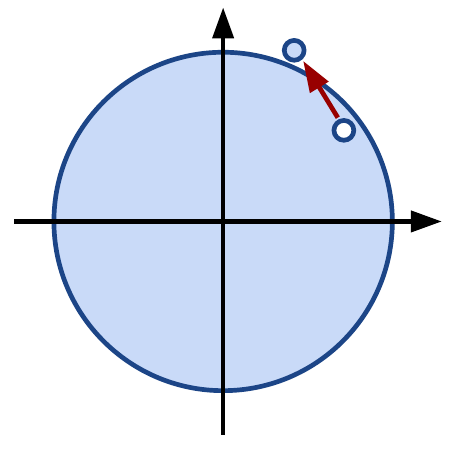}
		 \put (35,93) {$k_y$} 
		 \put (90,38) {$k_x$}
	\end{overpic}
	}
	\hfill
	\subfloat[]{
	\begin{overpic}[page=2,width=0.23\linewidth,tics=10]{particlehole}
		 \put (25,95) {$\Im \langle j j \rangle(\omega,k)$} 
		 \put (-3,82) {$\omega$} 
		 \put (85,-3) {$k$}
	\end{overpic}
	}
	\hfill
	\subfloat[]{
	\begin{overpic}[page=3,width=0.23\linewidth,tics=10]{particlehole}
		 \put (25,95) {$\Im \langle j j \rangle(\omega,k)$} 
		 \put (-3,82) {$\omega$} 
		 \put (85,-3) {$k$}
	\end{overpic}
	}\hfill\,
\caption{{\bf(a)} The particle-hole continuum of Fermi surfaces lead to {\bf(b)} low energy spectral densities of currents $j^\mu$ at finite wavevector $k\leq 2k_F$ (dark gray, with multi-particle-hole continuum shown in lighter gray). {\bf (c)} In contrast, QFT spectral densities in the vacuum only have support for $\omega\geq k$.
\label{fig_FS}
}
\end{figure}
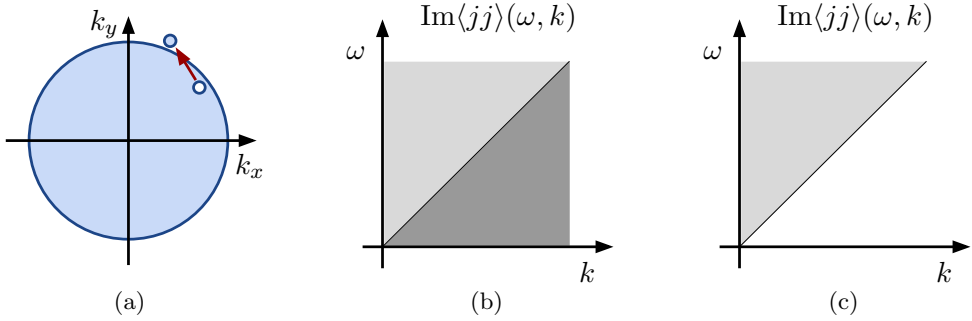

Compressible phases constitute some of the most intricate and diverse phenomena in quantum many-body systems. Their finite compressibility, or charge susceptibility, implies that they are necessarily gapless%
	\footnote{Indeed, a nonzero compressibility $\chi$ implies that the $\omega$ and $q\to 0$ limits do not commute in response functions, implying gaplessness:
	\begin{equation*}
	\lim_{q\to 0} \lim_{\omega \to 0} G^R_{\rho\rho}(\omega,q) 
	=\chi 
	\neq 0
	= \lim_{\omega \to 0} \lim_{q\to 0} G^R_{\rho\rho}(\omega,q) \, ,
	\end{equation*}
	where the last equation follows from the fact that the total charge $\rho_{q=0}$ commutes with itself. This simple argument is presumably well-known to many, but we could not find a textbook reference for it. We thank Nabil Iqbal for pointing it out to us.
	} 
-- examples include superfluids, Wigner crystals, pair density waves, electron smectics, extremal black holes, Fermi liquids and non-Fermi liquids. While many of these are symmetry-broken states that have simple effective field theory  (EFT) descriptions in terms of a handful of Nambu-Goldstone modes, those with Fermi surfaces are particularly challenging to describe with the conventional tools of quantum field theory: the presence of an extended Fermi surface leads to a continuum of low energy particle-hole excitations at finite wavevectors. This extreme gaplessness manifests in a number of striking ways: super-area law entanglement, large specific heat, and low frequency spectral densities with support at finite wavevector (see Fig.~\ref{fig_FS}). Despite textbook treatments of Fermi liquids using Landau's phenomenological approach \cite{pinesnozieres2018theory,pitaevskii1980statistical}, and the development of EFT approaches \cite{abrikosov2012methods,Polchinski:1992ed,Shankar:1993pf}, many challenges remain in establishing systematic descriptions of Fermi liquids, perhaps most clearly evidenced by the sparsity of controlled approaches to understand their strongly coupled cousins, non-Fermi liquids.

Fermi liquids are ubiquitous in nature, both in a non-relativistic context (Helium-3, metals, nuclear matter), and a relativistic one (high density quark matter, possibly in the interior of cold, dense neutron stars \cite{PhysRevC.58.1804, 1974AnPhy..83..491W, MATSUI1981365}). In any of these situations, the emergence of a Fermi liquid phase is not obvious or guaranteed from microscopics. Relatedly, if a Fermi liquid phase emerges, the data parametrizing it -- the Fermi momentum $p_F$ and velocity $v_F$, the Landau parameters $F_0,\,F_1,\,F_2,\,$ etc.\ -- can be difficult to relate to microscopics in general. However, UV/IR constraints can in some cases non-perturbatively constrain the possible emergent behavior at low energies. UV/IR constraints have seen a resurgence of interest in the context of EFTs both in the high-energy \cite{Paulos:2017fhb, Bellazzini:2020cot, Arkani-Hamed:2020blm, Caron-Huot:2020cmc, Caron-Huot:2021rmr, Tolley:2020gtv, Sinha:2020win}\footnote{See \cite{Kruczenski:2022lot} for a more comprehensive overview of progress in this direction in recent years.} and condensed matter literature, including for compressible phases \cite{Luttinger:1960zz,Oshikawa:2000lrt,Alberte:2020eil,Else:2020jln,Komargodski:2021zzy}. The simplest constraint is symmetry: microscopic symmetries must be respected by the low energy dynamics. In this paper, we study systematically the consequences of spacetime symmetries, namely boost and dilatation symmetry, on relativistic and non-relativistic Fermi liquids. We will find that even these simple constraints lead to new universal results in Fermi liquid theory. For relativistic Fermi liquids, we also explore constraints of microcausality, the vanishing of commutators outside of the lightcone. Our results are summarized below.

\subsection{Summary of results}
We consider translation invariant Fermi liquids, that have in addition several other symmetries such as Galilean boosts,%
	\footnote{Galilean boost symmetry is sometimes confused or conflated with translation invariance, even in classic books \cite{pinesnozieres2018theory}. These two symmetries are different.}
Lorentz boosts, relativistic or non-relativistic scale invariance. It is well-known that boost symmetries constrain the first Landau parameter of Fermi liquids \cite{pinesnozieres2018theory,pitaevskii1980statistical,BAYM1976527}:
\begin{subequations}\label{eq_boost}
\begin{align}
\hbox{Galilean symmetry:}&&1+F_1
	&= \frac{m_*}{m}\, , & \\
\hbox{Lorentz symmetry:}&&
1+F_1
	&= \frac{m_*}{\mu}\, ,
\end{align}
\end{subequations}
where the effective mass is defined by the Fermi velocity as $m_*\equiv p_F / v_F$, $m$ is the central charge of the Galilean group, and $\mu = \e_F$ the chemical potential or Fermi energy. These expressions hold in both $d=2$ and $d=3$ spatial dimensions.%
    \footnote{A different normalization $F_1 \to \frac13 F_1$ is sometimes used in $d=3$. We normalize all Landau parameters such that stability bounds read $F_\ell \geq -1$ in any dimension.}
We will show that similar constraints arise from dilation symmetries, both non-relativistic and relativistic systems:
\begin{subequations}\label{eq_dilation}
\begin{align}
\hbox{Schr{\"o}dinger symmetry:}&&
1+F_0
	&= 2 \frac{\mu}{v_Fp_F} \, , & \\
\hbox{Conformal symmetry:}&&
1+F_0
	&= \frac{\mu}{v_Fp_F} \, .
\end{align}
\end{subequations}
These results also hold in $d=2$ and $d=3$.

We emphasize that a Fermi surface {\em state} clearly breaks boost and dilation symmetry, because it has a finite density (or Fermi wavevector $p_F$). The constraints above apply to finite density states arising from microscopic systems respecting these symmetries. Relatedly, we will see these symmetries are nonlinearly realized on the EFT of Fermi liquids.

\begin{figure}
\centerline{
\subfloat[]{\label{sfig_collective}

	\begin{tikzpicture}[scale=2]

	\draw[thick,->] (-0.1,0) -- (3.2,0) node[anchor=north] {$q$};
	\draw[thick,->] (0,-0.1) -- (0,2.5) node[anchor=east] {$\omega$};

	\fill[gray!30, opacity=0.6] 
		(0,0) 
		-- (3,1.5) 
		-- (3,0) 
		-- cycle;

	\draw[thick, gray] (0,0) -- (3,1.5);
	\node[gray] at (2.4,0.7) {$\omega \leq v_Fq$};

	\draw[thick, blue] (0,0) -- (3,2.4);
	\node[blue,text width=2cm] at (1.2,2) {$\omega = v_0 q$};

	\draw[thick, ForestGreen] (0,0) -- (3,1.8);
	\node[ForestGreen,text width=2cm] at (1.2,1.7) {$\omega = v_\perp q $};

	\node[gray] at (2,0.2) {\small Particle-hole continuum};

	\end{tikzpicture}
}
\hspace{0.08\linewidth}
\subfloat[]{
    \begin{overpic}[width=0.4\linewidth,trim=0 -10 0 0,clip]{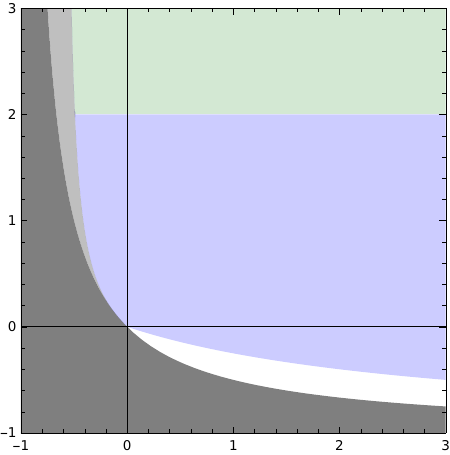}\label{sfig_exclusion}
	\put(45,90){\small\color{ForestGreen} shear sound}
	\put(55,85){\small\color{ForestGreen} +}
	\put(45,80){\small\color{ForestGreen} zero sound}
	\put(45,53){\small\color{blue} zero sound}
	\put(45,90){\small\color{ForestGreen} shear sound}
	\put(55,85){\small\color{ForestGreen} +}
	\put(45,80){\small\color{ForestGreen} zero sound}
	\put(45,53){\small\color{blue} zero sound}
	\put(-5,60){\small $F_1$}
	\put(57,0){\small $F_0$}
	\end{overpic}
}
}
\caption{\label{fig_causality_collective} (a) In Fermi liquids, the particle-hole continuum and collective excitations such as zero sound (blue) and shear sound (green) produce non-analyticities in $G^R_{\rho\rho}(\omega,q)$. (b) Causality constraints on conformal Fermi liquids in $d=3$, in the space of the first Landau parameters $F_0,F_1$. The dark gray region is excluded by Eq.~\eqref{3d_vF_causality}. Demanding the collective modes be causal leads to a stronger constraint, excluding the light gray region. The remaining allowed parameter space either features no collective excitation (white), a coherent zero sound mode (blue), or both zero sound and shear sound (green). See Sec.~\ref{sec_linear} and App.~\ref{app_dimgen} for details.}
\end{figure}
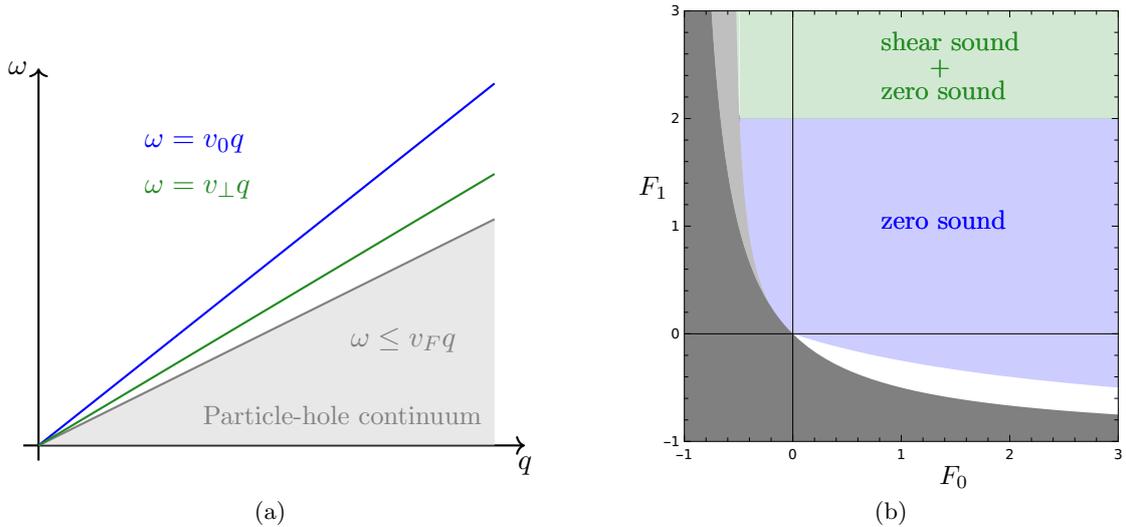

Relativistic systems are also subject to strict UV/IR constraints arising from microcausality, i.e.~the commutation of spacelike separated operators. This leads to interesting non-perturbative bounds on the marginal  parameters of a Fermi liquid. Consider for example a conformal field theory, which enters a Fermi liquid phase upon `doping' (i.e., turning on a chemical potential, or equivalently considering finite density states of the CFT). Combining Eqs.~\eqref{eq_boost} and \eqref{eq_dilation} implies that the Fermi velocity of a conformal Fermi liquid is entirely fixed in terms of the first two Landau parameters:
\begin{align}
\hbox{Conformal Fermi liquid:}&&
v_F^2
	&= \frac{1}{(1+F_0)(1+F_1)} \, . &
\end{align}
Demanding subluminality of the particle-hole continuum therefore leads to the universal bound 
\begin{equation}\label{3d_vF_causality}
F_0F_1 + F_0 + F_1 \geq 0\, .
\end{equation}
Fermi liquids can also harbor collective excitations beyond the particle-hole continuum, as illustrated in Fig.~\ref{sfig_collective}. Demanding that these be causal leads to stronger constraints on Landau parameters, as shown in Fig.~\ref{sfig_exclusion}. These bounds are potentially of phenomenological interest, particularly in the context of QCD at very high densities. It is interesting that they exclude most of the parameter space without collective excitations in Fig.~\ref{sfig_exclusion}; in other words, it is very likely that QCD at high density has a coherent zero sound excitation. It is also interesting that our bounds constrain marginal EFT parameters, whereas most recent results in UV/IR constraints in QFT apply to irrelevant parameters.

We then turn to the question of nonlinear response of Fermi liquids in Sec.~\ref{sec_nonlin_EFT}. This is an area of recent interest \cite{PhysRevLett.122.257401,Tam:2022tpy,Tam:2023scw,bradlyn2024spectraldensitysumrules, Beane:2022wcn} which however has not been treated systematically for interacting Fermi liquids before. We show that even at leading order in small momenta and frequencies $q,\omega/v_F\ll p_F$, nonlinear response depends on data beyond the Landau parameters $F_\ell$. Using the nonlinear EFT of Fermi liquids introduced in \cite{Delacretaz:2022ocm}, we parametrize the ``generalized Landau parameters'' that enter in nonlinear response, focusing on three-point functions. This allows us to provide a closed-form expression for the density three-point function of an interacting Fermi liquid in Eq.~\eqref{rrrexactf20f30}. 

The nonlinear EFT also allows for a systematic analysis of symmetries. In Sec.~\ref{sec_nonlin_symmetry}, we impose Galilean invariance, scale invariance, and Lorentz invariance, recovering the previous linear constraints \eqref{eq_boost} and \eqref{eq_dilation}, and deriving new nonlinear constraints on the generalized Landau parameters. In particular, our nonlinear Lorentz constraints can be stated as follows:
\begin{equation}\label{eq_Lorentz_nonlin_intro}
\begin{split}
\left\{2 \nabla^i_\p(\epsilon_p F^{(2,0)}(\p,\p'))  - 2 \int_{p''}F^{(2,0)}(\p'',\p') F^{(2,0)}(\p,\p'')\nabla^i_{\p''} f_{p''}^0 \right.\\
\left.-3 \int_{p''}\epsilon_{p''} F^{(3,0)}(\p,\p',\p'') \nabla^i_{p''} f_{p''}^0\right\}_{p_F} &=0\, .\\
\end{split}
\end{equation}
Here $F^{(2,0)}$ is the usual Landau interaction function, whereas $F^{(3,0)}$ denotes generalized Landau parameters that also contribute at leading order to the density three point function, and $f_{k}^0 = \Theta(p_F- |\vec{k}|)$ is the distribution function of the unperturbed Fermi surface state. The equation, to be evaluated at $|\p|=|\p'|=p_F$ performing the derivatives, leads to an infinite series of constraints on the harmonics of the Landau parameters. We emphasize the appearance of the generalized Landau parameters $F^{(3,0)}$ in this Lorentz constraint, which was missed in previous studies on relativistic constraints on Fermi liquids \cite{BAYM1976527}.

Finally, in Sec.~\ref{sec_micro_models}, we test our constraints in various microscopic models: a free Fermi gas of Dirac fermions,  Dirac fermions with a small four-Fermi interaction, and a strongly interacting Fermi liquid arising from a large $N$ Chern-Simons matter theory \cite{Geracie:2015drf}. We find that they are satisfied in each of these models. Along the way, we discuss how to extract EFT parameters from microscopics, and match microscopic operators to effective operators in the EFT. All our main results, Eqs.~\eqref{eq_boost}-\eqref{eq_Lorentz_nonlin_intro}, are valid both in $d=2$ and 3 spatial dimensions (and reduce to simpler but analogous expressions in $d=1$). We mostly focus on $d=2$ throughout the main text for clarity, and extend to $d=1,~3$ in App.~\ref{app_dimgen}.

\,

\,

\pagebreak

\section{Linear constraints on Fermi liquids}\label{sec_linear}

An efficient way to parametrize and study interacting Fermi liquids is through Fermi surface bosonization \cite{PhysRevB.19.320,Haldane:1994,PhysRevB.49.10877,Houghton:2000bn}. To leading order in derivatives and fields, the action for a (2+1)-dimensional Fermi liquid in this approach is
\begin{equation}\label{eq_S_gaussian}
S
	= -\frac{p_F}{2} \int \frac{dt d^2x d\theta}{(2\pi)^2}
	\nabla_n \phi \left( \dot \phi + v_F \nabla_n \phi + v_F \int \frac{d\theta'}{(2\pi)^2} F^{(2,0)}(\theta-\theta')\nabla_{n'}\phi'\right)+ \cdots\, ,
\end{equation}
where $\nabla_n = \hat n(\theta)\cdot \nabla$, with $\hat n = { \cos (\theta)\choose \sin (\theta)}$ the unit vector perpendicular to the Fermi surface.
The degree of freedom $\phi(t,\vec x,\theta)$, or its Fourier transform $\phi(t,\vec q,\theta)$, represents a particle-hole excitation with momentum $q\ll p_F$ near the Fermi surface at the angle $\theta$. These particle-hole pairs are described by chiral bosons at every angle $\theta$ propagating with Fermi velocity $v_F$ in the direction $\hat n$, which couple through the Landau interactions $F^{(2,0)}(\theta-\theta')$. We have assumed translation and rotation invariance. 

This action reproduces the bosonic response of Fermi liquids, in the approximation where only marginal interactions (forward scattering) are considered \cite{abrikosov2012methods,Benfatto:1990zz,Polchinski:1992ed,Shankar:1993pf}. There are several ways to arrive at this action, whose equation of motion is the linearization of Landau's kinetic theory. It can be motivated from the algebra of fermion bilinears near the Fermi surface \cite{PhysRevB.19.320,Haldane:1994}, normal ordered with respect to the Fermi surface state. Relatedly, it is the unique theory that nonlinearly realizes the (anomalous) $LU(1)$ symmetry \cite{Else:2020jln}. Finally, the coadjoint orbit formalism \cite{Delacretaz:2022ocm} allows to systematically extend this EFT to higher order in derivatives and fields. This last approach will be reviewed in Sec.~\ref{sec_nonlin_EFT}, so we do not further motivate Eq.~\eqref{eq_S_gaussian} here.

\subsection{Static susceptibilities and symmetry constraints}\label{ssasc}

The quadratic action for particle-hole fluctuations \eqref{eq_S_gaussian} features a number of parameters that characterize a Fermi liquid state: the Fermi wave-vector $p_F$ related to the density via Luttinger's theorem%
    \footnote{On the lattice, a nonperturbative derivation of Luttinger's theorem follows from understanding the response to a $U(1)$ flux \cite{Oshikawa:2000lrt}. The corresponding proof in the continuum uses the Lorentz force: any QFT with spacetime translation invariance and $U(1)$ symmetry satisfies $\nabla_\mu T^{\mu\nu} = F^{\nu\lambda}j_\lambda$, in the presence of a background field for the $U(1)$. This fixes the susceptibility between momentum density and current $\chi_{T_{0i}j_j} = \langle\rho\rangle \delta_{ij}$ (see App.~\ref{app_thermoward}). Reproducing this in the EFT then gives the Luttinger relation.}
\begin{equation}\label{eq_Luttinger}
\langle \rho\rangle = \frac{p_F^2}{4\pi}\, , 
\end{equation}
the Fermi velocity $v_F$, and the Landau parameters $F_\ell$, $\ell=0,1,2,\ldots$, defined as the Fourier components of the couplings appearing in \eqref{eq_S_gaussian}:
\begin{equation}\label{F20ththpharmonics}
F^{(2,0)}(\theta-\theta')
	= 2\pi\sum_{\ell}F_\ell e^{i\ell (\theta-\theta')}\, , \qquad
F_{-\ell} = F_\ell^* = F_\ell \, .
\end{equation}
We would like to find the constraints that symmetry imposes on these parameters. Symmetries in QFT are usually imposed directly at the level of the action. Boost and dilation symmetries are somewhat subtle to implement in this way, because they are nonlinearly realized on the fluctuations $\phi$. This will be done more systematically in Sec.~\ref{sec_nonlin_symmetry}; we will first follow a more pedestrian approach in this section, by considering a set of observables and imposing symmetries on these.

At the Gaussian level \eqref{eq_S_gaussian}, Fermi liquids have a conserved density at every patch, given by
\begin{equation}
\rho_\theta \equiv \frac{p_F}{2\pi} \nabla_n \phi\, , \qquad
\rho_\theta
	= \sum_\ell e^{i\ell \theta} \rho_\ell\, , \qquad
	\rho_{-\ell} = \rho_\ell^*\, .
\end{equation}
The two-point function of these densities is simple to evaluate perturbatively in the Landau parameters
\begin{equation}\label{eq_rr_theta_EFT}
\langle \rho_\theta \rho_\theta'\rangle (\omega,\vec q)
	= \frac{i\, p_F}{2\pi v_F} \frac{v_F q_n}{\omega - v_F q_n} \left[2\pi \delta(\theta - \theta') + \frac{1}{2\pi}F^{(2,0)}(\theta - \theta')  \frac{v_Fq_{n'}}{\omega - v_Fq_{n'}} + \cdots\right]\, .
\end{equation}
While the general expression for finite Landau parameters is complicated, it simplifies in the static limit $\omega\to 0$. The static susceptibilities for the harmonics $\rho_\ell$ are a simple sequence  of observables that measure the Landau parameters:
\begin{equation}\label{eq_chi_ell}
\chi_\ell 
	\equiv \lim_{q\to 0}\lim_{\omega\to 0} G^R_{\rho_\ell \rho_{-\ell}}(\omega,\vec q)
	= \frac{p_F}{v_F} \frac{1}{2\pi} \frac{1}{1+F_\ell}\, .
\end{equation}
This result can be found by expressing the Hamiltonian in terms of the harmonics $\rho_\ell$
\begin{equation}
H = \frac12 \frac{v_F}{p_F} 2\pi \sum_\ell \int d^2 x \,\rho_\ell\rho_{-\ell} (1+F_\ell)\, .
\end{equation}
and introducing static (time-independent) sources $H\to H -\sum_\ell \int d^2 x \mu_{\ell}\rho_{-\ell}$ before evaluating $\chi_\ell = \partial\rho_\ell / \partial\mu_\ell$.

In a Fermi liquid beyond the approximation \eqref{eq_S_gaussian}, all the densities $\rho_\ell$ are only approximately conserved, except for the $\ell=0,1$ harmonics which correspond to $U(1)$ charge density $\rho$ and momentum density $\pi^i$ \footnote{Momentum density is the $T^{0i}$ component of the stress tensor.}:
\begin{equation}
\rho = \rho_0 = \frac{p_F}{2\pi}\int \frac{d\theta}{2\pi} \nabla_n \phi\,, \qquad
\pi^i = p_F {\Re \rho_1\choose \Im \rho_1} 
	= \frac{p_F^2}{2\pi} \int \frac{d\theta}{2\pi} \hat n^i \nabla_n \phi\, .
\end{equation}
The susceptibilities associated with these densities, $\chi_{\rho\rho} = \chi_0$ and $\chi_{\pi\pi} = \frac12 p_F^2 \chi_1$, can in some cases be fixed by symmetries of the underlying microscopic (zero density) system. For example, scale invariant microscopics with dynamic critical exponent $z$ implies that the density depends on the chemical potential $\mu=\epsilon_F$ as $\rho \propto \mu^{d/z}$ in $d$ spatial dimensions, so that the charge susceptibility is
\begin{equation}\label{eq_chi0_scale}
\chi_0 = \frac{\partial\rho}{\partial\mu}
	= \frac{\rho}{\mu}\frac{d}{z}\, .
\end{equation}
Comparing with Eq.~\eqref{eq_chi_ell} and using the Luttinger relation Eq.~\eqref{eq_Luttinger}, this fixes the zeroth Landau parameter:
\begin{equation}\label{eq_F0sym}
1+F_0 = \frac{\mu}{v_Fp_F} z\, .
\end{equation}
This result applies in particular to CFTs ($z=1$) and non-relativistic (Schr\"odinger) CFTs ($z=2$) that become Fermi liquids upon doping. In these cases, one recovers \eqref{eq_dilation} as advertised. This expression holds in $d=3$ as well, see App.~\ref{app_dimgen}.

Boost invariance similarly fixes the momentum susceptibility:
\begin{equation}\label{eq_chipp_boost}
\chi_{\pi\pi}
	= 
	\begin{cases}
	m \rho &\hbox{Galileo},\\
	\varepsilon + P =\mu \rho & \hbox{Lorentz}.
	\end{cases}
\end{equation}
Here $m$ is the ``bare mass'', or more formally the central charge of the Galilean algebra.
While these relations are well-known, their derivation is somewhat subtle and is reviewed in App.~\ref{app_thermoward}. Comparing again with \eqref{eq_chi_ell}, this fixes the first Landau parameter:
\begin{equation}\label{eq_F1sym}
1+F_1 = \frac{p_F \rho}{v_F \chi_{\pi\pi}}
	= 
	\begin{cases}
	\frac{p_F}{v_F m} & \hbox{Galileo},\\
	\frac{p_F}{v_F \mu} &\hbox{Lorentz}.
	\end{cases}
\end{equation}
Expressed in terms of the effective mass $m_* \equiv p_F / v_F$, these take the form \eqref{eq_boost}.

\subsubsection*{Conformal Fermi liquids}
For conformal Fermi liquids, combining \eqref{eq_F0sym} and \eqref{eq_F1sym} one finds that the Fermi velocity is entirely fixed in terms of the first two Landau parameters
\begin{equation}\label{eq_vF_CFT}
v_F^2 = \frac{1}{(1+F_0)(1+F_1)}\, .
\end{equation}
Similarly, the Fermi wavevector is fixed in terms of the chemical potential (or Fermi energy) and first two Landau parameters
\begin{equation}\label{eq_pF_mu_CFT}
p_F^2 = \mu^2 \frac{1+F_1}{1+F_0}\, .
\end{equation}
This equation, together with the Luttinger relation between density $\rho$ and $p_F$, is the equation of state of the CFT at finite density. We will discuss an example of a CFT that becomes an interacting Fermi liquid in the presence of a chemical potential in Sec.~\ref{sec_micro_models}.

\subsubsection*{Non-relativistic Conformal Fermi liquids}

Non-relativistic CFTs can also become Fermi liquids upon doping by a chemical potential. When they do, the low energy dynamics will be constrained by \eqref{eq_F0sym} and \eqref{eq_F1sym}. Similarly to \eqref{eq_pF_mu_CFT}, the Landau parameters $F_0,F_1$ determine the deviation of the equation of state from that of a free fermion:
\begin{equation}
\frac{p_F^2}{2m\mu} = \frac{1+F_1}{1+F_0}\, .
\end{equation}
One perturbative NRCFT that is expected to have a Fermi liquid phase in a parametrically large temperature range is the unitary Fermi gas in $d=2+\epsilon$ spatial dimensions \cite{Nishida:2006eu, Nishida:2006rp, Nishida:2007} -- this model consists of weakly coupled fermions, which only condense below a BCS temperature much smaller than chemical potential $T_c/\mu \sim  e^{-1/{\epsilon}}\ll 1$, allowing for a Fermi liquid at temperatures $T_c < T \ll \mu$. It would be interesting to test our constraints in this model.%
	\footnote{Refs.~\cite{Rothstein:2017niq,Rothstein:2017twg,Pavaskar:2021jla} studied the implementation of non-relativistic Galilean and dilation symmetries in the fermionic EFT of Fermi liquids \cite{Polchinski:1992ed,Shankar:1993pf}, elegantly recovering the Landau relation \eqref{eq_F1sym}. However, they did not find the corresponding constraint from dilations \eqref{eq_F0sym}, and instead (incorrectly) concluded that Schr\"odinger symmetry precludes interacting Fermi liquid behavior. As discussed above, there are known  Schr\"odinger invariant models that become Fermi liquids at finite density. We suspect that the technical mistake in Ref.~\cite{Rothstein:2017twg} lies in dropping the dependence of the coupling $g(k_1,k_2,k_3,k_4)$ on radial momentum too early in the calculation that is implicit below their Eq.~(6.19).}
In contrast, the unitary Fermi gas in $d=4-\epsilon$ as well as large $N$ in $d=3$ \cite{Nikolic:2007zz} both effectively feature weakly coupled bosons, which Bose-condense at $T_c/\mu \sim 1$, leaving no room for a Fermi liquid phase.

%
\subsection{Collective modes and causality}

Causality places constraints on the dynamics that can emerge from relativistic systems. Even in the vacuum, these constraints are often not obvious from symmetry principles alone \cite{Pham:1985cr, Ananthanarayan:1994hf, Adams:2006sv}. Here, we are interested in such constraints beyond the vacuum, in finite density states that are not themselves Lorentz invariant. One way causality manifests itself throughout the spectrum of relativistic QFTs is from microcausality: $[\mathcal{O}_1,\mathcal{O}_2] = 0$ for space-like separated operators. This implies that the Fourier transform of retarded Green's functions must be analytic for momenta $p^\mu = (\omega,k^i)$ with imaginary part pointing in the forward light-cone \cite{itzykson2006quantum}
\begin{equation}\label{eq_causality_analyticity}
G^R(\omega,\vec q) \ \hbox{ analytic in $\omega$ for } 
\Im \omega > |\Im q|\, ,
\end{equation}
a condition which offers a simple, necessary UV/IR constraint on the emergence of non-trivial physics at low energies, in any state. Microcausality has been studied in Lorentz non-invariant states before (see, e.g., \cite{Dubovsky:2007ac,Hartman:2017hhp,Grall:2021xxm,Delacretaz:2021ufg,Creminelli:2022onn,Heller:2022ejw,Creminelli:2024lhd}). However, its implications for Fermi liquids arising from relativistic QFTs has to our knowledge not been explored; we will do so here.

The charge density two-point function features a host of non-analyticities in Fermi liquids. Already in a free Fermi gas (obtained by setting $F^{(2,0)} = 0$ in \eqref{eq_rr_theta_EFT} and integrating over $\theta$), it is given by
\begin{equation}\label{eq_GRnn_free}
G^R_{\rho\rho}(\omega,\vec q)
	= -i \frac{p_F}{2\pi v_F} \left[-1 + \frac{s}{\sqrt{(s+i0^+)^2-1}}\right]\, , \qquad
	s \equiv \frac{\omega}{v_F |\vec q|},
\end{equation}
and features a branch cut signalling the particle-hole continuum $\omega\leq v_F q$. In the presence of interactions, nonzero Landau parameters often lead to additional collective excitations above the particle-hole continuum, illustrated in Fig.~\ref{sfig_collective}, which we will turn to shortly.

Demanding that the branch point in Eq.~\eqref{eq_GRnn_free} satisfy \eqref{eq_causality_analyticity} requires the Fermi velocity to be subluminal $v_F\leq 1$. This is interesting to apply to a conformal Fermi liquid, where the velocity is fixed in terms of Landau parameters: in this case, Eq.~\eqref{eq_vF_CFT} leads to
\begin{equation}\label{eq_F0F1_CFTconstraint}
F_0F_1 + F_0 + F_1 \geq 0\, .
\end{equation}
In the absence of conformal symmetry, this constraint becomes 
\begin{equation}
(1+F_0)(1+F_1) \frac{d \log p_F}{d \log \mu}\geq 1\, , 
\end{equation}
and depends on the equation of state of the QFT $\rho(\mu)$ or $p_F(\mu)$.

Collective modes, illustrated in Fig.~\ref{sfig_collective}, correspond to poles in the Green's function which must also satisfy \eqref{eq_causality_analyticity}. As they are by definition faster than $v_F$, requiring them to be subluminal will lead to stronger constraints. The velocity of collective modes is a complicated nonlinear function of the $F_\ell$'s -- we will therefore make simplifying assumption to study them below.

\begin{figure}
\centerline{
	\begin{overpic}[width=0.5\linewidth,trim=0 -10 0 0,clip]{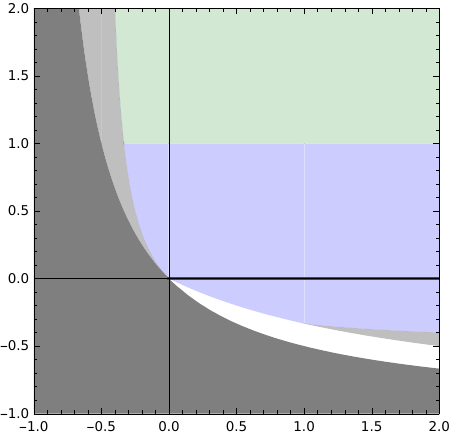}
	\put(50,88){\small\color{ForestGreen} shear sound}
	\put(60,81){\small\color{ForestGreen} +}
	\put(50,74){\small\color{ForestGreen} zero sound}
	\put(50,53){\small\color{blue} zero sound}
	\put(50,88){\small\color{ForestGreen} shear sound}
	\put(60,81){\small\color{ForestGreen} +}
	\put(50,74){\small\color{ForestGreen} zero sound}
	\put(50,53){\small\color{blue} zero sound}
	\put(-5,60){\small $F_1$}
	\put(57,0){\small $F_0$}
	\end{overpic}
}
\caption{\label{fig_causality_collective_2d} Causality constraints on conformal Fermi liquids in $d=2$ spatial dimensions, in the space of the first Landau parameters $F_0,F_1$. The dark gray region is excluded by \eqref{eq_F0F1_CFTconstraint}. Stronger constraints arise from demanding that collective excitations be causal; if higher Landau parameters are negligible, the constraint \eqref{eq_subluminal_ZS} leads to the light-gray exclusion region. The remaining parameter space either features no coherent sound mode (white), a zero sound mode when \eqref{eq_ZS_exists} is satisfied (blue) or both zero sound and shear sound when $F_1\geq 1$ (green). The red line $(F_0,F_1)\in 0\otimes \mathbb R_+$ shows the values realized in the class of CFTs considered in Sec.~\ref{ssec_CS}.}
\end{figure}
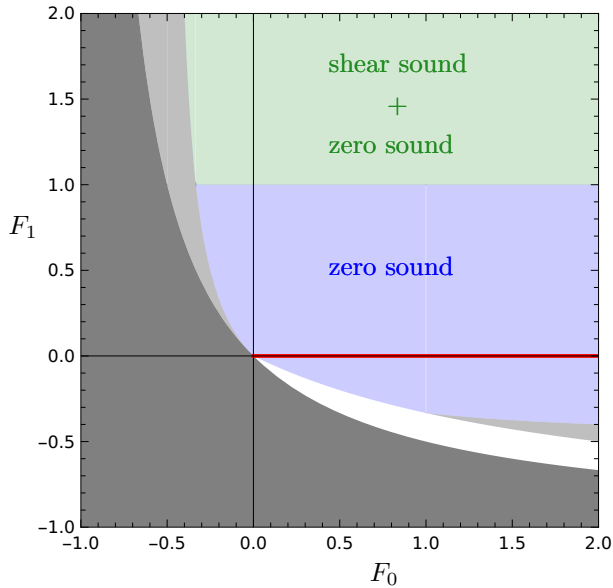

\subsubsection*{Collective modes with $F_{\ell \geq 2} = 0$}

These interesting symmetry constraints on $F_0$ and $F_1$ suggest investigating collective (zero sound) excitations in this system. For simplicity, we will assume all other Landau parameters are zero $F_{\ell \geq 2} = 0$. The equation of motion from \eqref{eq_S_gaussian} is
\begin{equation}
(s - \cos\theta) \rho_\theta
	= \cos\theta \int \frac{d\theta'}{(2\pi)^2} F^{(2,0)}(\theta-\theta')  \rho_{\theta'}\, , 
\end{equation}
where we Fourier transformed and set $s\equiv \omega/(v_Fq)$. Because we are looking for a collective excitation above the particle-hole continuum, we shall assume $s>1$. Dividing by the factor $s-\cos \theta$, one then obtains
\begin{equation}\label{eq_solve_for_collective}
\rho_\theta
	= \frac{\cos\theta}{s - \cos\theta}\int \frac{d\theta'}{(2\pi)^2} F^{(2,0)}(\theta-\theta')\rho_{\theta'}\, .
\end{equation}
A collective excitation will exist if and only if this equation has a solution. For our choices of Landau parameters, $F(\gamma) = F_0 + 2F_1 \cos\gamma$, the solution must take the form
\begin{equation}
\rho_\theta(s)
	= \frac{\cos\theta}{s-\cos\theta}
	\left(u_0(s) + e^{i\theta}u_1(s) + e^{-i\theta}u^*_1(s)\right)\, .
\end{equation}
Inserting this expression back in Eq.~\eqref{eq_solve_for_collective} leads to the linear system
\begin{equation}
\left(\begin{array}{c}
u_1\\
u_0\\
u^*_1
\end{array}\right)
=
\left(\begin{array}{ccc}
F_1 I_0 & F_1 I_1 & F_1 I_2\\
F_0 I_1 & F_0 I_0 & F_0 I_1\\
F_1 I_2 & F_1 I_1 & F_1 I_0
\end{array}\right)
\left(\begin{array}{c}
u_1\\
u_0\\
u^*_1
\end{array}\right)\, , 
\end{equation}
where
\begin{equation}
I_n \equiv \int \frac{d\theta}{2\pi} \frac{\cos \theta}{s-\cos \theta} e^{in\theta}
	= \frac{(s-\sqrt{s^2-1})^n}{\sqrt{s^2-1}} \left(s - \delta_n^0 \sqrt{s^2 - 1}\right)\, .
\end{equation}
A solution exists if the linear system is singular, which leads to a zero sound mode when
\begin{equation}\label{eq_ZS_exists}
F_0>-\frac{2F_1}{1+F_1}\, , 
\end{equation}
with speed
\begin{equation}
\frac{v_0}{v_F} = 
\frac{\sqrt{(F_1+1) \left(2 F_0 F_1+\sqrt{F_1+1}
   \sqrt{4 F_0 ((F_0+3) F_1+F_0+1)+9 F_1+1}-2
   F_0+3 F_1-1\right)}}{2 \sqrt{2 F_1}}
\end{equation}
Furthermore, when $F_1>1$, a second solution exists: shear sound \cite{Abrikosov_1959,pinesnozieres2018theory,pitaevskii1980statistical}%
	\footnote{See Refs.~\cite{Conti_1999, Khoo:2018nel, Valentinis_2021} for discussions of shear sound in metals. The possibility of collective excitations in non-Fermi liquids has also been explored in \cite{Mandal:2021fbg, Khveshchenko:2024vid}.}, with velocity 
\begin{equation}
\frac{v_\perp}{v_F} 
	=  \frac{1 + F_1}{2 \sqrt{F_1}}\, .
\end{equation}
This mode is always slower than zero sound. Requiring zero sound to be subluminal, for the case of a CFT where $v_F = 1/\sqrt{(1+F_0)(1+F_1)}$ leads to the constraint
\begin{equation}\label{eq_subluminal_ZS}
2F_0F_1 + F_0 + F_1 \geq 0\, ,
\end{equation}
which is a little stronger than Eq.~\eqref{eq_F0F1_CFTconstraint}; in particular it requires $F_0\geq -\frac12$. Note that \eqref{eq_subluminal_ZS} can only be imposed when the zero sound mode exists, i.e.~when Eq.~\eqref{eq_ZS_exists} is satisfied. The excluded region is therefore disconnected, and is shown in light gray in Fig.~\ref{fig_causality_collective_2d}.

\subsubsection*{Strong coupling limit}\label{ssSCL}

It is interesting to consider the limit where $v_F$ becomes small. For a CFT, this requires $F_0$ or $F_1$ to become large, i.e.~the Fermi liquid is strongly coupled. Taking them to infinity with a fixed ratio, one finds that $v_F\to 0$ and $v_{\perp}\to 0$ but 
\begin{equation}
\lim_{F_0,F_1\to \infty} v_0
	= \frac{1}{\sqrt{2}}\, , 
\end{equation}
which matches the conformal speed of sound of a superfluid. One can furthermore check that zero sound carries most of the spectral weight of the density two-point function in this limit. The system therefore behaves like a superfluid.
A similar conclusion holds if only $F_0\to \infty$. If one instead takes $F_1\to \infty$ with $F_0$ finite, then both modes retain a finite velocity sandwiching $1/\sqrt{2}$
\begin{equation}
\lim_{F_1\to \infty} v_0
	= \frac{1}{2} \sqrt{\frac{3+2F_0}{1+F_0}}\, , \qquad
	\hbox{and} \qquad v_{0}^2 = \frac12 + v_\perp^2\, .
\end{equation}
This second equation matches  the shear and longitudinal sound modes of a conformal solid \cite{Esposito:2017qpj}. In this limit, the Fermi liquid behaves like a solid. It is interesting that superfluids and solids appear to be in the `moduli space' of Fermi liquids.

\section{Nonlinear EFT of Fermi liquids}\label{sec_nonlin_EFT}

Our discussion so far has revealed certain symmetry and causality constraints on the familiar Landau parameters of Fermi liquids. Elevating this to a systematic analysis requires a controlled effective field theory (EFT), capable of capturing power-law corrections to Fermi liquids beyond Eq. ~\eqref{eq_S_gaussian}. Such effective field theories, written in terms of fermionic quasiparticles, were elegantly laid out by Polchinski and Shankar \cite{Polchinski:1992ed,Shankar:1993pf}. While the symmetry analysis we will perform could be carried out in that formalism, we will instead make use of another recently proposed approach: nonlinear bosonization of Fermi surfaces \cite{Delacretaz:2022ocm}, which provides the nonlinear completion to Eq.~\eqref{eq_S_gaussian}. The advantage of bosonization is that scaling is simpler: the momentum of bosonic excitations does not scale to the Fermi wavevector at low energies. Moreover, partial cancellations in fermion loops make nonlinearities smaller than expected \cite{metzner1997Fermisystemsstrongforward}, a feature that is manifest in nonlinear bosonization.

However, bosonization of Fermi surfaces comes with a cost: while classical effects are well-captured, quantum corrections in $d>1$ require subtle regularization to make sense of a quantum field living in phase space \cite{PhysRevB.49.10877,Houghton:2000bn} (instead, they are well understood in $d=1$ \cite{pereira2007dynamical,RevModPhys.84.1253}). We will avoid this issue by studying the leading behavior of correlation functions at small frequencies and wavevectors $\omega/v_F, q \ll p_F$.

\subsection{Landau's Kinetic Equation}

Fermi liquids, in the semiclassical limit, are described by a nonlinear version of the collisionless Boltzmann equation, also known as Landau's kinetic equation \cite{pinesnozieres2018theory,pitaevskii1980statistical}
\begin{equation}\label{eq_Landau_kinetic_eq}
\begin{split}
\d_t f &+ \{ f, \tilde\epsilon[f] \} = 0,\\
\tilde\epsilon[f] = \epsilon(\p) + &2\int \frac{d^d p}{(2\pi)^d} F^{(2,0)}(\p,\p') \delta f(t,\x,\p') + \cdots,
\end{split}
\end{equation}
where $\{F,G\} = \nabla_x F \cdot \nabla_p G - \nabla_p F \cdot \nabla_x G$ is the Poisson bracket, $f(t,\x,\p)$ is the single-particle distribution function, $\delta f$ is the fluctuation of this distribution function from its ground state value, $\epsilon(\p)$ is the free particle dispersion relation, and $F^{(2,0)}(\p,\p')$ is a phenomenological function, known as Landau's interaction function, which characterizes two-body interactions. Its superscript will be explained below as we generalize this equation.

We will see shortly that Eq.~\eqref{eq_Landau_kinetic_eq} can reproduce the linearized results obtained in Sec.~\ref{sec_linear}. In reality, however, Landau's Fermi liquid theory should be viewed as an expansion in fluctuations and spatial derivatives: indeed, no symmetry forbids $O(\delta f^2)$ terms or higher gradient terms to appear in $\tilde\epsilon[f]$. These are ignored in most treatments of Fermi liquids, because they only produce subleading corrections to linear response observables such as two-point functions. However, they contribute to the leading behavior of nonlinear observables, and should be incorporated in any systematic EFT of Fermi liquids. We will include them below after introducing an action principle for Eq.~\eqref{eq_Landau_kinetic_eq}.

\subsection{Fermi liquids in action}

Ref.~\cite{Delacretaz:2022ocm} showed that Landau's kinetic equation \eqref{eq_Landau_kinetic_eq} can be obtained as a variation principle from the action
\begin{equation}\label{eq_S2_coadj}
\begin{split}
S
	&= \int dt \Tr \left[f_0 U^{-1} \left(\partial_t - \epsilon\right)U\right]\\
	&- \int \frac{dt d^dx d^dp d^d p'}{(2\pi)^{2d}} F^{(2,0)}(\vec p,\vec p') \delta f(\vec x,\vec p)\delta f(\vec x,\vec p') + \cdots\, .
\end{split}
\end{equation}
In the first term, similar to the Berry phase term for ferromagnets or Wess-Zumino-Witten (WZW) terms more generally, we have used a matrix notation: objects in the trace are elements of the Lie algebra of canonical transformations, whose commutators are the Poisson brackets encountered above. The trace over an element of the algebra is defined as
\begin{equation}\label{trace_eq}
\Tr \left[ A\right]
	\equiv \int \frac{d^d x d^dp}{(2\pi)^d} A(\vec x, \vec p).
\end{equation}
We have also defined the group element $U$ that takes a chosen reference state $f_0$ to the true distribution function
\begin{equation}\label{eq_f_U}
f = U f_0 U^{-1}\, .
\end{equation}
With these definitions at hand, trace expressions can be handled similarly as in nonlinear sigma models. For example, the dispersion term $\epsilon$ can be written
\begin{equation}
\int dt \Tr \left[f_0 U^{-1}\epsilon U\right]
	= \int dt \Tr \left[f \epsilon\right]
	=\int \frac{dt d^dx d^dp }{(2\pi)^{d}} f(t,\vec x,\vec p) \epsilon(\vec p)\, .
\end{equation}
We have assumed translation invariance, so that $\epsilon(\vec x,\vec p) = \epsilon(\vec p)$.
These definitions find a more formal underpinning in the framework of coadjoint orbits, see \cite{Delacretaz:2022ocm,Mehta:2023cwi} for details.%
    \footnote{See also \cite{Khveshchenko_1995} for earlier work on the bosonization of Fermi surfaces using coadjoint orbits. Related work includes \cite{Das:1991uta,Park:2023coa}, where particle-hole excitations in the entire Fermi sea are kept in the description, and \cite{Sakita:1996ne,Karabali:2003bt}
 where similar constructions were considered in the context of quantum Hall states. }
The only point that is relevant for our purposes is that the equations of motion leads to Landau's kinetic equation \eqref{eq_Landau_kinetic_eq}. To establish this, it is important to note that the degree of freedom is not an arbitrary distribution function $f(t,\vec x,\vec p)$, but rather the subset that can be obtained from phase-space preserving deformations of an arbitrary reference state $f_0$. These can be parametrized as \eqref{eq_f_U}, so that the action must be varied with respect to $U \to e^{\delta \alpha}U $. The equation of motion $\delta S / \delta \alpha(t,\vec x,\vec p) = 0$ then leads to \eqref{eq_Landau_kinetic_eq}. 
Note that an (arbitrary) reference state $f_0$ must be specified to express the WZW term, even though this reference state does not enter in the equation of motion.%
	\footnote{As usual with WZW terms, the action can be made manifestly independent of $f_0$ when expressed in one higher dimension.}

At the classical level, the action \eqref{eq_S2_coadj} is therefore equivalent to Landau's kinetic theory \eqref{eq_Landau_kinetic_eq}. However, the action formulation has a number of advantages over the equation of motion. Already classically, it allows for a simpler implementation of symmetries, which we will make use of below. At the quantum level, the normalization of the action $S$ contains more information than its equation of motion $\delta S = 0$. In $d=1$, this reproduces nonlinear bosonization beyond Luttinger liquids \cite{RevModPhys.84.1253}. In higher dimensions, a full quantum treatment remains to be established (see \cite{PhysRevB.49.10877,Houghton:2000bn,Ye:2024osp} for partial results in that direction).

The generalization of Landau's Fermi liquid theory to higher orders in fluctuations and fields is now straightforward: one allows the action to take the most general form as a double expansion in fluctuations $\delta f = f - f_0$ as well as spatial derivatives $\nabla_x / p_F$, with $p_F$ the expected UV cutoff scale of the EFT. Assuming both translational and rotational invariance, the Hamiltonian takes the form
\begin{equation}\label{eq_general_H}
\begin{split}
H[f] &=\int_{\x\p} \epsilon(\p) f(\x,\p),\\
&+ \int_{\x\p\p'} F^{(2,0)}(\p,\p') \delta f(\x,\p) \delta f(\x,\p') + \vec{F}^{(2,1)}(\p,\p') \cdot \nabla_x \delta f(\x,\p) \delta f(\x,\p') + \ldots\, ,\\
&+ \int_{\x\p\p'\p''} F^{(3,0)}(\p,\p',\p'') \delta f(\x,\p) \delta f(\x,\p') \delta f(\x,\p'') + \ldots\, ,
\end{split}
\end{equation}
with $\int_{\vec x}\equiv \int d^dx$ and $\int_{\vec p} \equiv \int \frac{d^dp}{(2\pi)^d}$. The functions $\epsilon, F^{(m,n)}$ are the Wilson coefficients of this theory. $F^{(m,n)}$ denotes such a term entering at $m$th order in fluctuations, and $n$th order in derivatives, with $F^{(2,0)}$ being Landau's interaction function. These arbitrary functions are really ``Wilson coefficient functions'' and encode towers of Wilsonian couplings, as can be seen by expanding in fluctuations, which we turn to next. In summary, the action to nonlinear orders in the phase-space density is explicitly

\begin{align}\label{eq_action_f}
S=&\int dt \Tr \left[f_0 U^{-1} \left(\partial_t - \epsilon\right)U\right],\nonumber\\
	&- \int_{t\x\p\p'} F^{(2,0)}(\p,\p') \delta f(\x,\p) \delta f(\x,\p') + \vec{F}^{(2,1)}(\p,\p') \cdot \nabla_x \delta f(\x,\p) \delta f(\x,\p') + \ldots\, ,\nonumber\\
&- \int_{t\x\p\p'\p''} F^{(3,0)}(\p,\p',\p'') \delta f(\x,\p) \delta f(\x,\p') \delta f(\x,\p'') + \ldots\, .
\end{align}
Note that $F^{(2,1)}$ must break time-reversal symmetry, which acts on the distribution function as $f(t,\vec x,\vec p)\to f(-t,\vec x,-\vec p)$. It is interesting that this term enters at the same order in derivatives as the Berry curvature in $d=2$ \cite{Son:2012zy,Chen:2016fns}. For time-reversal invariant Fermi liquids, higher gradient corrections instead start at $O(\nabla^2)$ and take the schematic form $F^{(2,2)}_{ij} \partial_i \delta f \partial_j \delta f'$. Because we focus on the leading low momentum response functions below, we will not consider such gradient corrections in this paper.

\subsection{Expansion for fluctuations}

In order to study observables with the EFT, which will allow us to match its Wilsonian coefficients with microscopic models, it is necessary to expand it in fluctuations. We will proceed similarly to nonlinear sigma models. Since we are focusing on isotropic systems, we will expand around a spherical (or circular) Fermi surface
\begin{equation}\label{eq_f0}
f_0(p) = \Theta(p_F - |\vec p|)\, .
\end{equation}
The most general distribution function $f(t,\vec x,\vec p)$ that is reachable from \eqref{eq_f0} through canonical transformations can be written as Eq. \eqref{eq_f_U}. Parametrizing $U$ in terms of an element of the algebra of canonical transformations $\phi\in \mathfrak g_{\rm can}$ as $U\equiv e^{-\phi}$,\footnote{Note in particular that in the matrix notation we're using here, the exponential $e^{-\phi}$ is not to be confused with the function $k(\x,\p) = e^{-\phi(\x,\p)}$, but rather corresponds to the exponent map that takes an element of the Lie algebra of canonical transformations to a finite canonical transformation that it generates.} one can expand in fluctuations as
\begin{equation}\label{eq_fexpansion0}
\begin{split}
f &\equiv U f_0 U^{-1} 
	= f_0 - \{ \phi, f_0 \} + \frac{1}{2} \{ \phi, \{ \phi, f_0 \} \} + \cdots\,.
\end{split}
\end{equation}
One can check that $f$ is indeed a function that only takes values $0$ and $1$ by observing that the above series takes the form of a Taylor expansion. The orbit of $f_0$ under arbitrary canonical transformations is the set of sharp, deformed Fermi surface of a fixed volume. Indeed, the above series takes the form of a Taylor expansion of $\Theta \left(p_F(t,\vec{x}, \theta)- |\vec{p}|\right)$.

Not all canonical transformations generate different states from $f_0$ though, since the stabilizer subgroup or subalgebra of the ground state is nontrivial
\begin{equation}
\begin{split}
H &= \{ V ~|~ Vf_0V^{-1} = f_0 \} \subset G_{\rm can}\, , \\
\h &= \{ \alpha ~|~ \{ \alpha, f_0 \} = 0 \} \subset \g_{\rm can}\, .
\end{split}
\end{equation}
Here $G_{\rm can}$ is the group of canonical transformations, and $H$ the subgroup that leaves $f_0$ invariant. Similarly, $\h \subset \g_{\rm can}$ is the corresponding subalgebra.
The degree of freedom therefore lives in the coset (coadjoint orbit) $G_{\rm can} / H$. The equivalence relation on $\phi$ reads as
\begin{equation}
e^{-\phi} \sim e^\alpha e^{-\phi}, \qquad\hbox{or} \qquad
\phi \sim \phi - \alpha + \frac{1}{2} \{ \phi, \alpha \} + \ldots, \qquad \alpha \in \h\, .
\end{equation}
This gauge invariance allows us to restrict $\phi(t,\x,\p)$ to its value on the Fermi surface
\begin{equation}
\phi(t, \x,\p) = \phi (t, \x,\theta)\, ,
\end{equation}
where $\theta$ denotes the $d-1$ angles on the Fermi surface,
by choosing an appropriate $\alpha \in \h$. For example, to leading order in fields, the appropriate choice is $\alpha = \phi - \phi|_{p_F}$, which indeed satisfies $\{\alpha , f_0\} = 0$.

We are now ready to expand in fluctuations. The distribution function Eq.~\eqref{eq_fexpansion0} takes the form
\begin{equation}\label{eq_fexpansion}
\begin{split}
f 
	&= \Theta(p_F - |\vec p|) + \nabla_n \phi \delta(|\p|-p_F)\\
&+ \frac{1}{2}\left[\frac{1}{p_F}\left(\partial_{\theta^i}\phi \nabla^i_s\nabla_n \phi - \nabla^i_s \phi \partial_{\theta^i}(\nabla_n \phi)\right)\delta(|\p|-p_F) - (\nabla_n \phi)^2\delta'(|\p|-p_F)\right] \\
&+ O(\phi^3),\\
\end{split}
\end{equation}
where $\nabla_n\phi = \n\cdot\nabla_\x\phi, \nabla^i_s\phi = \s^i\cdot\nabla_\x\phi$ and $\n(\theta)$ and $\s^i(\theta)$ denotes the normal and $d-1$ tangent vectors to the Fermi surface. The action, up to cubic order in $\phi$ and leading order in derivatives (i.e., keeping only terms $F^{(n,m)}$ with $n\leq 3$ and $m=0$ in Eq.~\eqref{eq_action_f}), takes the following form
\begin{equation}\label{eq_action_phi}
\begin{split}
S = ~ &- \frac{p_F^{d-1}}{2} \int_{t\x\theta} \nabla_n\phi \left( \dot{\phi} + v_F \nabla_n \phi + v_F \int_{\theta'} F^{(2,0)}(\theta,\theta') (\nabla_n\phi)' \right)\\
&- \frac{p_F^{d-2}}{3!} \int_{t\x\theta} \nabla_n\phi \left[ (\nabla^i_s\phi) (\d_{\theta^i} \dot{\phi}) - (\nabla^i_s\dot{\phi}) (\d_{\theta^i} \phi) \right] + \left( \frac{d-1}{2}v_F + p_F \epsilon'' \right) (\nabla_n\phi)^3\\
&- \frac{p_F^{d-2}}{2}\int_{t\x\theta\theta'}\! v_F\biggl\{ F^{(2,0)}(\theta,\theta') \Big[ \nabla^i_s (\nabla_n\phi \d_{\theta^i}\phi) (\nabla_n\phi)' \Big]\\
& \qquad \quad + \d_{\theta^i} F^{(2,0)}(\theta,\theta') \Big[ (\nabla_n\phi \nabla^i_s\phi) (\nabla_n\phi)'\Big] +  F_1^{(2,0)}(\theta,\theta') \left[ (\nabla_n\phi)^2(\nabla_n\phi)'\right]\biggr\}\\
&- {p_F^{d-2}} \int_{t\x\theta\theta'\theta''} F^{(3,0)}(\theta,\theta',\theta'') (\nabla_n\phi)(\nabla_n\phi)'(\nabla_n\phi)'' + \cdots\, .
\end{split}
\end{equation}
where $\int_{t\vec x}\equiv \int dt d^dx$ and $\int_\theta\equiv \frac{d^{d-1}\theta}{(2\pi)^d}$, and the $'$ superscript indicates evaluation of all fields at angle $\theta'$, e.g.~$(\nabla_n\phi)' = \hat n(\theta')\cdot\nabla\phi(t,\vec x,\theta')$. The Wilsonian coefficients appearing in the action are derivatives of the dispersion $v_F \equiv \epsilon'(p_F)$ and $\epsilon''(p_F)$, as well as generalized Landau parameters: $F^{(2,0)}(\theta,\theta') = F^{(2,0)}(\theta-\theta') \equiv 2p_F^{d-1} F^{(2,0)}(p_F \n, p_F \n')/v_F$ and $F^{(2,0)}_1 \equiv \frac{2p_F^{d}}{v_F} \d_{|\p|} \left( F^{(2,0)}(\p,\p') \right)_{p_F}$. $F^{(3,0)}$ has also similarly been evaluated at the Fermi surface and rescaled as $F^{(3,0)}(\theta,\theta',\theta'') \equiv p_F^{2d-1} F^{(3,0)}(p_F \n, p_F \n', p_F \n'')$.  The ellipsis denotes higher derivative and $O(\phi^4)$ terms. Each term in this expanded action is scale-covariant, and one can use simple power counting to determine the classical scaling dimension of all the interactions. Scaling $\omega\sim q$ and $\phi\sim q^{(d-1)/2}$, one finds that the entire Gaussian action, in the first line is marginal\footnote{This is different from engineering dimensions of the parameters necessary for making the action a dimensionless phase. In particular, $p_F$ has has the same engineering dimensions as momenta but doesn't scale under RG.}. This is the action that we already considered in Sec.~\ref{sec_linear} -- in particular the usual Landau parameters $F^{(2,0)}$ are marginal as expected. The remaining cubic terms are irrelevant and scale as $S^{(3)}/S^{(2)}\sim \nabla\phi \sim q^{(d+1)/2}$. The other generalized Landau parameters $F^{(m,n)}$ are therefore irrelevant, but some will contribute to the leading order nonlinear response. For example, $F^{(3,0)}$ will contribute to the leading density three-point function of a Fermi liquid, as evidenced by the fact that it is no more irrelevant than other cubic terms in the EFT. These higher Landau parameters are allowed by symmetry, and are therefore generically non-zero. In fact some of them can be inferred, say in helium-3, through the non-trivial density and pressure dependence of regular Landau parameters. How the density dependence of familiar parameters are related to some generalized Landau parameters is discussed in App.~\ref{3ptcalceft} and we quote the result below (see Eq.~\eqref{change_F20_mu} for the precise derivation) 
\begin{equation}\label{change_F20_mu_2}
\begin{split}
\frac{\partial v_F}{\partial \mu}=\frac{ \epsilon''_{F}}{v_F}(1-F_0) + \frac{\tilde{F}_0}{p_F}\\
\partial_\mu\left[v_F p_F F^{(2,0)}(\theta, \theta')\right]&=\left[2 F^{(2,0)}(\theta, \theta')+2 F_1^{(2,0)}(\theta,\theta')+\frac{6}{v_F} \int_{\theta''} F^{(3,0)}(\theta,\theta',\theta'')\right].
\end{split}
\end{equation}

Let us briefly comment on some advantages of this EFT compared to the more conventional fermionic EFTs for Fermi liquids \cite{Benfatto:1990zz, Shankar:1993pf, Polchinski:1992ed}. The latter is formulated in momentum space in terms of a fermionic (quasiparticle) degree of freedom $\psi(t,\p)$. The scaling of momentum under RG is nonlinear since it scales radially towards the Fermi surface instead of towards the origin of momentum space. Due to this, individual interactions need to be made nonlocal in position space for an efficient scaling analysis, for example, the four fermion interaction is written as
\begin{equation}
    \int_{\p_1\p_2\p_3\p_4} V\left(\p_{F1},\p_{F2},\p_{F3},\p_{F4}\right) \psi^\dagger(\p_1)\psi^\dagger(\p_2)\psi(\p_3)\psi(\p_4) \delta\left(\sum_i\p_i\right)\, .
\end{equation}
In particular, the interaction potential $V$ isn't evaluated at the momenta $\p_i$, but instead at the intersections $\p_{Fi}$ of those momenta with the Fermi surface in momentum space, making the term non-local. Even with this approximation, this interaction is not scale-covariant. It's scaling dimension depends crucially on whether $\sum_i \p_{Fi}$ vanishes or not. When it does, the interaction is classically marginal and corresponds to either the 1-loop marginal quadratic interaction function $F^{(2,0)}(\theta,\theta')$, or a marginally (ir)relevant BCS interaction, depending on the sign of $V$. For other configurations of $\p_{Fi}$, the interaction is strictly irrelevant, making a power counting approach highly nontrivial, especially when applied to higher order interactions. Eventually the BCS interactions destabilize the Fermi liquid phase so that none of this applies at very low $\omega, p$. So our results (as any other Fermi liquid results) apply in the intermediate phase where the system is a Fermi liquid and not a superfluid.

An improvement to this EFT was provided recently in \cite{Borges2023,Ma2024} where the couplings functions are no longer restricted to the Fermi surface, which seems to reveal other instabilities to the theory that the Shankar-Polchinski EFT does not capture. The consequences of fully momentum-dependent couplings at the level of EFT are also captured by the coadjoint orbit EFT in Eq.~\eqref{eq_action_f} via the momentum dependence in the various interaction functions $F^{(m,n)}(\p_1,\ldots,\p_m)$, resulting in additional couplings such as $F^{(2,0)}_1(\theta,\theta')$ in Eq.~\eqref{eq_action_phi}.

Furthermore, the fermion EFT suffers from subtle cancellations in nonlinear correlation functions of bosonic operators such as the $U(1)$ current, rendering power counting impossible \cite{metzner1997Fermisystemsstrongforward}. The coadjoint orbit EFT, however, makes the scaling of correlation functions transparent, which simplifies the calculation of certain correlation functions.

\subsection{Operator matching in the EFT}\label{EFTops}

The primary observables in our effective field theory (EFT) are the correlation functions of operators. In the interacting EFT, these correlation functions depend on the Landau parameters, which encode our limited understanding of the ultraviolet (UV) physics. As is typical in EFT, one can compute the correlation functions of the EFT operators to extract information about these parameters by comparing with analogous microscopic calculations. However, determining the EFT counterpart of a given microscopic operator is generally nontrivial, except in special cases such as conserved currents. In this section, we identify the EFT counterpart of several microscopic operators of physical relevance to our study, and discuss their leading correlators at small external momenta. 

\subsubsection*{$U(1)$ current}
The first operator we will be interested in is the conserved $U(1)$ current $j_\mu$. The simplest way to obtain $j_\mu$ in the EFT is to gauge the symmetry, as described in \cite{Delacretaz:2022ocm, Mehta:2023cwi}:
\begin{align}\label{eftbgf}
S_{\rm WZW}(A)&= \int dt \Tr  \left[f_0 U^{-1}(\partial_t+A_0)U\right],\\ \notag
S_H[f,A] &=\int_{t,\x,\p} \epsilon(\p) \delta f_A(t, \x,\p) + \int_{t,\x,\p,\p'} F^{(2,0)}(\p,\p') \delta f_A(t, \x,\p) \delta f_A(t, \x,\p') + \cdots\, ,
\end{align}
where $\delta f_A(t, \x,\p)= f(t,\x,\p+\vec{A})- f_0(\p)$ and $f_0(\vec p)$ given by Eq.~\eqref{eq_f0}. Viewing $A_\mu(t,\vec x)$ as a background gauge field and differentiating the action with respect to it yields the $U(1)$ current: 
\begin{equation}\label{u1eftops}
\begin{split}
\rho \equiv j^0 &= \int_{\p} f(t,\x,\p), \\
j^i & = \int_{\p} \left[\nabla^i_\p \epsilon(|\p|) f(t,\x,\p) -2\int_{\p'} F^{(2,0)}(\p, \p')\nabla^i_\p f_0(\p) \delta f(t,\x,\p')\right.\\
&\left.+ \int_{\p'}\left( 2\nabla^i_\p F^{(2,0)}(\p, \p')- 3\int_{\p''}  F^{(3,0)}(\p, \p',\p'')\nabla^i_{\p''} f_0(\p'')\right)\delta f(t,\x,\p) \delta f(t,\x,\p')\right]\\
& + O(\nabla^2\delta f, \delta f^3)\, .
\end{split}
\end{equation}
{The currents and densities constructed in this manner satisfy the $U(1)$ charge conservation equation $\partial_\mu j^\mu=0$ upon using the equations of motion.} 
These can also be expanded in fluctuations, using Eq.~\eqref{eq_fexpansion}. For example, the density  operator  up to quadratic order in $\phi$ is given by,
\begin{equation}\label{rhoeft}
\rho(t,\x)= {p_F^{d-1}}\int \frac{d^{d-1}\theta}{(2\pi)^d} \left[
	 \nabla_n \phi +  \frac{1}{2p_F}\nabla_{s^i} \left(\partial_{\theta_i} \phi \nabla_n \phi\right)
\right] + O(\phi^3)\, .
\end{equation}
%

\subsubsection*{Stress tensor}\label{Tmunu_EFT}

In principle, the stress tensor $T_{\mu\nu}$ can be similarly obtained by coupling the EFT to a background metric. We have found it simpler to instead obtain the stress tensor using the Noether method. We start with spatial translations, whose action on fields can be conveniently implemented with a canonical transformation 
\begin{equation}\label{noethert}
W_x= e^{ \vec{a}(t,\x)\cdot \p},\qquad  f \rightarrow  W_x\, f\, {W_x}^{-1}.
\end{equation}
This allows us to directly implement this transformation at the level of the action in terms of the phase space density $f$ in Eq.~\eqref{eq_action_f}, without explicitly expanding in fluctuations $\phi$. 
Under this canonical transformation, the phase-space density changes as follows
\begin{equation}
\begin{split}
\delta_a f_p \equiv W_x\, f_p\, {W_x}^{-1}-f_p
    &\simeq \{\vec{a}\cdot \p, f(t,\p,\x)\}\\
&=\partial_i\vec{a} \cdot \vec p\,  \partial_{p_i} (\delta f_p+f_p^0)-\vec{a}\cdot \nabla_{\x} \delta f_p ,\\
\end{split}
\end{equation}
We are keeping the $p$ dependence of $f$ explicit as it will be useful below, i.e., $f_{p}\equiv f(t,\x,\p),~\delta f_{p}\equiv f(t,\x,\p)- \Theta(p_F-|\p|)$ and $f^0_p \equiv f_0(\p)$.

As a result, the change in the Hamiltonian and the WZW term are
\begin{subequations}\label{nothert_H}
\begin{align}
\delta_a S_H &= -\int_{t\x\p} \left[\epsilon_p \delta_a f_p + 2\int_{\p'} F^{(2,0)}(\p,\p') \delta_af_p \delta f_{p'} + 3\int_{\p'\p''} F^{(3,0)}(\p,\p',\p'') \delta_af_p \delta f_{p'} \delta f_{p''}\right]\nonumber\\
&\quad +O(\delta f^3),\nonumber\\
\delta_a S_{\rm WZW} &=\int_{t,\x,\p} \partial_t \vec{a}\cdot \p f_p.   
\end{align}
\end{subequations}
The stress tensors $T^{ij}$ and $T^{ti}$ (where $i,j$ are spatial indices) generating the corresponding Noether currents are obtained using standard Noether procedure
\begin{equation}\label{Tijnoether}
\begin{split}
T^{ij}&=\int_{\p} \left\{p^j \nabla^i_{\p} \epsilon_p f_p+\int_{\p'}\left( -2{p}^j F^{(2,0)}(\p,\p')  \nabla^i_{\p}\left(\delta f_p+f^0_{p}\right) - \delta^{ij} F^{(2,0)}(\p,\p')\delta f_p \right.\right.\\
&\left.\left. -3\int_{\p''} p^j F^{(3,0)}(\p,\p'\p'') \nabla^i_\p f^0_p\delta f_{p''}\right)\delta f_{p'}\right\} + O(\delta f^3),\\
T^{ti}&= \int_\p p^i f_p  \, .
\end{split}
\end{equation}
We now turn to time translations. Unlike spatial translations, implementing time translations as canonical transformations is not possible. Therefore, we explicitly examine the transformation of the field $\phi$ under an infinitesimal time translations as follows:
\begin{equation}
\phi(t,\x,\theta_i) \rightarrow  \phi(t,\x,\theta_i) + a(t,x) \partial_t \phi(t,\x,\theta_i) \, .
\end{equation} 
For convenience, we restrict ourselves to change of the cubic action in  Eq.~\eqref{eq_action_phi} to $O(\phi^2)$. The Noether stress tensors corresponding to time translations are
\begin{equation}\label{noethertimealt}
\begin{split}
{\mathcal T}^{it}&=-\frac{p^{d-1}_F}{2}\int \frac{d^{d-1}\theta}{(2\pi)^d}\hat{n}\left[\dot \phi\left(\dot \phi + 2v_F \nabla_n \phi +2v_F \int \frac{d^{d-1}\theta'}{(2\pi)^d} F^{(2,0)}(\theta,\theta') \nabla_{n'} \phi'\right)\right]+ O(\phi^3)\,, \\
{\mathcal T}^{tt}&=-\frac{p^{d-1}_F }{2}\int \frac{d^{d-1}\theta}{(2\pi)^d} \left[(\nabla_n \phi) \dot \phi\, - L  \right] + O(\phi^3)\, .
\end{split}
\end{equation}
where $L$ denotes the integrand of the Gaussian action with interactions. Along with the gaussian expansion of Eq.~\eqref{Tijnoether}, they satisfy the conservation equations $\partial_\mu T^{\mu \nu}=0$, using the equations of motion. Denote the Hamiltonian density appearing in Eq.~\eqref{eq_general_H} by ${\mathcal H}$. From its expansion in terms of $\phi$, it is evident that 
\begin{equation}
{\mathcal H} = {\mathcal T}^{tt}+ \mu j^0.
\end{equation}
Hence we work with an equivalent set of stress tensors, satisfying the conservation equations
\begin{equation}\label{improved_noether}
T^{tt} = {\mathcal H},\qquad T^{it} = {\mathcal T}^{it}+ \mu j^i\, .
\end{equation}

For imposing symmetry constraints on our action, it will be more convenient to obtain the nonlinear version of $T^{it}$ component of the stress tensor, i.e., as an expansion in $\delta f$ rather than $\phi$. This can be achieved by identifying $T^{ti}$ from  the conservation equation $\partial_\mu T^{\mu t}=0$ with $T^{tt}= \mathcal{H}$; one finds
\begin{equation}\label{Titnoether}
\begin{split}
T^{it}&=  \int_p \left\{\left(\epsilon_p \nabla^i_\p \epsilon_p - 2\int_{\p'} \epsilon(|\p'|) F^{(2,0)}(\p,\p')  \nabla^i_{p'} f^0_{p'}\right)\delta f_p\right.\\
&\left.+\int_{ p'} \left(2 \nabla^i_\p(\epsilon_p F^{(2,0)}(\p,\p'))  -3 \int_{p''}\epsilon_{p''} F^{(3,0)}(\p,\p',\p'') \nabla^i_{p''} f_{p''}^0\right.\right.\\
&\left.\left.- 2 \int_{p''}F^2(\p'',\p') F^{(2,0)}(\p,\p'')\nabla^i_{\p''} f_{p''}^0\right)\delta f_{p'} \delta f_p\right\} + O(\delta f^3)\, ,
\end{split}
\end{equation}
where we use the equation of motion\footnote{One can verify that the expansion in $\phi$ agrees with the Noether construction Eq.~\eqref{improved_noether} up to the relevant order, using equation of motion of $\phi$ from the linearised action. } in terms of the phase space density $f_p$ obtained from the action \eqref{eq_action_f} to leading order in derivatives ($F^{(m,n>0)}=0$)
\begin{equation} \label{eomint}
\begin{split}
\partial_t f_p + \nabla^i_\p \epsilon_p  \nabla_\x f_p &=2 \int_{\p'} \left(F^{(2,0)}(\p,\p') \nabla^i_\x \delta f_{p'}  \nabla^i_\p (\delta f_p+ f^0_p) - \nabla^i_\p F^{(2,0)}(\p,\p') \delta f_{p'}  \nabla^i_\x \delta f_p\right)\\
&+3 \int_{\p', \p''} \left(F^{(3,0)}(\p,\p',\p'') \nabla^i_\x \left(\delta f_{p'} \delta f_{p''}\right)  \nabla^i_\p (\delta f_p+f^0_p) \right.\\
&\left.- \nabla^i_\p F^{(3,0)}(\p,\p',\p'') \delta f_{p'} \delta f_{p''} \nabla^i_\x \delta f_p\right) + O(\delta f^3)\, .
\end{split}
\end{equation}
Since currents are only conserved on-shell, i.e., upon using equations of motion, this last construction leaves a potential ambiguity in their definition. At the level of correlation functions, such ambiguity manifests itself in the form of contact terms which are analytic in the external momenta and frequencies. In this work we will focus on non-analytic contributions to correlation functions at leading order in small external momenta and frequencies and as such these ambiguities will play no role.\footnote{In \cite{Kourkoulou:2022ajr}, it was shown how to resolve this ambiguity by a modification of the usual Noether procedure- it would be interesting to see if their method can be modified to our case.} 

Similar stress tensors have also been obtained in \cite{BAYM1976527} in a different way. We point out the similarities and differences of our results with them. To quadratic order in fluctuations $\delta f_p$ our Hamiltonian density and $T^{ti}$ component of the stress tensors agree with theirs (Note $T^{ti}_{\text{here}}=T^{it}_{\text{there}}$). We note however their stress tensor components $T^{it}$ and $T^{ij}$ differs from ours due to the absence of contribution from the generalized landau parameter $F^{(3,0)}$ at the quadratic level.     

\subsubsection*{Generic scalar operator}

We can now consider a generic local operator in the EFT. Assuming that it transforms as a scalar under spatial rotations, it must take the form
\begin{eqnarray}\label{genscopeft}
\mathcal{O}(t,\x)= \int_p  \gamma_1(|p|, \p\cdot\nabla_\x, \nabla^2_x) f_p + O(\delta f^2)\, , 
\end{eqnarray} 
where $\gamma_1$ is in principle an arbitrary function of its arguments. Note that any time derivative can be substituted for spatial derivatives using the equations of motion. Indeed, two operators differing by terms proportional to the equations of motion have identical correlation functions, up to contact terms. As explained above, we will focus on non-analytic parts of correlation functions and therefore can freely make use of the equations of motion here. 

As in the rest of this paper, we will focus on observables at leading order in small momenta. We can therefore ignore the dependence of $\gamma_1$ on $\nabla_x$. In summary, the general form of a scalar operator, to leading order in gradients is
\begin{equation}\label{genopf2}
\begin{split}
\mathcal{O}(t,\x)&= \int_\p \gamma_1(|\p|) f_p+ \int_{\p,\p'} \gamma_2\left(\p,\p'\right) \delta f_p \delta f_{p'} + \cdots\, , 
\end{split}
\end{equation}
where we slightly generalized the above to include the most general $O(\delta f^2)$ at zeroth order in gradients. Notice that if one sets $\gamma_1(p)=\epsilon(p)$, $\gamma_2 (\vec p,\vec p') = F^{(2,0)}(\vec p,\vec p')$, this operator corresponds to the Hamiltonian density $\mathcal{O} = \mathcal H$. This is no accident -- the Hamiltonian density was precisely constructed as the most general scalar operator. 

We now expand this operator in fluctuations. Using Eq.~\eqref{eq_fexpansion}, one finds that the operator takes the form
\begin{equation}\label{genopphiexp}
\begin{split}
\mathcal{O}(t,\x)&=p_F^{d-1}\int \frac{ d^{d-1}\theta}{(2\pi)^{d}} \gamma_1 \left(\nabla_n \phi + \frac1{2p_F} \nabla^i_s (\nabla_n\phi\partial_{\theta^i} \phi)\right) + \frac12 \gamma_1'  (\nabla_n \phi)^2\\
&+p_F^{2(d-1)}\int \frac{ d^{d-1}\theta d^{d-1}\theta'}{(2\pi)^{2d}} \gamma_2(\theta - \theta') \nabla_n \phi \nabla_{n'} \phi' + O(\phi^3),
\end{split}
\end{equation}
where $\gamma_1 = \gamma_1(p_F)$, $\gamma_1' = \frac{d}{dp_F}\gamma_1(p_F)$, and we used rotation invariance to set $\gamma_2(p_F\hat n(\theta), p_F\hat n(\theta')) = \gamma_2 (\theta-\theta')$. Integrating over space, this has the identical form as the quadratic Hamiltonian \eqref{eq_S_gaussian}. It also reduces to the density operator Eq.~\eqref{rhoeft} upon setting $\gamma_1(p) = 1,\,\gamma_2 = 0$.

To illustrate these results for a nontrivial scalar operator, we study in Sec.~\ref{freefermionseft}, the microscopic operator $\mathcal{O}=\bar\psi \psi$ for a free (2+1)d Dirac fermion of mass $m$ at finite density. We find that for this operator, $\gamma_2 = 0$, and
\begin{equation}
\gamma_1(p_F) = \frac{1}{\sqrt{1 + v_F^2}}\, ,
\end{equation}
with $v_F = p_F/m$.

\subsection{Nonlinear density correlators}\label{den_corr}

The main observable we will use to match EFT to microscopics will be correlation functions of the charge density $\rho = j^0$. As discussed in the previous section, unlike generic operators the charge density operator and other Noether currents are uniquely fixed in the EFT, which simplifies the matching procedure.

At the Gaussian level, the EFT \eqref{eq_action_phi} reduces to the effective action traditionally used in Fermi surface bosonization, Eq.~\eqref{eq_S_gaussian}. This approximation is sufficient to capture linear response, or two-point functions, at leading order in the long wavelength limit, as was studied in Sec.~\ref{sec_linear}. Matching this two-point function with a given microscopic model allows to identify its Landau parameters. In this section, we turn to three-point functions of the density, whose comparison with microscopics will allow to fix the generalized Landau parameters appearing in Eq.~\eqref{eq_action_phi}. These are in principle straightforward to obtain using Feynman diagrams: cubic vertices arise from the WZW term and dispersion relation (second line in Eq.~\eqref{eq_action_phi}), as well as generalized Landau parameters  (third and fourth lines in Eq.~\eqref{eq_action_phi}). At tree-level---which captures the observable at leading order in small frequencies and momenta---these cubic vertices will only enter diagrams once. The only challenge in establishing a closed form expression to all orders in Landau parameters is then the same as the one encountered in linear response: because regular Landau parameters $F^{(2,0)}$ are marginal (they enter the Gaussian action without derivative suppression), they must be kept to all orders even to obtain the leading observable at low energies.

In the following, we review the calculation of the density three-point function of a free Fermi gas, before turning to the general Fermi liquid. We focus on $d=2$ spatial dimensions for simplicity, but this analysis can carried out similarly in any dimension. 

\subsubsection*{Free Fermi gas}

The density three-point function for a free Fermi gas was obtained using nonlinear bosonization in \cite{Delacretaz:2022ocm}, we summarize their result here. The $\phi$ propagators from the Gaussian action are
\begin{equation}\label{eq_Sphi_free}
\mathcal{S}^{\phi}_{\theta,\theta'}(\omega,q)
	\equiv \int dt d^2x \, e^{i(\omega t - \vec q\cdot \vec x)} \langle \phi(t,\vec x, \theta)\phi(0,0,{\theta'})\rangle 
	= \frac{1}{p_F}\frac{i (2\pi)^2 \delta(\theta - \theta')}{q_n (\omega - v_F q_n)}\, .
\end{equation}
Using the definition of the density Eq.~\eqref{rhoeft}, and the cubic vertices from the second line of Eq.~\eqref{eq_action_phi}, one finds two sets of diagrams shown in fig.~\ref{figfreeeft} that contribute to the connected three-point function, given by
\begin{equation}\label{freepartsrrr}
\begin{split}
\langle \rho_p \rho_q \rho_{-p-q} \rangle_{(0)} &= \int_\theta \left( \langle\rho\rho\rho\rangle^\theta_\text{WZW} + \langle\rho\rho\rho\rangle^\theta_{\rho^{(2)}} + \langle\rho\rho\rho\rangle^\theta_H \right),\\
\langle\rho\rho\rho\rangle^\theta_\text{WZW} &= \frac{1}{3!} \frac{p_n}{\omega_p-v_F p_n} \frac{q_{s}}{\omega_q-v_F q_n} \partial_{\theta} \frac{\omega_p+2\omega_q}{(\omega_p+\omega_q)-v_F (p+q)_n} + \text{Perm},\\
\langle\rho\rho\rho\rangle^\theta_H &= - \left(\frac{1}{2}v_F + p_F \epsilon''\right) \frac{p_n}{\omega_p-v_F p_n} \frac{q_n}{\omega_q-v_F q_n} \frac{(p+q)_n}{(\omega_p+\omega_q)-v_F (p+q)_n},\\
\langle\rho\rho\rho\rangle^\theta_{\rho^{(2)}} &= - \frac{1}{2} \frac{p_n(p+q)_{s}}{\omega_p-v_F p_n} \partial_{\theta} \frac{1}{\omega_q-v_F q_n} + \text{Perm.}\\
\end{split}
\end{equation}
The WZW and the $H$ pieces arise from the terms in the second line of Eq.~\eqref{eq_action_phi}, 
and the third contribution arises from the $O(\phi^2)$ part of the density operator in Eq.~\eqref{rhoeft}. ``$\text{Perm}$" denotes the $S_3$ permutation of the external momenta $\{p,q,-p-q\}$.  Schematically these correspond to diagrams in fig.~\ref{figfreeeft}, where the red dot denotes the interaction and the boson propagator is denoted by the dotted line. 

\begin{figure}
\begin{adjustwidth}{1cm}{}
\begin{tikzpicture}[scale=0.75]
\draw[dotted] (0,0) -- (2.5,0); 
\fill[black] (2.5,0) circle (2pt);  
\node at (3,0) {$q$};
\draw[dotted] (0,0) -- (-2,2);  
\fill[black] (-2.0,2) circle (2pt);  
\node at (-2.3,2) {$p$};
\draw[dotted] (0,0) -- (-2,-2); 
\fill[black] (-2.0,-2) circle (2pt);  
\node at (-2.2,-2.2) {$-p-q$};
\fill[red] (0,0) circle (3pt);  

\node at (4,0) {$+$}; 


\node at (8.2,2.3) {$p$};  
\node at (8,1.3) {$\rho^{(2)}$}; 
\draw[dotted] (6,-1) -- (8,2);   
\fill[black] (6,-1) circle (2pt);
\node at (6,-1.3) {$q$};
\draw[dotted] (10,-1) -- (8,2);   
\fill[black] (10,-1) circle (2pt);
\node at (10,-1.3) {$-p-q$};
\fill[red] (8,2) circle (3pt);
\node at (4,0) {$+$}; 
\node at (11,0) {$+$};  
\node at (12,0) {Cyclic};

\node at (1,0.5) {\small WZW, $H$};
\end{tikzpicture}
\end{adjustwidth}
\caption{Density three-point function in a free Fermi gas.}
	\label{figfreeeft}
\end{figure}
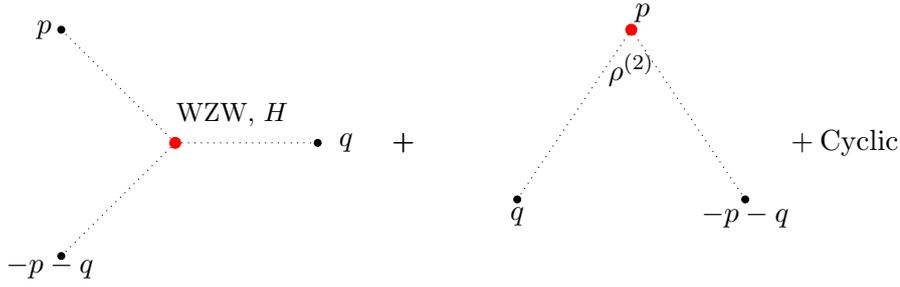

\begin{figure}
\begin{adjustwidth}{1cm}{}
\begin{tikzpicture}[scale=0.75]
 \draw[thick] (0,0) -- (2.5,0); 
 \fill[black] (2.5,0) circle (2pt);  
 \node at (3,0) {$q$};
 \draw[thick] (0,0) -- (-2,2);  
\fill[black] (-2.0,2) circle (2pt);  
\node at (-2.4,2) {$p$};
 \draw[thick] (0,0) -- (-2,-2); 
 \fill[black] (-2.0,-2) circle (2pt);  
 \node at (-2.2,-2.2) {$-p-q$};
 \fill[red] (0,0) circle (3pt);  

\node at (4,0) {$+$}; 


\node at (8,2.3) {$p$};  
\node at (8,1.1) {$\rho^{(2)}$}; 
\draw[thick] (6,-1) -- (8,2);   
\fill[black] (6,-1) circle (2pt);
\node at (6,-1.3) {$q$};
\draw[thick] (10,-1) -- (8,2);   
\fill[black] (10,-1) circle (2pt);
\node at (10,-1.3) {$-p-q$};
\fill[red] (8,2) circle (3pt);
\node at (4,0) {$+$}; 
\node at (11,0) {$+$};  
\node at (12,0) {Cyclic};
 \draw[thick] (1,-4)--(3,-4);
  \node at (3.5,-4) {$=$};
  \draw[dotted] (3.6,-4)--(4.6,-4);
  \node at (5,-4) {$+$}; 
  \draw[dotted] (5.5,-4)--(6.5,-4);
  \fill[blue] (6.5,-4) circle (2pt);
  \draw[dotted] (6.5,-4)--(7.5,-4); 
  \node at (8,-4) {$+$}; 
  \draw[dotted] (8.5,-4)--(9.5,-4);
  \fill[blue] (9.5,-4) circle (2pt);
  \draw[dotted] (9.5,-4)--(10.5,-4);
\fill[blue] (10.5,-4) circle (2pt);
\draw[dotted] (10.5,-4)--(11.5,-4);
 \node at (12,-4) {$+$};
\node at (13,-4) {$\cdots$};

\node at (1,0.5) {\small WZW, $H$};
\node at (1.1,-0.8) {\small $F^{(2,0)}\!, \,F^{(3,0)}$};
\node at (6.5,-4.5) {$F^{(2,0)}$};
\node at (2,-4.5) {\small $\mathcal{S}^\phi_{\theta,\theta'}$};
 \end{tikzpicture}
 \end{adjustwidth}
\caption{Density three-point function in a general Fermi liquid.}
	\label{figfreeeft2}
\end{figure}
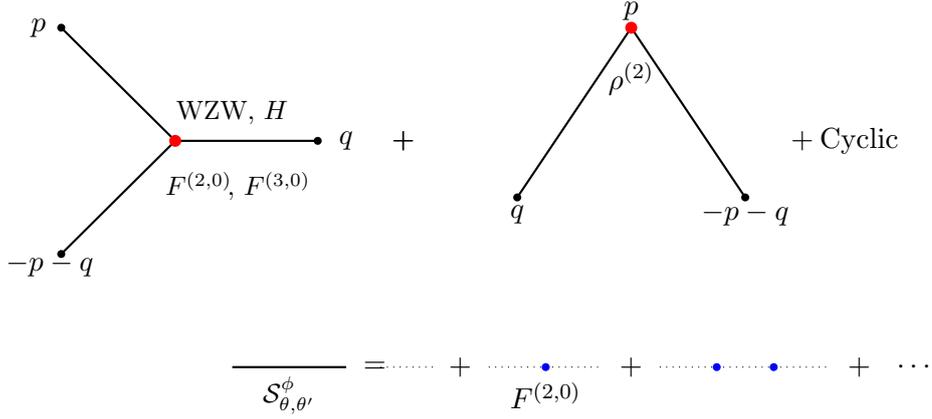

\subsubsection*{General Fermi liquid}

We now turn to interacting Fermi liquids. Even linear response at low momenta, which is captured by a Gaussian action \eqref{eq_S_gaussian}, is complicated: there is no simple closed form expression for the propagator $\mathcal{S}^{\phi}_{\theta,\theta'}(\omega,q)$ in Eq.~\eqref{eq_Sphi_free}  in the presence of general Landau parameters. Interestingly, nonlinear response is not qualitatively more complicated: one can establish a closed form expression for higher-point functions of densities (or other operators), up to expressions involving this Gaussian propagator $\mathcal{S}^{\phi}_{\theta,\theta'}(\omega,q)$.

There are two qualitatively different contributions to the three-point function. The first contribution is analogous to the free fermion answer, but now  with exact propagators $\mathcal{S}^{\phi}_{\theta,\theta'}(p)$. The diagrams for this part is given by fig.~\ref{figfreeeft2} with exact propagators.  In equations:
\begin{equation}\label{rrrexact1}
\begin{split}
\langle \rho_p \rho_q \rho \rangle^1_{\rm Exact} &= \Pi_{i=1}^3\int_{\theta_i}\left( \langle\rho\rho\rho\rangle_\text{\bf WZW} + \langle\rho\rho\rho\rangle_{\bf \rho^{(2)}} + \langle\rho\rho\rho\rangle_{\bf H} \right)\, , \\
\langle\rho\rho\rho\rangle_\text{\bf WZW} &=\frac{ip^3_F}{3!}\int_\theta\left[\left(p_{n_1}q_{n_2}(p+q)_{n_3}(p+q)_nq_s(\omega_p-\omega_q)\right)\right. \\
&\left.\times \mathcal{S}^{\phi}_{\theta,\theta_1}(p+q)\mathcal{S}^{\phi}_{\theta,\theta_2}(q)\partial_\theta \mathcal{S}^{\phi}_{\theta,\theta_3}(p)+ \text{Perm.}\right]\, , \\
\langle\rho\rho\rho\rangle^\theta_{\bf H} &= \frac{-ip^3_F}{3!} \int_\theta\left( \frac{v_F}{2}+p_F \epsilon_{p_F}''\right)\left[\left(p_{n_1}q_{n_2}(p+q)_{n_3}(p+q)_nq_np_n\right)\right. \\
&\left.\times 
\mathcal{S}^{\phi}_{\theta,\theta_1}(p)\mathcal{S}^{\phi}_{\theta,\theta_2}(q)\mathcal{S}^{\phi}_{\theta,\theta_3}(p+q)+ \text{Perm.}\right]\,, \\
\langle\rho\rho\rho\rangle^\theta_{\bf \rho^{(2)}} &= \frac{-p^2_F}{2} \left(p_{s_1}(p+q)_{n_1}q_{n_2}(p+q)_{n_3}\right)\partial_{\theta_1}\mathcal{S}^{\phi}_{\theta_1,\theta_2}(q)\mathcal{S}^{\phi}_{\theta_1,\theta_3}(p+q) + \text{Perm},
\end{split}
\end{equation}
where Perm ~denotes the $S_3$ permutations of $\{(p,\theta_1), (q,\theta_2), (-p-q,\theta_3)\}$.

The second contribution instead involves new cubic vertices of interacting Fermi liquids that are absent in Fermi gases: specifically, the cubic terms in Eq.~\eqref{eq_action_phi} involving $F^{(2,0)}$ and  $F^{(3,0)}$.  As before, the Wick contractions involve the exact propagators $\mathcal{S}^{\phi}_{\theta,\theta'}(p)$. The relevant diagrams are the left most figure in fig.~\ref{figfreeeft2} and the red dot replaced by these relevant interactions. These contributions evaluate to, 
\begin{equation}\label{rrrexact2}
\begin{split}
\langle \rho_p \rho_q \rho_{-p-q} \rangle^2_{\rm Exact} &= \left( \langle\rho\rho\rho\rangle_\text{\bf $F^{(2,0)}$} + \langle\rho\rho\rho\rangle_\text{\bf  $F^{(3,0)}$} \right),\\
\langle\rho\rho\rho\rangle_\text{\bf $F^{(2,0)}$}&= \frac{-i v_F p^3_F}{2}\Pi_{i=1}^3\int_{\theta_i}\int_{\theta, \theta'} \left\{p_{n_1} q_{n_2} (p+q)_{n_3}\right.\\
&\left.\left[(p+q)_sp_n(p+q)_{n'}F^{(2,0)}(\theta,\theta')\mathcal{S}^{\phi}_{\theta,\theta_1}(p)\partial_\theta \mathcal{S}^{\phi}_{\theta,\theta_2}(q) \mathcal{S}^{\phi}_{\theta',\theta_3}(p+q) \right.\right.\\
&\left. \left. + p_n q_s (p+q)_{n'}\partial_\theta F^{(2,0)}(\theta,\theta') \mathcal{S}^{\phi}_{\theta,\theta_1}(p) \mathcal{S}^{\phi}_{\theta,\theta_2}(q) \mathcal{S}^{\phi}_{\theta',\theta_3}(p+q)\right. \right.\\
&\left. \left. +p_nq_n(p+q)_{n'} F^{(2,0)}_1(\theta,\theta') \mathcal{S}^{\phi}_{\theta,\theta_1}(p)\mathcal{S}^{\phi}_{\theta,\theta_2}(q) \mathcal{S}^{\phi}_{\theta',\theta_3}(p+q)\right]+ \text{Perm.}\right\}\,,\\
\langle\rho\rho\rho\rangle_\text{\bf $F^{(3,0)}$}&=-i p^3_F\Pi_{i=1}^3\int_{\theta_i}\int_{\theta, \theta',\theta''} \left\{p_{n_1} q_{n_2} (p+q)_{n_3}\right.\\
&\left.\left[p_nq_{n'}(p+q)_{n''} F^{(3,0)}(\theta,\theta',\theta'')\mathcal{S}^{\phi}_{\theta,\theta_1}(p)\mathcal{S}^{\phi}_{\theta',\theta_2}(q) \mathcal{S}^{\phi}_{\theta'',\theta_3}(p+q)\right]+\text{Perm.}\right\}
\end{split}
\end{equation}
The asymptotically exact three-point density correlator of a Fermi liquid (i.e., in the limit small frequency and momenta) is then given by
\begin{equation}\label{rrrexactf20f30}
\langle \rho_p \rho_q \rho_{-p-q} \rangle_{\rm Exact}=\langle \rho_p \rho_q \rho_{-p-q} \rangle^1_{\rm Exact}+\langle \rho_p \rho_q \rho_{-p-q} \rangle^2_{\rm Exact}\, .
\end{equation}
We emphasize that the only piece that cannot be readily evaluated for general values of the Landau parameters is the  propagator $\mathcal{S}^{\phi}_{\theta,\theta'}(p)$ that already appears in the linear response of Fermi liquids. However, it can be evaluated in certain limits, such as static limits: see Eq.~\eqref{eq_rr_theta_EFT}, which holds to all orders in Landau parameters. Alternatively, this expression can be expanded at weak coupling, which we turn to next.

\subsubsection*{Weak coupling limit}\label{weak_cpl_rrr}

We now consider the limit of small Landau parameters, in which case the exact result obtained above can be expanded and has a closed form expression. Specifically, we will focus here on  weakly coupled Fermi liquids and obtain the first perturbative correction to the free density three point function. We relegate the detailed computation of the first correction to the App.~\ref{3ptcalceft}, presenting just the analytic result in the main text:
\begin{equation}\label{rho3eft}
\begin{split}
&\langle\rho_p\rho_q\rho_{-p-q}\rangle_{(1)} =\\
& v_F\int_{\theta,\theta'} \left[F^{(2,0)}(\theta, \theta') \frac{p_{n'}}{(\omega_p- v_F p_{n'})}\langle\rho_p\rho_q\rho_{-p-q}\rangle^{\theta}_\text{WZW}+\frac{\partial_{\theta}F^{(2,0)}(\theta', \theta)}{3!}  \frac{p_{n'}g^\theta(p|q,-p-q)}{(\omega_p - v_F p_{n'})}\right.\\
&\left.+ \text{Cyclic} \right]+ v_F\int_{\theta,\theta'} \left[F^{(2,0)}(\theta, \theta') \frac{p_{n'}}{(\omega_p- v_F p_{n'})}\langle\rho_p\rho_q\rho_{-p-q}\rangle^{\theta}_\text{H}+ \text{Cyclic} \right]\\
&+v_F\int_{\theta,\theta'} \left[F^{(2,0)}(\theta, \theta') \frac{p_{n'}}{(\omega_p- v_F p_{n'})}\langle\rho_p\rho_q\rho_{-p-q}\rangle^{\theta}_{\rho^{(2)}}+ \frac{p_{n'}\partial_{\theta}F^{(2,0)}(\theta',\theta)}{2(\omega_p- v_F p_{n'})}h^{\theta}(p|q,-p-q)\right.\\
&\left.-\frac12\int_{\theta,\theta'}\partial_\theta F^{(2,0)}(\theta,\theta')\frac{p_{n'}}{(\omega_{p}-v_F p_{n'})} j^{\theta}(p|q,-q-p)+ \text{Cyclic}\right]\\
&-\frac{v_F}2\int_{\theta,\theta'} \frac{
F_1^{(2,0)}(\theta, \theta')p_nq_n (p+q)_{n'}}{(\omega_{p+q}-v_F (p+q)_{n'})(\omega_q-v_F q_n)(\omega_p-v_F p_n)}+\text{Perm}\\
&-6 \int_{\theta,\theta',\theta''}\frac{  F^{(3,0)}(\theta,\theta',\theta'')p_{n}q_{n'}(p+q)_{n''}}{(\omega_p-v_F p_{n})(\omega_q-v_F q_{n'})(\omega_{p+q}-v_F (p+q)_{n''})},
\end{split}
\end{equation}
where the functions $g,h$ and $j$ are given in Eqs. \eqref{gdefinition}, \eqref{hdefinition} and \eqref{jdefinition} respectively and ``$\text{Cyclic}$" denotes the cyclic permutations, i.e., the $Z_3$ permutation of the external momenta $\{p,q,-p-q\}$. 

One important qualitative point is that even at the leading order in small frequencies and wavevectors, this correlation function is sensitive to data beyond the familiar Landau parameters $F^{(2,0)}(\theta,\theta')$. Specifically, the new generalized Landau parameters entering are $F^{(2,0)}_1(\theta,\theta')$ and $F^{(3,0)}(\theta,\theta',\theta'')$. These can therefore be measured by experiments probing higher-point functions of densities, which should be accessible in cold atoms \cite{Tam:2023scw}.

\section{Symmetry constraints on the EFT}\label{sec_nonlin_symmetry}
    
The Landau parameters $F^{(m,n)}$ in our EFT  are already subject to simple constraints arising from translation and rotation symmetry.  In this section, we derive additional more subtle constraints that the coefficients parametrizing Fermi liquids must obey when the underlying microscopic theories exhibit certain other symmetries. Specifically, we systematize the constraints derived in Sec.~\ref{sec_linear} for Galilean, Lorentz and scale invariance, presenting them in a manner that facilitates their extension to the generalized Landau parameters appearing in the EFT of Fermi liquids.

\subsection{Galilean invariance}
We begin this section by examining Galilean invariance. The presence of a Fermi surface breaks Galilean boosts, implying that for a microscopic theory with Galilean invariance and a Fermi liquid IR phase, this symmetry must be realized nonlinearly in the EFT.
To illustrate the advantage of our EFT formulation in this context, let us derive the consequences of  Galilean invariance on our action. As outlined in App.~D of \cite{Delacretaz:2022ocm}, in the EFT action this is obtained by implementing Galilean boosts (with boost velocity $\vec{v}$) as a canonical transformation
\begin{equation}
W= e^{B_v}, \qquad B_v= \vec{v}\cdot(\p t-m \x)\, .
\end{equation}
To linear order in the boost velocity, the Galilean boost symmetry then acts nonlinearly on $\delta f_p$:
\begin{equation}\label{gboostf}
\begin{split}
\delta_v \delta f_p \equiv W\, f_p\, W^{-1}- f_p&\simeq \{B_v, f_p\},\\
&= - t \vec{v}\cdot \nabla_{\x} \delta f_p- m \vec{v}\cdot \nabla_{\p} (\delta f_p+ f^0_p)\, .
\end{split}
\end{equation}
Let us illustrate implications of Galilean invariance on the free EFT first. The change of the free action under Galilean transformation, to linear order in boost parameter, takes the form
\begin{equation}\label{freeGalilean}
\begin{split}
\delta_v S^\text{free}&=\int_{t,\x,\p}\left[\vv\cdot \p\, - m\vv\cdot\nabla_{\p}\epsilon_p\right](\delta f_p+ f^0_p)\, ,
\end{split}
\end{equation}
where $\epsilon_p\equiv \epsilon(|\p|)$ due to rotational invariance and we have ignored a term which is total derivative in $\x$. The term proportional to $f^0_p$ vanishes because of rotational symmetry and Galilean invariance therefore tells us that the variation in $\delta f_p$ must be zero. We emphasize that this equation holds in any dimension $d$. Now we specialize to $d=2$, considering the fluctuations about a spherical Fermi surface
\begin{equation}\label{deltapf}
    \begin{split}
        \delta f_p &= \Theta (-|\p|+p_F-\delta p_F(t,\vec{x},\theta))-\Theta (-|\p|+p_F)\\
        &\sim -\delta p^\theta_F~ \delta(-|\p|+p_F)+ \frac{1}{2} {\delta p^\theta_F}^2 \partial_{|\p|}\delta(-|\p|+p_F) + O((\delta p^\theta_F)^3)\, , 
    \end{split}
\end{equation}
where $\delta p^\theta_F \equiv \delta p_F(t,\x,\theta)$ denotes the fluctuation about the Fermi surface. Rotational invariance and a systematic expansion of our constraint equation \eqref{freeGalilean} in $\delta p^\theta_F$  translates to
\begin{equation}
\begin{split}
\epsilon'(p_F) \equiv v_F=\frac{p_F}{m},\qquad \epsilon''({p_F})\equiv \epsilon''_F=\frac{1}{m},\qquad \partial^{n\geq 3}_{p_F}\epsilon(p_{F})=0\, .
\end{split}
\end{equation}
Since our variation truncates beyond the linear term in $\delta f_p$, these constraints fix the entire function $\epsilon_p$ to be $\frac{p^2}{2m}$. This changes once interactions are introduced because of the nonlinear, inhomogeneous nature of the Galilean transformation on $\delta f_p$.
Up to quadratic order in the fluctuations, the change in the action with interactions takes the following form
 \begin{equation}\label{galileo_change_action}
\begin{split}
\delta_v S&=\int_{t,\x,\p}\left[\vv\cdot \p\, - m\vv\cdot\nabla_{\p}\epsilon_p + 2m \int_{\p'} F^{(2,0)}(\p, \p')  \vv\cdot\nabla_{\p'} f_{p'}^0 \right]\delta f_p \\
&+\int_{t,\x,\p, \p'} \left[2m F^{(2,0)}(p,p') \vv\cdot\nabla_{\p} \delta f_p \delta f_{p'} + 3m\int_{\p''} F^{(3,0)}(p,p',p'') \vv\cdot\nabla_{\p''} f^0_{p''} \delta f_p \delta f_{p'}\right]\\
&+O(\delta f^3)  \, , 
\end{split}
\end{equation}
where $O(\delta f^3)$ terms involve higher order Landau parameters and their derivatives. This also highlights the nonlinear nature of the variation, which leads to a mixing of Landau parameters that appear in the action at different orders in $\delta f_p$. Hence, contrary to the free theory, the infinite tower of constraints involving $\partial^n_{p_F}\epsilon({p_F})$ now also involve various (generalized) Landau parameters along with their derivatives. As before, this variation holds true in any dimension $d$ and we now specialize to $d=2$. To leading order in $\delta p^\theta_F$ we recover the familiar Galilean constraint on the effective mass
\begin{equation}\label{lingalileo}
m^*\equiv \frac{p_F}{v_F}=(1+F_1) m,
\end{equation}
as was found in Eq.~\eqref{eq_F1sym}. Going beyond leading order to ${\delta p^\theta_F}^2$, $\epsilon''_{F}$ is constrained by the derivative of the Landau parameter $F^{(2,0)}$ and we also obtain constraints between the generalized Landau parameters that appear in our action \eqref{eq_action_phi}

\begin{equation}\label{nonlingallileo} 
\begin{split}
&0 =\int_\theta \left[-p_F\hat{n}_\theta + \frac{m}{2}\left(v_F+p_F\epsilon''_{F}\right)\hat{n}_\theta+m\int_{\theta'}\left(\frac{v_F}{2}F^{(2,0)}(\theta,\theta')+\frac{v_F}{2}F_1^{(2,0)}(\theta,\theta')\right)\hat{n}_{\theta'}\right]{\delta p^\theta_F}^2\\
+&\int_{\theta,\theta'}\left[-2m\left(\frac{v_F}{2}F^{(2,0)}_1(\theta,\theta') \hat{n}_{\theta}+\frac{v_F}{2}\partial_\theta F^{(2,0)}(\theta,\theta') \hat{s}_\theta\right)-3m\int_{\theta''} F^{(3,0)}(\theta,\theta',\theta'')\hat{n}_{\theta''}\right]\delta p^\theta_F\delta p^{\theta'}_F
.
\end{split}
\end{equation}
It is convenient to present the constraints in Eq.~\eqref{nonlingallileo} in terms of new harmonic functions $\tilde{F}_\ell$ and $G_{\ell,\ell'}$. The generalized Landau parameter $F^{(3,0)}$ is real and symmetric under permutations of $\{\theta, \theta',\theta''\}$. The only constraint on $F^{(2,0)}_1$ is that it must be real. These conditions impose the following relations on $G_{\ell,\ell'}$ and $\tilde{F}_\ell$
\begin{equation}\label{F30harmonicdecomp}
\begin{split}
F^{(2,0)}_1(\theta,\theta')&=2\pi \sum_{\ell} \tilde{F}_{\ell} e^{i \ell(\theta-\theta')},\qquad F^{(3,0)}(\theta,\theta',\theta'')=2\pi^2 \sum_{l,l'} G_{\ell,\ell'} e^{i \ell(\theta-\theta')+i \ell'(\theta-\theta'')},\\
 G_{\ell,\ell'}&=G^*_{\ell,\ell'}=G_{-\ell,-\ell'}=G_{\ell',\ell}=G_{\ell,-\ell-\ell'}=G_{-\ell-\ell',\ell'},\qquad \tilde{F}^*_{\ell}= \tilde{F}_{-\ell}.
\end{split}
\end{equation}
Assuming a Fourier series expansion of our fluctuations $\delta p_F^\theta$, the nonlinear Galilean constraints Eq. \eqref{nonlingallileo} give rise to three new sets of constraints on the harmonics of $F_{\ell}$ and $G_{\ell,\ell'}$
\begin{equation}\label{galnnharmonics}
\begin{split}
&\tilde{F}_ 1 m v_F+m p_F \epsilon''_F-p_F=0,\qquad \tilde{F}_\ell= \tilde{F}_{-\ell},\\
& -3 G_{1,\ell}+ \left(\ell F_{\ell}-(\ell+1) F_{\ell+1}\right) v_F- v_F \left(\tilde{F}_\ell+\tilde{F}_{\ell+1}\right)=0.
\end{split}
\end{equation} 
The first line is simply a derivative of the standard Galilean constraint Eq.~\eqref{lingalileo} with respect to $p_F$. This constraint arises from the quadratic expansion of the term linear in fluctuation $\delta f_p$ in Eq.~\eqref{galileo_change_action}, around the Fermi surface,
\begin{equation}
\left\{\partial_{|\p|}\left[p^i-m\nabla^i_\p \epsilon_p +2m\int_{\p'} F^{(2,0)}(\p,\p') \nabla_{\p'} f^0_{p'}\right] \right\}_{p_F}=0
\end{equation}
where the equation is evaluated at $|\p|=p_F$ after performing the derivative. The second line in Eq.~\eqref{galnnharmonics} denotes a new set of constraints on the generalized Landau parameters of any Galilean invariant Fermi liquid.

As a consistency check, these results can be derived in an alternative way by recalling that in Galilean invariant theories \cite{Landau1987Fluid}, the momentum and current densities are related
\begin{equation}
T^{ti}= m j^i\, .
\end{equation}
$T^{ti}$ and $j^i$ are given by Eqs. \eqref{Tijnoether} and \eqref{u1eftops} respectively.
Expanding to quadratic fluctuations around the spherical Fermi surface then reproduces the linear and nonlinear constraints given by Eqs. \eqref{lingalileo} and \eqref{galnnharmonics}.

\subsection{Scale invariance}\label{sec_scale_inv}

If the microscopic theory is scale invariant, the scale transformation will also be nonlinearly realized in the EFT. Similar to Galilean boosts, the nonlinear realizations of scaling symmetries are incorporated into our EFT in part through canonical transformations of the EFT fields. We will derive this transformation for dilatations here. We will work with an arbitrary dynamical scaling exponent $z$, which determines the ratio of the scaling dimension of time and space, and then briefly comment on two special values $z=1$ and $z=2$ corresponding respectively to relativistic and non-relativistic conformal invariance.

Analogously to Galilean transformations, the first step is to identify the representation of the operator $D_b$ that generates dilatations on functions of phase space. Under an infinitesimal dilatation, parameterized by $b$, coordinates transform as,
\begin{equation}
    \delta_b \x = b \x, \qquad \delta_b \p = - b \p, \qquad \delta_b t = zb t.
\end{equation}
The canonical transformation that implements the dilatation in phase space is given by the function $D_b= - b \x \cdot p$. It's easy to check that $D_b$ generates the correct infinitesimal transformations of $\x$ and $\p$ and has the following action on phase space functions $F(\x,\p)$
\begin{equation}
  W_b=e^{D_b},\qquad  W_b F(\x,\p) W_b^{-1} = F(e^b \x, e^{-b} \p).
\end{equation}
However, canonical transformations do not have a natural action on the time coordinate and as such if the function $F$ depends on $t$ as well, the representation of infinitesimal dilatations needs to be modified to the following differential operator
\begin{equation}
    \mathcal{D}_b(~\cdot~) \equiv - b \{ \x\cdot\p, ~\cdot~ \} + bz t ~ \d_t (~\cdot~) = b \left( \x\cdot\nabla_\x - \p\cdot\nabla_\p + zt~\d_t \right).
\end{equation}
What a finite dilatation does in practice is implement the canonical transformation generated by $D_b = - b \x\cdot\p$, followed by a rescaling of time. An infinitesimal dilatation acts on the distribution function $f(t,\x,\p)$ and its fluctuation $\delta f = f - f_0$ from the ground state via
\begin{equation}\label{eq_dil_transf}
    \begin{split}
        \delta_b \delta f_p \equiv e^{\mathcal{D}_b} f_p - f_p \simeq  b \left( zt \d_t + \x\cdot\nabla_\x - \p\cdot\nabla_\p \right) (\delta f_p+f^0_p).
    \end{split}
\end{equation}
Similar to Galilean boosts, the action of dilatations on $\delta f$ is nonlinear. The change in the WZW term can be implemented as left action of $D_b$ on $U$ along with the rescaling of time coordinate of $\phi$ to $t'=e^{z b}t$:
\begin{equation}
S_{\rm WZW} \rightarrow \int dt \Tr\left[f_p^0 e^{\phi\left(t'\right)}e^{-D_b}\partial_t \left(e^{D_b}e^{-\phi\left(t'\right)}\right)\right]=\int dt' e^{-zb} \Tr\left[f_p^0 e^{\phi\left(t'\right)}e^{zb}\partial_{t'}e^{-\phi\left(t'\right)}\right],
\end{equation}
where in the first line, we have used the fact that $D_b$ is independent of time and then rescaled the time coordinate. This implies that the WZW term is invariant under a scaling transformation with any $z$, since it is independently invariant under time reparametrization and canonical transformations. Scale invariance constraints hence come purely from the Hamiltonian part of the action. Let us warm up first with just the free Fermi gas for which $S_H^\text{free} = - \int_{t\x\p} \epsilon_p f_p$. The transformation of this term under the dilatation in Eq.~\eqref{eq_dil_transf} is given by,
\begin{equation} \label{scale_ham_free}
    \begin{split}
        \delta_b S_H^\text{free} &= - b \int_{t\x\p} \epsilon_p \left( z t \d_t + \x\cdot\nabla_\x - \p\cdot\nabla_\p \right) f_p,\\
        &= b \int_{t\x\p} \left[ \p\cdot\nabla_\p \epsilon_p - z \epsilon_p \right]f_p
    \end{split}
\end{equation}
Using the expansion of $f$ as in Eq.~\eqref{deltapf} and rotational invariance, we obtain a tower of constraints on all derivatives of $\epsilon_p$ evaluated at the Fermi surface. Assuming  that the dispersion is an analytic function, this is equivalent to solving the following equation
\begin{equation}
    \p\cdot\nabla_\p \epsilon_p = z \epsilon_p, \qquad\implies \epsilon_p \propto p^z,
\end{equation}
in agreement with the expected results for $z=1$ and $z=2$.

Next, turning on interactions and we expand the change in the action to quadratic order in fluctuations $\delta p_F^\theta$. Disregarding a $f_p$ independent constant shift in the action,\footnote{More precisely, this constant shift is given by $\sim b \int_{t\x\p} \left[ \p\cdot\nabla_\p \epsilon_p - z \epsilon_p \right]f^0_p$, which does not affect our observables, i.e., neither the EFT operators nor their correlation functions. Stated differently, the finite density state is obviously not scale invariant but our constraints concern only the fluctuations about the ground state.}
we obtain the following constraints. The first of these is the linear constraint on $F_0$ that was obtained from thermodynamic arguments in Eq.~\eqref{eq_F0sym}:
\begin{equation}\label{eq_scale_constr}
    \begin{split}
        z \epsilon_F - p_F v_F (1+F_0) = 0, \qquad
         &\left[ p_F \epsilon''_{F} - (z-1) v_F \right] + v_F\tilde{F}_0 = 0,\\
        3 G_{\ell,0} + v_F \left( \tilde{F}_\ell + \tilde{F}_{-\ell} \right)& +  (z-2) v_F F_{\ell} = 0
    \end{split}
\end{equation}
The second constraint can be understood as the derivative of the linear constraint (analogous to the Galilean case) while the third truly involves the cubic couplings. The nonlinear realization of the scale invariance manifests itself in our third constraint by relating $F^{(2,0)}$ with the higher-order generalized (irrelevant) Landau parameters.  

For $z=2$, the dilatations form a part of a larger algebra which includes boosts, spacetime translations and one special conformal transformation called Schr\"odinger's algebra (see \cite{Hagen:1972,Nishida:2007,Baiguera:2023fus} for reviews and related developments). The scale invariance constraints for this take the form
\begin{equation}\label{eq_schrodinger_constr}
        2 \epsilon_{F} - p_F v_F (1+F_0)= 0,~~~ \left( p_F \epsilon''_{F} - v_F \right) + v_F \tilde{F}_0 = 0,~~~  3 G_{\ell,0} +  v_F \left( \tilde{F}_\ell + \tilde{F}_{-\ell} \right) = 0.
\end{equation}
Apart from dilatations, the Schr{\"o}dinger group also features a special conformal transformation which can be implemented via the following canonical transformation, 
\begin{equation}
    \mathcal{D}_c(~\cdot~) \equiv  c \left\{\frac{m\x^2}{2}-t \x\cdot\p, ~\cdot~ \right\} + c t^2 \partial_t (~\cdot~) = c \left(m \x \cdot \nabla_p +t\left\{\x\cdot\nabla_\x - \p\cdot\nabla_\p\right\} + t^2~\partial_t \right).
\end{equation}
We find that it leads to no new constraints other than the Galilean constraints in Eqs. \eqref{lingalileo}, \eqref{galnnharmonics} and the dilatation constraints in Eq. \eqref{eq_schrodinger_constr}. Hence, 
for Schr\"odinger invariant field theories that enter a Fermi liquid IR phase, the Landau parameters must obey  these three sets of constraints. Consequently, the criterion for Schrödinger invariance can also be derived from the conformal Ward identity,\footnote{See Refs \cite{Nakayama:2009ww} for subtleties concerning this ward identity in non-relativistic CFTs.}
$2 T^t_t + T^i_i = 0$ in conjunction with Galilean invariance. We will comment on $z=1$ in the next subsection after imposing Lorentz invariance on our EFT.

\subsection{Lorentz invariance}\label{FLEFT}

We turn our attention to symmetry constraints imposed on Fermi liquids arising from Lorentz invariant microscopics. It might seem reasonable to impose Lorentz invariance in a manner analogous to Galilean boosts and scale invariance but this approach quickly runs into difficulties. Lorentz boosts are not canonical transformations, complicating the direct generalization from Galilean or scale invariance. We now obtain a systematic derivation of the Lorentz constraints by considering the constraints imposed on the stress tensors of the EFT due to these symmetries. 

Translation invariance implies that the canonical stress-energy tensor in a QFT is conserved, $\partial_\mu\mathcal{T}^{\mu\nu}(x)=0.$ In general, $\mathcal{T}^{\mu\nu}$ is not symmetric. However one can always define an ``improved" stress tensor $T^{\mu\nu}$ which is symmetric and furnishes a representation of the Lorentz generators \cite{BELINFANTE1940449}. Hence Lorentz invariance, together with translation invariance, implies a conserved and symmetric stress tensor. Conversely, given a conserved and symmetric stress tensor one can always construct generators of the Lorentz algebra from them. We have already constructed the conserved stress tensor from the EFT in Sec.~\ref{Tmunu_EFT}; in this section we impose that the stress tensor is symmetric in its indices which leads to a necessary and sufficient condition for Lorentz invariance of the EFT. A similar method has been used to arrive at linear Lorentz constraints in \cite{Son:2012zy}.

As warm up, let us investigate the consequences of Lorentz invariance on the free EFT (i.e., the Landau parameters have been set to zero). From our Eqs.~ 	
\eqref{Tijnoether} and \eqref{Titnoether} we have, 	
\begin{equation}\label{freetmunu} 
\begin{split}
&T^{tt}=  \int_\p~ \epsilon_p f_p, \qquad T^{it}= \int_\p~  \epsilon_p \nabla^i_\p \epsilon_p f_p,\\ 
&T^{ij}= \int_\p~ p^j \nabla^i_\p \epsilon_p f_p, \qquad T^{ti}=\int_\p~ p^i f_p.\\
\end{split}
\end{equation}
Our stress tensors are conserved and when our quasi-particle dispersion is rotationally invariant (i.e., $\epsilon(\p) \equiv \epsilon(|\p|)$), our stress tensors are symmetric in the spatial indices. However they are explicitly not symmetric in the all the indices, especially when one of the indices is along the time direction. We impose Lorentz invariance by demanding the stress tensor must be symmetric in all of its indices, 
\begin{equation}
\begin{split}
T^{ti}=T^{it}, \implies \epsilon_p \nabla^i_\p \epsilon_p= p^i.
\end{split}
\end{equation}
We recover the dispersion relation for a free relativistic particle $\epsilon_p = \sqrt{p^2+\rm{constant}}$. We now apply the same approach to the conserved stress tensors derived from the fully interacting EFT in Sec.~ \ref{Tmunu_EFT}. Utilizing the expansion in Eq.~\eqref{deltapf}, we systematically examine the constraints arising from the symmetry properties of the stress tensor.
At linear order in the fluctuations we recover the Lorentz constraint of \cite{BAYM1976527}, that we had already found in Eq.~\eqref{eq_F1sym}
\begin{equation}\label{linconstLharm}
p_F \hat{n}^i_\theta = \left[ \epsilon_Fv_F \hat{n}^i_\theta + \epsilon_F v_F\int_{\theta'} F^{(2,0)}(\theta, \theta')\hat{n}^i_{\theta'} \right],\qquad \hbox{or}\qquad \frac{p_F}{\epsilon_F v_F} =(1+F_1) \, , 
\end{equation}
where equation on the right is a result of expanding the Landau parameters in terms of its harmonics in order to express these constraints in a more familiar form. The nonlinear constraints at $O({\delta p^\theta_F}^2)$ take the following form in terms of harmonics of the landau parameters $F^{(2,0)}$, $F_1^{(2,0)}$ and $F^{(3,0)}$
\begin{equation}\label{nonlin3f2harmonics}
\begin{split}
\tilde{F}_ 1 v_F \epsilon_F + p_F \left(\epsilon_F \epsilon''_F+v_F^2-1\right)=0,\qquad \tilde{F}_\ell=\tilde{F}_{-\ell} &,\\
3 \pi^2 \epsilon_F G_{1,\ell}+ v_F \left[ \pi^2 F_{\ell+1} \left(p_F v_F+\ell \epsilon_F+\epsilon_F\right)+\pi^2\left(\tilde{F}_\ell+\tilde{F}_{\ell+1}\right) \epsilon_{F}\right.\\
\left.+F_{\ell} \left( \pi^2\left(p_F v_F-\ell \epsilon_F\right)+F_{\ell+1} p_F v_F \right)\right]=0&,
\end{split}
\end{equation}
The first equation can be thought of as a derivative of the known linear constraint. Instead, the second and third equations are new constraints on the generalized Landau parameters of relativistic Fermi liquids.

For Fermi liquids which are obtained as IR phase of a CFT, the Landau parameters now must obey  scale invariance constraints in Eq.~\eqref{eq_scale_constr} along with constraints due to Lorentz invariance in Eqs.~\eqref{linconstLharm} and \eqref{nonlin3f2harmonics}. This is achieved by $z=1$ in our general scale invariance constraint, for which dilatations are part of the larger conformal group $SO(3,2)$, resulting in the following,
\begin{equation}\label{eq_rel_conf_constr}
    \begin{split}
        \epsilon_F - p_F v_F (1+F_0) = 0, ~~~  p_F \epsilon''_{F} + v_F \tilde{F}_0= 0,~~~
        3 G_{\ell,0} + v_F \left( \tilde{F}_\ell + \tilde{F}_{-\ell} \right) + v_F F_{\ell} &= 0
    \end{split}
\end{equation}
The same constraints are obtained if we impose the ward identity $T^{\mu}_{\mu}=0$\footnote{See refs \cite{1971AnPhy..67..552C, CALLAN197042, 1360292620544287104} for subtleties concerning this Ward identity in CFTs.} on our stress tensors.

A nonlinear Lorentz constraint was obtained in Ref.~\cite{BAYM1976527}. However, neither of their constraints -- whether under the weak coupling approximation or not -- account for the new Landau parameter $F^{(3,0)}$. In comparison our nonlinear constraints (in general $d$) can also be stated as, \begin{equation}
\begin{split}
\left\{2 \nabla^i_\p(\epsilon_p F^{(2,0)}(\p,\p'))  - 2 \int_{p''}F^{(2,0)}(\p'',\p') F^{(2,0)}(\p,\p'')\nabla^i_{\p''} f_{p''}^0 \right.\\
\left.-3 \int_{p''}\epsilon_{p''} F^{(3,0)}(\p,\p',\p'') \nabla^i_{p''} f_{p''}^0\right\}_{p_F} &=0\, .\\
\end{split}
\end{equation}
The equation when evaluated at $|\p|=|\p'|=p_F$, leads to the nonlinear constraint in Eq.~\eqref{nonlin3f2harmonics} for $d=2$. The generalized Landau parameter $F^{(3,0)}$, which is generically present in Fermi liquids, is crucial for ensuring Lorentz invariance, as demonstrated in our systematic treatment. In Sec.~ \ref{nnlincons} we show that for certain microscopic models at weak coupling, our constraints reduce to that of \cite{BAYM1976527}  since the contribution to $F^{(3,0)}$ occurs at sub-leading order. However, beyond weak coupling or for models where there is a leading order microscopic contribution  to $F^{(3,0)}$, we believe that our constraints involving $F^{(3,0)}$ are the appropriate ones. 
\section{Microscopic models}\label{sec_micro_models}

We now shift our focus to relativistic QFTs that enter a Fermi liquid phase at finite density, allowing for explicit tests of our symmetry and causality constraints. We begin by considering free Dirac fermions at finite density, which, in addition to setting the stage, will allow us to work out the mapping of microscopic operators to effective operators of the EFT. Next, we consider a small quartic interaction, which perturbatively activates the Landau parameters and generalized Landau parameters in the Fermi liquid EFT. We will demonstrate how to extract the Wilson coefficients from density correlation functions instead of more traditional methods of quasi-particle scattering. Finally, we turn to a strongly coupled Fermi liquid, obtained by considering a Chern-Simons--matter CFT at finite density, and verify our symmetry and causality constraints there.

\subsection{Free fermions}\label{freefermionseft}

We start with free Dirac fermions, the simplest relativistic theory exhibiting a Fermi surface at finite density. For simplicity, we will work in (2+1)-dimensions, which slightly simplifies the treatment since the Fermi surface only carries one spin species. The microscopic action for the free relativistic Dirac fermion at finite chemical potential is:
\begin{equation}\label{dirac}
\begin{split}
\mathcal L_{\rm free} = \bar\psi \left( i \cancel \partial - m + \mu \gamma^0\right)\psi\, .
\end{split}
\end{equation}
 We will use the following conventions for the gamma matrices and metric:
\begin{equation}\label{gammamatrices}
\begin{split}
\gamma^0 &=\begin{pmatrix}
0 & 1\\
1& 0
\end{pmatrix}, \quad \gamma^1 =\begin{pmatrix}
0 & 1\\
-1& 0
\end{pmatrix}, \quad \gamma^2 =\begin{pmatrix}
i & 0\\
0& -i
\end{pmatrix}, \\
&-\frac12\{\gamma^\mu, \gamma^\nu\} = \eta^{\mu\nu}= \text{diag}\{-1,1,1\}\, .
\end{split}
\end{equation}
The free-fermion propagator at finite density is given by (see \cite{Podo:2023ute} for a derivation),
\begin{equation}\label{finitedenprop}
\begin{split}
\langle \Theta| T(\psi_\alpha (\omega, \vec{p}) \bar{\psi}_\beta (\omega', \vec{p}')) |\Theta\rangle &= S_{\alpha \beta} (\omega, \p)(2\pi)^3 \delta(\omega-\omega')\delta^2(\p-\p'),\\
S(\omega, \p) &= \frac{i \left[\left(\omega+\mu\right)\gamma^0-\p\cdot\vec{\gamma}+m\right]}{\left[\omega-\left(\epsilon_{k}-\mu\right)+ i \eta \,\text{Sgn}\left(\epsilon_{k}-\mu\right)\right]\left[\omega+\left(\epsilon_{k}+\mu\right)- i \eta\right]}\,,
\end{split}
\end{equation}
where $\epsilon_{k}= \sqrt{\kv^2+m^2}$ and we suppress the spinor indices for notational convenience. The background chemical potential $\mu$ defines the Fermi momentum via $\epsilon_{p_F}= \mu$. The theory has $U(1)$ symmetry with currents given by 
\begin{equation}\label{microconsu1}
j^\mu = \bar\psi \gamma^\mu \psi\, .
\end{equation}
These directly map to the corresponding $U(1)$ current  of the free EFT $j^\mu=(\rho,j^i)$ 
\begin{equation}
\begin{split}
\rho  &= \int_{\p} f(t,\x,\p), \qquad j^i  = \int_{\p} \nabla^i_\p \epsilon(|\p|) f(t,\x,\p)\, ,
\end{split}
\end{equation}
where $\epsilon(|\p|)=\sqrt{p^2+m^2}$. One can easily check that their correlators match explicitly, see App.~\ref{doOm}. In other words, the Wilsonian function $\epsilon(p)$ of the EFT is simply given by the microscopic dispersion $\epsilon_p$ of the Dirac fermions.

Consider now the operator $\mathcal{O}= \bar{\psi}\psi$ which is not a conserved current. In general, microscopic operators corresponding to non-conserved quantities are non-trivial to match with the EFT -- in particular they are subject to corrections from interactions and will include terms that are nonlinear in $\delta f$. However, the situation is simpler in the case of free fermions, as all microscopic fermion bilinears match exactly with an EFT operator that is linear in $f$. Indeed, in this case the collisionless Boltzmann Eq.~\eqref{eq_Landau_kinetic_eq} immediately follows from the Heisenberg equation for fermion bilinears $\psi_\sigma^\dagger(t,\vec x_1) \psi_{\sigma'}(t,\vec x_2)$, upon taking the Wigner transform and the semiclassical limit. This procedure provides a straightforward map between fermion bilinears and EFT operators. In the context of relativistic Dirac fermions, there is one additional step, since the antiparticles must be integrated out to produce the low energy EFT. One must diagonalize the finite density Hamiltonian arising from Eq.~\eqref{dirac}. However, since this change of basis is a linear process, the conclusion remains that microscopic bilinears map to EFT operators linear in $f$ (for free fermions). Returning to the operator $\mathcal{O}= \bar{\psi}\psi$, we therefore expect this microscopic operator to map to a generic scalar operator in the EFT as constructed in Eq.~\eqref{genopphiexp}, with $\gamma_2=0$
\begin{equation}\label{rhotilde2}
\begin{split}
\mathcal{O} (t,\x)&=p_F\gamma_1(p_F)\int \frac{ d\theta}{(2\pi)^2} 
\nabla_n \phi + O(\phi^2),
\end{split}
\end{equation}
where we have kept terms to linear order in $\phi$. To fix the function $\gamma_1$, we microscopically compute the two-point function of the $\mathcal{O}$ operator with the conserved density $\rho = j^0$ for free Dirac fermions, matching microscopics and EFT in App.~\ref{doOm}. Comparing the computations in Eqs.~\eqref{rhotmicro} and \eqref{rhorhoteft} one finds 
\begin{equation}\label{rhotilde}
\gamma_1(p_F) = \frac{m}{\mu}\, ,
\end{equation} 
where the chemical potential depends on the density and $p_F$ as $\mu(p_F) = \sqrt{p_F^2 + m^2}$. Because this matching applies for any $\mu$ or $p_F$, this in fact fixes the entire function $\gamma_1(|\p|)=\frac{m}{\sqrt{p^2+m^2}}$ for this operator in the free Dirac theory. One can check that correlation function of $\mathcal O$ with $j^i$ instead of $j^0$ then also automatically matches between microscopics and EFT.

\subsection{Fermi liquid at weak coupling}
We now add  a quartic interaction to the model
\begin{align}\label{yukawa}
\mathcal L_4 = \bar\psi \left( i \cancel \partial - m + \mu \gamma^0\right)\psi+\lambda   (\bar\psi\psi)^2 \, .
\end{align}
Let us test our Lorentz invariance constraints in this model at weak coupling. These involve not only the  Landau parameter $F^{(2,0)}$ but also the generalized Landau parameters $F^{(2,0)}_1$, $F^{(3,0)}$ as well as corrections to the Fermi momenta and velocity due to interactions. The objective for this section is two-fold: to test Lorentz constraints, and to show how an infinite tower of Wilsonian coefficients (the Landau parameters) can be extracted from a single correlator, the density two and three point functions, thanks to their nontrivial dependence on the dimensionless ratio $\omega/(v_F q)$.

\subsection*{Linear Lorentz constraints}

The observables we focus on are the correlation functions of the density operators at weak coupling. On the EFT side, we note from Eq.~\eqref{eq_rr_theta_EFT} that the leading correction to the density two-point function must take the form  
\begin{equation}\label{eq_rr_EFT}
\begin{split}
\delta \langle \rho_p\rho_{-p}\rangle^{\rm {EFT}}& = ip_F 
\left[\frac{\delta p_F}{p_F}\int_\theta \frac{q_n}{\omega-v_F q_n} \right.\\
&\left.+ 
\int_{\theta,\theta'} \frac{q_{n}}{\omega-v_F q_{n}} \left(v_F F^{(2,0)}(\theta-\theta') + (2\pi)^2\delta v_F \delta (\theta-\theta')\right)\frac{q_{n'}}{\omega-v_F q_{n'}}\right],
\end{split}
\end{equation}
where  $\langle \rho_p \rho_{-p} \rangle \equiv \langle \rho \rho \rangle (\omega, \p)$ for notational convenience. $\delta p_F, \delta v_F$ and $F^{(2,0)}$ are $O(\lambda)$ correction to these quantities from their free fermion values due to the four-Fermi interaction. 

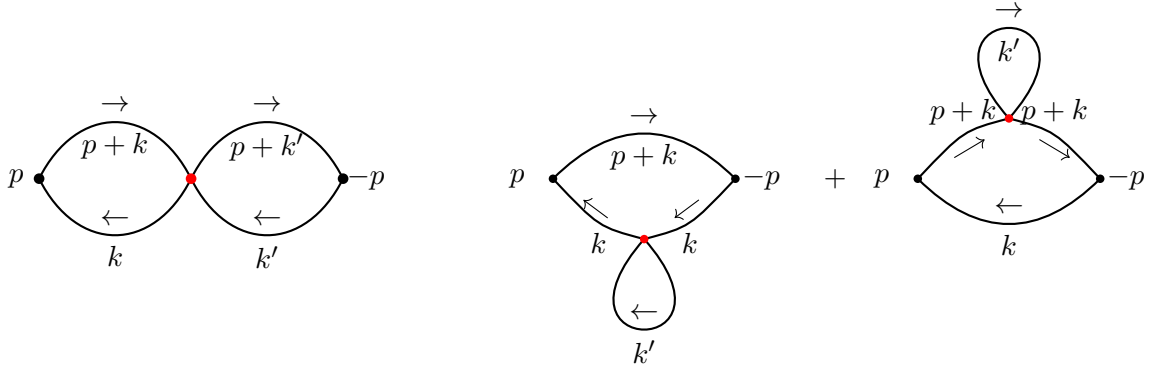
\begin{figure}
\begin{tikzpicture}
\node (a) at (0,0)
{
			\begin{tikzpicture}
	
	\draw[thick] (1,0) .. controls (0.5,1) and (-0.5,1) .. (-1,0) node[midway, above] {$\rightarrow$} node[midway, below] {$p+k$};
	\draw[thick] (1,0) .. controls (1.5,1) and (2.5,1) .. (3,0) node[midway, above] {$\rightarrow$} node[midway, below] {$p+k'$};
	\draw[thick] (1,0) .. controls (0.5,-1) and (-0.5,-1) .. (-1,0) node[midway, above] {$\leftarrow$} node[midway, below] {$k$};
	\draw[thick] (1,0) .. controls (1.5,-1) and (2.5,-1) .. (3,0) node[midway, above] {$\leftarrow$} node[midway, below] {$k'$};

	\fill[red] (1,0) circle (2pt);
	\fill[black] (-1,0) circle (2pt);
	\fill[black] (3,0) circle (2pt);
	\node at (-1.3,0) {$p$};
	\node at (3.3,0) {$-p$};
	\end{tikzpicture}
};
\node (b) at (a.east) [anchor=east,xshift=10cm]
{
\begin{tikzpicture}[scale=0.80]

\draw[thick] (8,0) .. controls (7,1) and (6,1) .. (5,0) node[midway, above] {$\rightarrow$} node[midway, below] {$p+k$};
\draw[thick] (8,0) .. controls (7.25,-0.8) .. (6.5,-1)  node[midway, below] {$k$};
\draw[->] (7.4,-0.35) -- (7,-0.65);
\draw[thick] (6.5,-1) .. controls (5.75,-0.8).. (5,0) node[midway, below] {$k$};
\draw[->] (5.9,-0.65) -- (5.5,-0.35);
\draw[thick] (6.5,-1) .. controls (4.75,-3) and (8.25,-3) .. (6.5,-1) node[midway, above] {$\leftarrow$} node[midway, below] {$k'$};
\fill[red] (6.5,-1) circle (2pt);
\fill[black] (8,0) circle (2pt);
\fill[black] (5,0) circle (2pt);

\node at (4.7,0) {$p$};
\node at (8.9,0) {$-p$};

\node at (10,0) {$+$};

\draw[thick] (14,0) .. controls (13,-1) and (12,-1) .. (11,0) node[midway, above] {$\leftarrow$} node[midway, below] {$k$};
\draw[thick] (14,0) .. controls (13.25,0.8) .. (12.5,1) node[midway, above] {$p+k$};
\draw[->] ( 13.0, 0.65) -- (13.5,0.35);
\draw[thick] (12.5,1) .. controls (11.75,0.8).. (11,0) node[midway, above] {$p+k$};
\draw[->] ( 11.6, 0.35) -- (12.1,0.65);
\draw[thick] (12.5,1) .. controls (10.75,3) and (14.25,3) .. (12.5,1) node[midway, above] {$\rightarrow$} node[midway, below] {$k'$};
\fill[red] (12.5,1) circle (2pt);
\fill[black] (14,0) circle (2pt);
\fill[black] (11,0) circle (2pt);

\node at (10.7,0) {$p$};
\node at (14.9,0) {$-p$};

\end{tikzpicture}
};
\end{tikzpicture}
\caption{Contributions to Landau parameter $F^{(2,0)}$ is captured by the diagrams on the left which correspond to $\langle \rho \rho \rangle^{\rm a}$ + $\langle \rho \rho \rangle^{\rm b}$. Self energy corrections $\langle \rho \rho \rangle^{\rm c}+\langle \rho \rho \rangle^{\rm d}$ are captured by the diagram on the right. Red dot indicates the quartic interaction. These correlations are listed in Eq.~\eqref{rhorhoabc}.}\label{directandexchange2pt}
\end{figure}

We determine these Wilson coefficients through the corresponding microscopic calculation of two point density correlator. At weak coupling, two possible diagrams contribute, as shown in fig.~\ref{directandexchange2pt}. The Wilson coefficients $\delta p_F$ and $\delta v_F$ are obtained from the self-energy diagrams, while the Landau parameter $F^{(2,0)}$ arises from the left diagram, as we show in App.~\ref{domtpcf}. For $\delta p_F$ and $\delta v_F$, it is somewhat convenient to identify them by evaluating the correction to the quasiparticle propagator near the Fermi surface. Comparing with Eqs.~\eqref{eq_rr_EFT}, \eqref{deltadepsinfM} and \eqref{rhoab} we find the Wilsonian coefficients 
\begin{equation}\label{f20infM}
\begin{split}
F^{(2,0)}(\theta, \theta') &= \frac{\lambda p^3_F}{\e_F^2v_F}\left[  1- \cos\left(\theta-\theta'\right)\right],\qquad \delta p_F =-\frac{\lambda\left(p_F^2-2 m^2\right) \epsilon_F}{4 \pi  p_F },\\
\delta v_F &=\frac{\lambda \left(p_F^2 m^2+2 m^4\right)}{4 \pi  p_F \epsilon^2_F},\qquad \epsilon_F= \sqrt{p_F^2+m^2}\, .
\end{split}
\end{equation}
Decomposing $F^{(2,0)}$ into harmonics we obtain,
\begin{equation}\label{F20harmonicsmicro}
\begin{split}
F_0=\frac{\lambda p_F^2 }{2\pi\epsilon_F},\qquad F_1=-\frac{\lambda p_F^2 }{4\pi\epsilon_F},\qquad F_{l\geq 2}=0.
\end{split}
\end{equation}
Intuitively the vanishing of the Landau parameter for $\ell\geq 2$ can be understood by considering quasi-particle scattering which is traditionally used to compute $F^{(2,0)}$. Since we are scattering two spin $1/2$ particles which interact via a zero derivative coupling, the maximum spin that can be exchanged in such a process is just the sum of angular momentum of external states, which for this case is spin 1. 

We are now in a position to check our linear Lorentz constraint involving $F^{(2,0)}$, given by Eq.~\eqref{linconstLharm}, 
\begin{equation}
p_F= v_F\epsilon_F(1+F_1).
\end{equation}
To leading order in the coupling ($\delta \epsilon_{F}=\delta \mu=0$, since the chemical potential is an external parameter in any system), the constraint becomes, 
\begin{equation}\label{linconstraintyukawa}
\left[\frac{p_F}{\epsilon_Fv_F}+\frac{\delta p_F}{\epsilon_Fv_F}- \frac{p_F \delta v_F}{\epsilon_Fv_F^2} \right]=(1+ F_1),
\end{equation}
where $\delta p_F$ and $\delta v_F$ denote the $O(\lambda)$ corrections of the EFT parameters from free fermion values due to the quartic interaction. Using the one-loop results, we find  the linear constraints are satisfied. A similar analysis of the linear constraint has been performed in \cite{MATSUI1981365}.

	\subsection*{Nonlinear Lorentz constraints}\label{nnlincons}

We now test our nonlinear constraints, Eq.~\eqref{nonlin3f2harmonics}. As with the linear constraints, we wish to verify them to leading order in the coupling constant of our microscopic theory. At weak coupling, the nonlinear constraints become
\begin{equation}
\begin{split}
\tilde{F}_ 1 v_F \epsilon_F + \left[\delta p_F \left(\epsilon_F \epsilon _ {F}''+v_F^2-1\right)+p_F \left(\epsilon_F \delta \epsilon''_F+2 \delta v_F v_F \right)\right]&=0,\\
3 G_ {1,1} \epsilon_F+ F_ 1 v_F \left(p_F v_F-\epsilon_F\right)+ v_F  \left(\tilde{F}_ 1+\tilde{F}_ 2\right) \epsilon_F&=0,\\
3 G_ {0,1} \epsilon_F+ F_ 1 v_F \left(p_F v_F+\epsilon_F\right)+ F_ 0 p_F v_F^2+v_F   \left(\tilde{F}_ 0+\tilde{F}_ 1\right) \epsilon_F&=0,\\
\end{split}
\end{equation}
where in these equations, $\delta p_F, \delta v_F, \delta \epsilon''_{F}$ and the Landau parameters $F_{\ell}, \tilde{F}_\ell$ and $G_{\ell,\ell'}$ are evaluated at $O(\lambda)$ while all the other terms correspond to values associated with the theory of free relativistic Dirac fermions considered above. For $\ell\geq 2$, we have an additional tower of constraints 
\begin{equation}
3  G_{1,\ell}+ v_F  \left(\tilde{F}_\ell+\tilde{F}_{\ell+1}\right)=0,
\end{equation}
where we have used that for our model $F_{\ell\geq 2}=0$. One observable that is sensitive to all of these Wilson coefficients is  the three point function of density. The relevant one loop contribution to the microscopic three point function is given by fig.~\ref{microscopic} 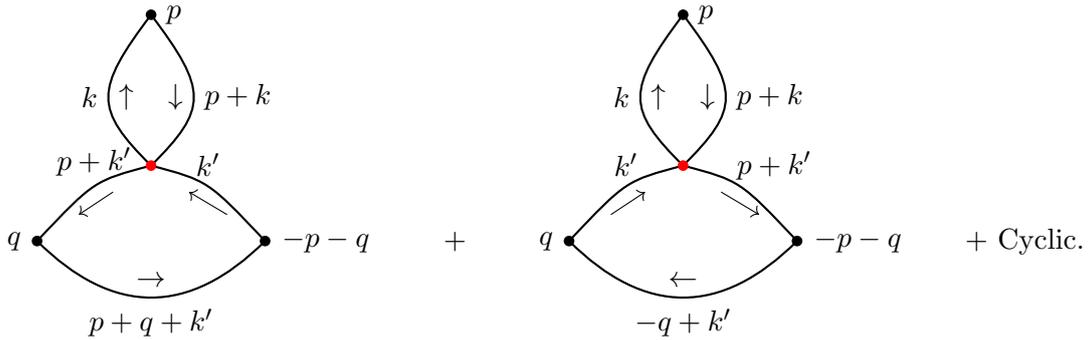
\begin{figure}[h!]
	\centering
	\begin{tikzpicture}
	\draw[thick] (14,0) .. controls (13,-1) and (12,-1) .. (11,0) node[midway, above] {$\rightarrow$} node[midway, below] {$p+q+k'$};
	\draw[thick] (14,0) .. controls (13.25,0.8) .. (12.5,1) node[midway, above] {$k'$};
	\draw[->] (13.5,0.35) -- (13,0.65);
	\draw[thick] (12.5,1) .. controls (11.75,0.8).. (11,0) node[midway, above] {$p+k'$};
	\draw[->] (12,0.65) -- (11.55,0.35);
	\draw[thick] (12.5,1) .. controls (11.75,1.75) and (11.75, 2).. (12.5,3) node[midway, right] {$\uparrow$} node[midway, left]  {$k$};
	\draw[thick] (12.5,1) .. controls (13.25,1.75) and (13.25, 2).. (12.5,3) node[midway, left] {$\downarrow$} node[midway, right]  {$p+k$};
	\fill[red] (12.5,1) circle (2pt);
	\fill[black] (14,0) circle (2pt);
	\fill[black] (11,0) circle (2pt);
	\fill[black] (12.5,3) circle (2pt);
	
	\node at (12.8,3) {$p$};
	\node at (10.7,0) {$q$};
	\node at (14.8,0) {$-p-q$};
	
	\node at (16.5,0) {$+$};

		\draw[thick] (21,0) .. controls (20,-1) and (19,-1) .. (18,0) node[midway, above] {$\leftarrow$} node[midway, below] {$-q+k'$};
	\draw[thick] (21,0) .. controls (20.25,0.8) .. (19.5,1) node at (20.7,1) {$p+k'$};
	\draw[->]  (20,0.65) -- (20.5,0.35);
	\draw[thick] (19.5,1) .. controls (18.75,0.8).. (18,0) node[midway, above] {$k'$};
	\draw[->] (18.55,0.35) -- (19,0.65);
	\draw[thick] (19.5,1) .. controls (18.75,1.75) and (18.75, 2).. (19.5,3) node[midway, right] {$\uparrow$} node[midway, left]  {$k$};
	\draw[thick] (19.5,1) .. controls (20.25,1.75) and (20.25, 2).. (19.5,3) node[midway, left] {$\downarrow$} node[midway, right]  {$p+k$};
	\fill[red] (19.5,1) circle (2pt);
	\fill[black] (21,0) circle (2pt);
	\fill[black] (18,0) circle (2pt);
	\fill[black] (19.5,3) circle (2pt);
	
	\node at (19.8,3) {$p$};
	\node at (17.7,0) {$q$};
	\node at (21.8,0) {$-p-q$};
	
	\node at (24,0) {$+$ Cyclic.};
	
	\end{tikzpicture}
		\caption{One loop correction to the three point function of densities which contribute to $F^{(2,0)}, F^{(3,0)}$. Red dot indicates the quartic interaction. The correlation function is listed in Eq.~\eqref{rrrmicroL}}
	\label{microscopic}
\end{figure}
where we have ignored the bubble diagrams which give corrections to $p_F, v_F$ and $\epsilon''_{F}$ since they have different analytic structure than the diagrams in fig.~\ref{microscopic}. We relegate the details of the microscopic computation to App.~ \ref{fmt} where the final answer is recorded in Eq.~\eqref{3ptfnmicroeftblc}. Using, $F^{(2,0)}$  in Eq.~\eqref{f20infM}, we now compare with the analogous EFT computation in Eq.~\eqref{rho3eft} to obtain the generalized Landau parameters at weak coupling,
\begin{equation}\label{f3f201match}
\begin{split}
F^{(3,0)}(\theta,\theta',\theta'')=O(\lambda^2),\qquad F_1^{(2,0)}(\theta,\theta')=\frac{\lambda  p_F^2}{\epsilon _F^3} \left[m^2-\epsilon _F^3 \epsilon ''_F \cos (\theta -\theta_p)\right]\,.
\end{split}
\end{equation}
Additionally, we obtain $\delta \epsilon''_{F}, \delta v_F$ and $\delta p_F$ from  \eqref{deltadepsinfM} and find that the constraints are satisfied. 

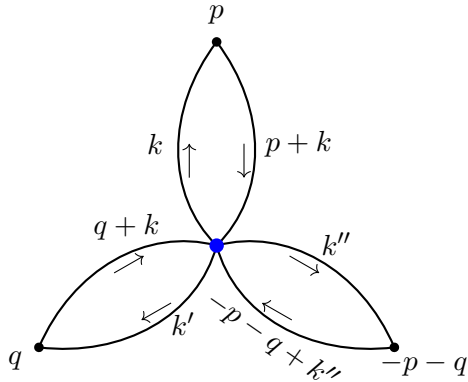
\begin{figure}
\begin{adjustwidth}{4cm}{}
	\begin{tikzpicture}[scale=0.9]
	\draw[thick] (0,0) .. controls (-.75,.75) and (-.75, 2).. (0,3);
	\node at (-0.9,1.5)  {$k$};  
	\draw[->] (-0.4,1) -- (-0.4,1.5);
	\draw[thick] (0,0) .. controls (.75,.75) and (.75, 2).. (0,3);  \node at (1.2,1.5) {$p+k$};
	\draw[->] (0.4,1.5) -- (0.4,1);
	\node at (0.0,3.4) {$p$};
	\fill[black] (0,3) circle (2pt);
	\begin{scope}[rotate=120]
	\draw[thick] (0,0) .. controls (-.75,.75) and (-.75, 2).. (0,3);
	\node at (-0.75,1)  {$k'$};  
	\draw[->] (-0.4,1) -- (-0.4,1.5);
	\draw[thick] (0,0) .. controls (.75,.75) and (.75, 2).. (0,3);  \node at (0.9,1) {$q+k$};
	\draw[->] (0.4,1.5) -- (0.4,1);
	\node at (0.0,3.4) {$q$};
	\fill[black] (0,3) circle (2pt);
	\end{scope}
	\fill[red] (0,0) circle (2pt);
	\begin{scope}[rotate=240]
	\draw[thick] (0,0) .. controls (-.75,.75) and (-.75, 2).. (0,3);
	\node at (-0.9,1.5)  {$k''$};  
	\draw[->] (-0.4,1) -- (-0.4,1.5);
	\draw[thick] (0,0) .. controls (.75,.75) and (.75, 2).. (0,3);  \node at (0.8,1.4) [rotate=-35] {$-p-q+k''$};
	\draw[->] (0.4,1.5) -- (0.4,1);
	\node at (0.0,3.5) {$-p-q$};
	\fill[black] (0,3) circle (2pt);
	\end{scope}
	\fill[blue] (0,0) circle (3pt);
	\end{tikzpicture}
    \end{adjustwidth} \caption{One-loop correction to the three-point density function in a theory with a $\left(\bar{\psi}\psi\right)^3$ interaction (blue dot) which generates $F^{(3,0)}$.}
	\label{F30_sextic}
\end{figure}

To leading order in the coupling $\lambda$, the nonlinear constraints in fact reduce to those of Ref.~\cite{BAYM1976527}. At subleading order we expect a non-trivial contribution to $F^{(3,0)}$ in our microscopic model. Alternatively, we could have also considered a weakly coupled model of Dirac fermions with a sextic interaction $(\bar{\psi}\psi)^3$. In this case, we expect a non-trivial contribution to all the generalized Landau parameters appearing in the three point density function including $F^{(3,0)}$ already at {\it leading} order in the coupling. In particular, the leading contribution to $F^{(3,0)}$ is obtained by evaluating the diagram in fig.~\ref{F30_sextic}.
In general, we expect that the Landau parameters wil not satisfy the nonlinear constraint of Ref.~\cite{BAYM1976527} but will instead obey our constraint in Eq.~\eqref{nonlin3f2harmonics}, which takes into account the generalized Landau parameters. The fact that irrelevant generalized Landau parameters also enter in this constraint would have been difficult to anticipate from the fermionic approach to Fermi liquid EFTs, as it requires recognizing that cancellations in fermion loops allow for naively subleading diagrams to compete with the leading ones.

\subsection{Chern Simons theories coupled to matter}\label{ssec_CS}

Our constraints for scale invariance find natural application in Fermi liquids with a conformal field theory as its UV completion. One such well studied example is $U(N)$ Chern-Simons theories at level $\kappa$ coupled to either bosons or fermions in the fundamental representation \cite{Giombi:2011kc, Aharony:2011jz}. In the large $\kappa, N$ limit, these theories can be solved to all orders in the finite 't Hooft coupling $\lambda=N/\kappa$. 

In Ref.~\cite{Geracie:2015drf}, these CFTs were studied at finite chemical potential where they were found to exhibit characteristics of Fermi liquid; we refer the reader to this paper for the relvant details, quoting only the necessary results here for our purpose. The Landau parameters to all orders in 't \,Hooft coupling and leading order in large $N$ were obtained as
\begin{equation}
F_0= \frac{\lambda^2}{1-\lambda^2}+O\left(\frac1N\right),\qquad F_{\ell\geq 1}= O\left(\frac1N\right),\qquad |\lambda|\leq 1\, .
\end{equation} 
The Fermi surface is characterised by Fermi momentum $p_F= \mu \sqrt{1-\lambda^2}$ (the finite density equation of state of a conformal Fermi liquid is trivial, except for the dimensionless coefficient $p_F/\mu$). The Fermi velocity is given by $v_F = p_F/\mu = \sqrt{1-\lambda^2}\leq 1$ (or, $m_* = \mu$), and is subluminal as expected. 

These values satisfy both constraints from Lorentz and conformal invariance found in Sec.~\ref{sec_linear}, namely
\begin{equation}
\frac{p_F}{\mu }=v_F(1+F_1), \qquad \frac{\mu}{p_F}=v_F (1+F_0)\, .
\end{equation}
As $\lambda$ is tuned between 0 and 1, this CFT carves a one-parameter family of conformal Fermi liquids shown in red in Fig.~\ref{fig_causality_collective_2d}. Their only collective excitation is zero sound, with subluminal velocity as required by causality
\begin{equation}
v_0 = v_F \frac{1+F_0}{\sqrt{1+2F_0}} = \frac{1}{\sqrt{1+\lambda^2}} \leq 1\, . 
\end{equation}
In the limit of strong coupling, $F_0 \rightarrow \infty$ and we recover the speed of sound of a conformal superfluid as expected from our general analysis in Sec.~\ref{ssSCL}:
\begin{equation}
    v_0 \xrightarrow{\lambda\to 1} \frac{1}{\sqrt{2}}\, .
\end{equation}
In this limit, the system behaves like a conformal superfluid, even though it does not strictly order. This is due to an order of limits: order parameters are exactly static, zero frequency observables, while the system only behaves like a superfluid for frequencies $\omega\gg v_F q  \simeq \sqrt{2 (1-\lambda)}q$; the $\omega\to 0$ and $\lambda\to 1$ limits therefore do not commute. In this Chern-Simons--matter theory, this superfluid-like phase realized at strong coupling may be simpler to understand from the dual bosonic Chern-Simons--matter description, which becomes weakly coupled \cite{Geracie:2015drf}.

Note that in addition to Poincar\'e and conformal symmetry, these CFTs have a higher-spin symmetry \cite{Maldacena:2011jn}. We expect this tower of symmetries will lead to Ward identities similar to App.~\ref{app_thermoward}, relating the other Landau parameters $1+F_{\ell}$ to the expectation values of the higher spin currents $\langle J_{\ell+1} \rangle$, beyond the $\ell=0,1$ cases we consider ($U(1)$ current and stress tensor). However, these expectation values are not universally related to the energy and charge density.

\section{Discussion} 

We have seen how relativistic and non-relativistic boost and scaling symmetries place non-perturbative constraints on the Landau parameters of Fermi liquids, and, for relativistic systems, how analyticity properties of Green's functions further bound the remaining parameter space. Beyond the possible applications to QCD at large baryon density, and many-body systems near a quantum phase transition, our results have formal implications for the landscape of compressible phases. A general approach to realizing compressible phases in QFT is to dope a CFT: turning on a chemical potential for a CFT (in more than 1+1 dimensions) that has a $U(1)$ symmetry \cite{Sachdev:2012tj}. Can any compressible phase be reached in this manner, or are some ruled out? We have shown that the parameter space for conformal Fermi liquids is tightly constrained, see Fig.~\ref{fig_causality_collective_2d}. 
Our results may also help guide holographic constructions of Fermi surfaces, which has proven particularly challenging \cite{Lee:2008xf,Karch:2008fa,Kulaxizi:2008kv,Kulaxizi:2008jx,Hartnoll:2009ns,Faulkner:2009wj,Faulkner:2010tq,Huijse:2011ef,Davison:2022vqh,Else:2023xgk}. 
They can also be interpreted as results on the large charge spectrum of CFTs -- while such states are sometimes assumed to be in a superfluid phase \cite{Hellerman:2015nra,Monin:2016jmo}, many other possibilities exist, including Fermi liquids \cite{Komargodski:2021zzy,Dondi:2022zna}. The landscape of compressible phases seems significantly more complex than the landscape of CFTs; for example, we have seen that the moduli space of Fermi liquids {\em includes} superfluids and solids. Relatedly, it would also be interesting to find CFTs that have a large $F_1$ (instead of the large $F_0$ discussed in Sec.~\ref{ssec_CS}), which could realize a conformal solid.

Finally, there are several ways in which our results could be improved. It would be interesting if Lorentz invariance, say, could be made {\em manifest} in an EFT of Fermi liquids, similar to how this is possible for superfluids \cite{Son:2002zn}. Relatedly, from spin-statistics relations one may expect that a spin degeneracy is inevitable for a 3+1d relativistic Fermi liquid. Lastly, one could imagine using dispersion relations to further constrain the Wilsonian coefficients of Fermi liquids \cite{Bellazzini:2020cot, Caron-Huot:2024lbf, Creminelli:2022onn}.

\subsection*{Acknowledgements}

We thank Jing-Yuan Chen, Aditya Hebbar, Zohar Komargodski, Petr Kravchuk, Gautam Mandal, Ira Rothstein and Dam Thanh Son for helpful discussions and comments on the draft. LD is supported by an NSF award No.~PHY2412710. SDC is supported by NSF Grant No. PHY2014195 (PI: S.~Sethi), Kadanoff fellowship at the University of Chicago and ``Exotic High Energy Phenomenology'' (X-HEP), a project funded by the European Union -- Grant Agreement n.~101039756 (PI: J.~Elias~Mir\'o). Views and opinions expressed are however those of the author(s) only and do not necessarily reflect those of the European Union or the ERC Executive Agency (ERCEA). Neither the European Union nor the granting authority can be held responsible for them. UM is supported by a Gordon and Betty Moore Foundation Grant GBMF10279 (PI: A.~Lucas) as well as a Simons Postdoctoral Fellowship in Ultra-Quantum Matter.

\appendix

\section{Symmetry and Causality constraints in other dimensions}\label{app_dimgen}

While the nonlinear EFT in \eqref{eq_action_f} is expressed in general spatial dimension $d$, we focused on $d=2$ for simplicity when expanding in fluctuations. We generalize some of our results here to $d=1$ and $d=3$. 

\subsection{$d=1$}

The bosonized EFT used throughout the text reduces to conventional bosonization in one spatial dimension, where it captures Luttinger liquids. Spacetime symmetry constraints on Luttinger liquids were obtained before \cite{Cazalilla_2004,Nishida:2009pg}. We will show here how our approach reproduces these constraints.
Expanding Eq.~\eqref{eq_S2_coadj} in fluctuations for a 1d Fermi surface $f_0(p) = \Theta(p_F - |p|)$ gives
\begin{equation}\label{eq_Luttinger_liquid}
\begin{split}
S 
	&=- \frac{1}{4\pi} \int \sum_\sigma\sigma \dot \phi_\sigma\partial_x \phi_\sigma+ v_F^0 \left[\sum_{\sigma} (\partial_x \phi_\sigma)^2 +  \sum_{\sigma\sigma'} F_{\sigma\sigma'} \partial_x \phi_{\sigma}\partial_x \phi_{\sigma'}  \right]\\
	&=- \frac{1}{4\pi} \int \sum_\sigma\sigma \dot \phi_\sigma\partial_x \phi_\sigma+ v_F \left[\sqrt{1+g^2}\sum_\sigma\left(\partial_x \phi_\sigma\right)^2 + 2g \partial_x \phi_+ \partial_x\phi_-\right]\, , 
\end{split}
\end{equation}
where $\phi_{\pm}$ denotes the field at the Fermi point $\pm p_F$. In the second line, we noted that $F_{++} = F_{--}$ can be absorbed in $v_F$. There is therefore a single marginal ``Landau parameter'' $g$, which in fermionic language would correspond to an interaction $\delta S = 2\pi v_F g \int \psi^\dagger_L\psi_L\psi^\dagger_R\psi_R$.%
	\footnote{Note that to preserve continuity with the rest of the paper, we let $v_F$ denote the physical velocity of particle-hole excitations, as in higher dimensions. Most references on 1d bosonization use $v_F$ to denote the bare Fermi velocity (denoted by $v_F^0$ in Eq.~\eqref{eq_Luttinger_liquid}).}
The action \eqref{eq_Luttinger_liquid} can be diagonalized to produce a compact boson with Luttinger parameter
\begin{equation}
K = \sqrt{1+g^2} + g\, .
\end{equation}
Letting $\rho_\sigma = \frac1{2\pi} \partial_x \phi_\sigma$, the Hamiltonian in terms of the charge and momentum densities
\begin{equation}
\rho = \rho_+ - \rho_-\, , \qquad
\pi_x = p_F (\rho_+ + \rho_-)\, 
\end{equation}
is
\begin{equation}
H = \frac{\pi v_F}2\int dx  \, K (\pi_x/p_F)^2 + \frac{1}{K}\rho^2\, , 
\end{equation}
from which we can read off the susceptibilities
\begin{equation}
\chi = \frac{1}{\pi v_F}  K\, , \qquad
\chi_{\pi\pi} = \frac{p_F^2}{\pi v_F} \frac{1}{K}\, .
\end{equation}
Comparing to the higher-dimensional expressions, one can loosely identify the correction to the Luttinger parameter $K-1$ as a Landau parameter: $K = 1+F_1 = \frac1{1+F_0}$. Boost invariance leads to:
\begin{subequations}
\begin{align}\label{seq_Lorentz}
\hbox{Lorentz:}&
	& K = \frac{p_F}{\mu v_F} = \frac{m_*}{\mu}\, ,  && \\
\hbox{Galileo:}& \label{seq_Galileo}
	& K = \frac{p_F}{m v_F} = \frac{m_*}{m}\, . && 
\end{align}
\end{subequations}
The Galilean constraint is discussed in Ref.~\cite{Cazalilla_2004}. Dilation invariance leads to 
\begin{equation}\label{eq_scale}
n\propto \mu^{1/z} \quad \Rightarrow \quad 
	\frac{1}{K} = z \frac{\mu}{v_F p_F}\, .
\end{equation}

\paragraph{CFT:} imposing both the Lorentz and the $z=1$ dilation constraint leads to 
\begin{equation}\label{eq_CFT_constraint}
v_F = 1\, , \qquad c_J =  \pi \chi =  \frac{1}{K}\, .
\end{equation}
The first relation is not surprising: 1+1d CFT observables at finite density are completely fixed in terms of zero density observables by the extended algebra, leaving no room for the emergence of novel IR physics. The second relation shows that the $U(1)$ level of the CFT $c_J = k$ can be changed by turning on interactions $K\neq 1$.

\paragraph{NRCFT:} imposing both the Galilean and $z=2$ dilation constraints, one finds that the equation of state and effective mass are entirely fixed by the Luttinger parameter
\begin{equation}\label{eq_1d_NRCFT_constraint}
\frac{p_F^2}{2m\mu} = K^2 = \left(\frac{m_*}{m}\right)^2\, .
\end{equation}
Ref.~\cite{Nishida:2009pg} studies an example of an interacting NRCFT in $d=1$ spatial dimension that becomes a Luttinger liquid upon doping, with EFT parameters satisfying \eqref{eq_1d_NRCFT_constraint}.

\paragraph{Nonlinear constraints:}
in one spatial dimension, the bosonized EFT of a Fermi (Luttinger) liquid is identical to that of a superfluid. It is well-known (and much simpler) to impose spacetime symmetries in the superfluid EFT. For example, Lorentz or Galilean boost invariance implies that the EFT must take the form (to leading order in derivatives, but all order in fluctuations) \cite{Son:2002zn, Son:2005rv}
\begin{equation}
\mathcal L = P(\bar \mu)\, , \qquad
\bar\mu \equiv
	\begin{cases}
	 \sqrt{-(\mu \delta_\nu^0 + \partial_\nu \phi)^2} & \hbox{Lorentz},\\
	\mu + \dot \phi + \frac1{2m}(\nabla \phi)^2 &\hbox{Galileo}.
	\end{cases}
\end{equation}
Scale invariance further restricts $P(\mu) \propto \mu^{1 + \frac{d}z}$. One can check that linearizing this action reproduces the constraints found above in $d=1$.

\subsection{$d=3$}

We now turn to Fermi liquids in three spatial dimensions. Because we will focus on linear response of density and momentum to leading order at small momenta, the fermion spin simply adds a degeneracy factor. 
We will consider spinless fermions for simplicity and comment on where this factor enters below. Expanding Eq.~\eqref{eq_S2_coadj} for a spherical Fermi surface gives
\begin{equation}\label{eq_3d_S_gaussian}
S
	= -\frac{p_F^2}{2} \int \frac{dt d^3x d^2\Omega}{(2\pi)^3}
	\nabla_n \phi \left( \dot \phi + v_F \nabla_n \phi + v_F \int \frac{d^2\Omega'}{(2\pi)^3} F^{(2,0)}(\Omega,\Omega')\nabla_{n'}\phi'\right)+ \cdots\, ,
\end{equation}
with $\nabla_n\equiv \hat n(\theta,\varphi)\cdot\nabla $ where now $n(\theta,\varphi) = (\sin\theta\cos\varphi, \sin\theta\sin\varphi,\cos\theta)$. The density at each patch, which is conserved in this Gaussian approximation, can be decomposed into spherical harmonics:
\begin{equation}
\rho(\theta,\varphi) = \frac{p_F^{2}}{(2\pi)^3}\nabla_n \phi= \frac{1}{\sqrt{4\pi}} \sum_{\ell=0}^\infty \sum_{m=-\ell}^l \rho_{\ell m} Y_{\ell m}(\theta,\varphi)\, ,
\end{equation}
In this normalization, the charge and momentum densities are given by
\begin{equation}
\rho = \rho_{00}\, , \qquad
\pi_i
	= \frac{p_F}{\sqrt{3}} 
\left(\begin{array}{c}
\sqrt{2} \Re \rho_{11}\\
\sqrt{2} \Im \rho_{11}\\
\rho_{10}
\end{array}\right)\, .
\end{equation}
By rotation invariance, the Landau parameter $F^{(2,0)}(\Omega,\Omega')=F^{(2,0)}(\gamma)$ is only a function of the angle between $\Omega$ and $\Omega'$, which satisfies $\cos \gamma = \cos\theta\cos\theta' + \sin\theta\sin\theta' \cos(\varphi-\varphi')$. It can therefore be expanded in terms of Legendre polynomials as
\begin{equation}
\frac{1}{(2\pi)^3}F^{(2,0)}(\gamma)
	\equiv \frac{1}{4\pi}\sum_\ell (2\ell + 1) F_{\ell} P_\ell(\cos \gamma)
	= \sum_\ell F_{\ell} \sum_{m=-\ell}^\ell Y_{\ell m}^*(\Omega)Y_{\ell m}(\Omega')\, , 
\end{equation}
where the last step made use of the addition theorem for spherical harmonics. We are following the normalization of Ref.~\cite{pitaevskii1980statistical} for the Landau parameters $F_{\ell}$ -- another common normalization \cite{pinesnozieres2018theory} is to define instead $F_{\ell}^{\rm there} = (2\ell+1)F_{\ell}^{\rm here}$. The Hamiltonian is then given by
\begin{equation}
H = \frac12 \frac{2 \pi^2 v_F}{p_F^2} \int dt d^3x \sum_{\ell, m} |\rho_{\ell m}|^2(1+F_{\ell})\, , 
\end{equation}
from which one can read off the charge susceptibilities
\begin{equation}\label{eq_chi_lm}
\chi_{\rho_{\ell m}\rho_{\ell m}^*}
	= \frac{3 \rho}{p_F v_F} \frac{1}{1+F_{\ell}}\, ,
\end{equation}
where we used the Luttinger relation $\rho = \frac{1}{3} \frac{p_F^3}{2\pi^3}$.  Accounting for spin would produce an extra factor of $2$ in each of the susceptibilities, as well as in the total density $\rho$; Eq.~\eqref{eq_chi_lm} therefore holds both for spinless and spinful Fermi surfaces. These susceptibilities must be positive, which imposes $F_{\ell}>-1$.

We are now ready to impose symmetries: scale invariance \eqref{eq_chi0_scale} again leads to
\begin{equation}
\chi_{0,0} = \frac{3\rho}{\mu} \frac{1}{z} \qquad \Rightarrow \qquad
1+F_0 = \frac{\mu}{p_F v_F} z\, .
\end{equation}
Boost invariance \eqref{eq_chipp_boost} also again leads to \eqref{eq_F1sym}. For a conformal Fermi liquid in (3+1)-dimensions, the Fermi velocity is therefore still given by \eqref{eq_vF_CFT}, and Eq.~\eqref{eq_pF_mu_CFT} holds as well. Imposing $v_F\leq 1$ again leads to the bound \eqref{eq_F0F1_CFTconstraint}.

Collective excitations for the simple model where $F_{l\geq 2} = 0$ were studied in Ref.~\cite{Abrikosov_1959}, where it was found that the zero sound mode is the solution to
\begin{equation}\label{eq_zerosound_3d}
\frac{s}{2} \log \frac{s+1}{s-1} - 1
	= \frac{1+F_1}{F_0 + F_0 F_1 + 3 F_1 s^2}\, ,
\end{equation}
where $s=\omega/v_Fq$.
This mode exists when 
\begin{equation}\label{3d_zero_exists}
F_0 + F_0 F_1 + 3 F_1 > 0
\end{equation}
a condition similar to the $d=2$ case Eq.~\eqref{eq_ZS_exists}, which can be written $F_0 + F_0 F_1 + 2 F_1 > 0$. The shear mode satisfies a similar equation
\begin{equation}
\frac{s}{2} \log \frac{s+1}{s-1} - 1
	= \frac{F_1-2}{3F_1 (s^2 - 1)}\, ,
\end{equation}
which has a solution when $F_1 > 2$. Combining these conditions leads to the exclusion plot in Fig.~\ref{fig_causality_collective}. We do not have an analytic expression for the upper boundary of the light gray shaded region, which is obtained by demanding that zero sound \eqref{eq_zerosound_3d} be subluminal. This condition can be studied analytically at large $F_1$, and leads to the constraint $F_0 \geq -\frac35$ (the analog of the $d=2$ condition $F_0 \geq -\frac12$ found in Sec.~\ref{sec_linear}). This is a stronger condition than the well-known stability condition $F_0\geq -1$.

In the limit of large Landau parameters, the Fermi liquid behaves again like a superfluid or solid, as in $d=2$. Considering a conformal Fermi liquid, which must satisfy \eqref{eq_vF_CFT}, one finds that when $F_0,F_1 \to \infty$ with fixed ratio (as well as when only $F_0\to \infty$), the speed of zero sound is given by the conformal value
\begin{equation}
v_0 \to \frac{1}{\sqrt{3}}\, .
\end{equation}
Instead, if only $F_1\to \infty$, one finds that both the zero sound and shear sound velocities remain finite, and satisfy the relation
\begin{equation}
\lim_{F_1\to \infty} v_\perp = \frac1{\sqrt{5(1+F_0)}}\, , \qquad \hbox{and} \qquad
v_0^2 = \frac13 + \frac43 v_\perp^2\, , 
\end{equation}
as expected for a conformal solid in $d=3$ \cite{Esposito:2017qpj}.

\section{Relativistic Ward identities at finite density}\label{app_thermoward}

Let us study some consequences of Lorentz invariance and $U(1)$ symmetry in a general finite density phase. We follow an approach similar to Refs.~\cite{Policastro:2002tn,Herzog:2009xv}, by coupling the system to sources $g_{\mu\nu},\,A_\mu$, and using gauge and diffeomorphism invariance of the path integral $Z[g,A]$ (a slightly different approach is considered in \cite{Alberte:2020eil,Komargodski:2021zzy}). This leads to the following Ward identities
\begin{equation}\label{eq_WI}
\nabla_\mu T^{\mu\nu} = F^{\nu\lambda} j_\lambda\, , \qquad
\nabla_\mu j^\mu = 0\, ,
\end{equation}
which can be viewed as expectation values in the presence of sources. Our goal is to derive the relativistic momentum susceptibility \eqref{eq_chipp_boost}: $\chi_{T^{0i}T^{0j}} = (\varepsilon+P)\delta_{ij}$. As a warm-up exercise, we will first show that the cross susceptibility between current and momentum density is set by the charge density: $\chi_{T^{0i} j^j} = \rho\delta^{ij}$. 
Taking the derivative of \eqref{eq_WI} with respect to $A^\alpha$ and then setting the sources to zero gives, after analytic continuation to Minkowski space,
\begin{equation}
p_\mu G^R_{T^{\mu\nu} j^\alpha}(p)
	= p^\nu \langle j^\alpha\rangle - p_\mu \eta^{\alpha\nu} \langle j^\mu\rangle\, .
\end{equation}
Using $\langle j^\mu\rangle = \delta^\mu_0 \rho$ and sending $p_i\to 0$, one finds
\begin{equation}
G^R_{T^{0\nu} j^\alpha}(\omega,\vec k=0)
	= -\rho \left(\delta^\nu_0 \delta^\alpha_0 + \eta^{\nu\alpha} \right)\, .
\end{equation}
This should vanish, given that the charges commute! We therefore learn that a contact term must be added for the analytic continuation of the Euclidean Green's function to have the interpretation of a retarded Green's function. Considering%
	\footnote{The most general counterterm satisfying symmetry of the stress tensor, rotation symmetry of the finite density state, and that restores vanishing of commutators is $\tilde G^R_{T^{\mu\nu} j^\alpha}
	= G^R_{T^{\mu\nu} j^\alpha} + \rho C^{\mu\nu\alpha}$ with $C^{\mu\nu\alpha} = \delta^\mu_0 \eta^{\alpha\nu} + \delta^\mu_0\delta^\nu_0\delta^\alpha_0 + a (\delta^\alpha_0 \eta^{\mu\nu} + \delta^\mu_0\delta^\nu_0\delta^\alpha_0) + (\mu \leftrightarrow \nu)$.}
\begin{equation}
\tilde G^R_{T^{0\nu} j^\alpha}
	= G^R_{T^{0\nu} j^\alpha} + \rho \left(\delta^\nu_0 \delta^\alpha_0 + \eta^{\nu\alpha} \right)\, , 
\end{equation}
taking now $p_0\to 0$ in the Ward identity one finds (setting $\nu=0,\,\alpha=j$)
\begin{equation}
p_i\tilde G^R_{T^{0i} j^j}(\omega=0,\vec k)
	= \rho \delta^{ij} p_i\, .
\end{equation}
Assuming that the $\vec k\to 0$ limit is regular (i.e., the static susceptibility is well-defined), one finds the cross susceptibility between current and momentum 
\begin{equation}
\chi_{T^{0i} j^j} \equiv
\lim_{k\to 0}\lim_{\omega\to 0} \tilde G^R_{T^{0i} j^j}(\omega,\vec k) =  \rho \delta^{ij}\, .
\end{equation}
Matching this result in the EFT leads to the Luttinger theorem.

One can similarly obtain the momentum susceptibility. Differentiating the diffeomorphism Ward identity now with respect to the metric one finds
\begin{equation}
p_\mu G_R^{\mu\nu\rho\sigma}(p)
	= -p_\mu \left(\eta^{\rho\nu} \langle T^{\mu\sigma}\rangle + \eta^{\sigma\nu} \langle T^{\mu \rho}\rangle - \eta^{\mu\nu} \langle T^{\rho\sigma}\rangle\right).
\end{equation}
First setting $\omega=0$ as before, one finds
\begin{equation}
p_0 G_R^{0i0j}(p^0,0)
	= -p_0\delta^{ij} \langle T^{00}\rangle \qquad \Rightarrow \qquad
\lim_{k\to 0}G_R^{0i0j} = -\delta_{ij} \varepsilon\, , 
\end{equation}
i.e. we must add the counterterm $\tilde G_R^{0i0j} =  G_R^{0i0j}  +\delta_{ij} \varepsilon$. After doing so, setting $\vec k=0$ in the Ward identity gives
\begin{equation}
k_i G_R^{i0j0}
	= k_i \langle T^{ij}\rangle = k_i \delta^{ij} P\, .
\end{equation}
The corrected Green's function therefore satisfies
\begin{equation}
\chi_{P^i P^j}
	\equiv \lim_{k\to 0} \lim_{\omega\to 0}
	\tilde G_R^{i0j0}(\omega,k) = \delta^{ij} (\varepsilon + P)\, .
\end{equation}
%

\section{Operator matching in free EFT}\label{doOm}
In this appendix we compute the two point function of operators in free fermion theory at finite chemical potential and compare it with analogous computation in the EFT to extract EFT data.	The density two point function in the free fermion theory takes the following form 
\begin{equation}
\begin{split}
\langle \rho_p \rho_{-p} \rangle^{\rm free} &=-\int_k  {\rm Tr}\left[\gamma^0 S(\omega_k + \omega, \kv+\p)\gamma^0 S(\omega_k, \kv)\right],	\end{split}
\end{equation}
where $\int_k \equiv \frac{d^3k}{(2\pi)^3}$, the propagator is defined in Eq.~\eqref{finitedenprop} and $\langle \rho_p \rho_{-p} \rangle \equiv \langle \rho \rho \rangle (\omega, \p)$ for notational convenience. We are interested in the leading order result of this integral in the limit $\omega, p \ll p_F$.

In order to compute the internal loops, we analyse the poles of $\omega$ from the fermion propagator.  Since the position of the poles in the complex $\omega$ plane depend on the sign of $\epsilon_{k+p} - \mu$ and  $\epsilon_{k} - \mu$, it is useful to examine their behaviour for the small external momenta and $k\simeq p_F$ (since $\epsilon_{p_F}=\mu$).  
\begin{equation}\label{epsilonexp}
\epsilon_{k+p} \sim \mu +  \left(\delta k + p_n \right) \epsilon'_{p_F} + \cdots, \qquad \delta k= |\kv|-p_F\ll p_F,\qquad p_n\equiv \p \cdot \hat{k}= p \sin \theta,
\end{equation}
where we $p= p \hat{y}$ using rotational symmetry. Here, we have retained only the leading-order contributions to $\epsilon_{k+p}$ in the limit of small external momenta. However, in principle, subleading terms in the external momenta could also be considered, though they are irrelevant to our analysis.

A careful analysis shows that UV divergences in this integral are sub-leading in external momenta. The dominant finite contribution to the correlation functions in this kinematic regime comes when the particle-hole pairs are very close to the Fermi surface. The finite part of the integral is given by 
\begin{equation}\label{EFTmatching_free_finite}
\begin{split}
\langle \rho_p \rho_{-p} \rangle_{\rm finite}^{\rm free}  &=-\int\limits_{-\infty}^\infty \frac{d \omega_k}{(2\pi)^3}\left[\int\limits_0^\pi d\theta  \int\limits_{p_F-p_n}^{p_F}\, k dk\,+ \int\limits_\pi^{2\pi} d\theta \int\limits_{p_F}^{p_F-p_n}\, k dk\,\right]\\
&\times \frac{2 \left(k^2+k p_n +\mu ^2+2 \mu  \omega_k+\mu  \omega+m^2+\omega_k^2+\omega_k \omega\right)}{\left[\omega_k-\left(\epsilon_{k}-\mu\right)+ i \eta\,\text{Sgn}(\epsilon_{k}-\mu)   \right]\left[\omega_k+\left(\epsilon_{k}+\mu\right)- i \eta\right]}\\
&\times \frac{1}{{\left[\omega +\omega_k-\left(\epsilon_{k+p}-\mu \right)+ i \eta\,\text{Sgn}(\epsilon_{k+p}-\mu) \right]\left[\omega+\omega_k+\left(\epsilon_{k+p}+\mu\right)- i \eta\right]}}.
\end{split}
\end{equation}
We can close the $\omega$ contours in the lower half plane for both the $\omega$ integrals and perform the rest of the integrals to finally obtain 
\begin{equation}\label{rhoimicro}
\begin{split}
\lim_{\omega, p \ll p_F}\langle \rho_p \rho_{-p} \rangle^{\rm free} &=\frac{ i   p_F }{2 \pi \epsilon'_{p_F}}\left[\frac{s}{\sqrt{(s+i 0^+)^2-1}}-1\right],\qquad s=\frac{\omega}{\epsilon'_{p_F} q}.
\end{split}
\end{equation}
 We compare this with the two point function computed in the free EFT. For this, we only need the linear  expansion in $\phi$ for our EFT operators and the free Gaussian action,
\begin{equation}\label{rhorhoiphiexp}
    \begin{split}
        \rho&=\frac{p_F}{(2\pi)^2}\int d\theta\, \nabla_n \phi +
        + O(\phi^2),\qquad j^i=\frac{v_F
        p_F}{(2\pi)^2}\int d\theta\, n^i \nabla_n \phi+ O(\phi^2)
    \end{split}
\end{equation}
leading to the following identification for free relativistic fermions.
\begin{equation}
 \langle \rho_p \rho_{-p} \rangle^{\rm EFT}= \frac{ i   p_F }{2 \pi v_F}\left[\frac{s}{\sqrt{(s+i 0^+)^2-1}}-1\right],~~~
 v_F= \epsilon'_{p_F} = \frac{p_F}{\sqrt{p_F^2+m^2}}
\end{equation}
since the function $v_F$ exhibits state dependence only through $|\p|=p_F$, the entire function $\epsilon(|\p|)$ in the free EFT is fixed to be $\sqrt{\p^2+m^2}$ as expected.

We now demonstrate that working to leading order in the expansion of $\epsilon_p$, as shown in Eq.~\eqref{epsilonexp}, is sufficient and that no terms with the same analytic structure, potentially arising from a subleading analysis, have been overlooked. In order to do so let us expand Eq.~\eqref{epsilonexp} to subleading order,
\begin{equation}\label{subleading_epsilon_exp}
\epsilon_{k+p} \sim \mu +  \left(\delta k + p_n \right) \epsilon'_{p_F} +\frac{1}{2 p_F} \left[p_F p_n^2 \epsilon_{p_F}''+(p^2-p^2_n) \epsilon_{p_F}'\right]+ \cdots
\end{equation}
After a few lines of algebra, we see that the finite part of the integral is again given by Eq.~\eqref{EFTmatching_free_finite} but now the limit of the integral changes according to Eq.~\eqref{subleading_epsilon_exp}.
\begin{equation}
p_F-p_n \rightarrow p_F-\left(p_n+ \frac{1}{2 \epsilon'_{p_F} p_F} \left[p_F p_n^2 \epsilon_{p_F}''+(p^2-p^2_n) \epsilon_{p_F}'\right]\right)
\end{equation}
Following the same process we arrive at the same result to leading order in small $\omega, \p$. We can also obtain similarly the two point function of the density operator with the current which agrees with the EFT computation as well.
\begin{equation}
\langle \rho_p j_{-p}^y\rangle^{\rm free}=
\frac{i p_F s }{2 \pi}\left[\frac{s}{\sqrt{(s+i 0^+)^2-1}}-1\right]
\end{equation}

Finally we turn to the operator $\mathcal{O}= \bar{\psi} \psi$, for which after a similar computation in the free fermion theory one obtains to leading order in small frequencies and external momenta,
\begin{equation}\label{rhotmicro}
\langle \rho_p \mathcal{O}_{-p} \rangle^{\rm free} = \frac{i m\, p_F }{2 \pi  \epsilon_{p_F}  v_F}\left[\frac{s}{\sqrt{(s+i 0^+)^2-1}}-1\right].
\end{equation}
The EFT computation using the Gaussian action gives us, 
\begin{equation}\label{rhorhoteft}
\begin{split}
\langle \rho_p \mathcal{O}_{-p} \rangle^{\rm EFT} &=\frac{i \gamma (p_F) p_F}{2 \pi  v_F} \left[\frac{s}{\sqrt{(s+i 0^+)^2-1}}-1\right].
\end{split}
\end{equation}
This leads to the identification $\gamma(p_F)=\frac{m}{\mu}$ for a theory of free Dirac fermions.

Our analysis thus far suggests that, in principle, the EFT parameters can be identified at the integrand level, without explicitly performing the angular integrals, as the analytic structures of these integrals are distinct. This observation will be important when computing the Landau parameters from the microscopic theory.

\section{Two point density correlation: One loop contribution}\label{domtpcf}
In this appendix we  evaluate the one loop contribution to the density two point function in \eqref{yukawa}, without explicitly using the finite density fermion propagator. This utilizes the power of the  EFT and the operator matching that we discuss in the main text. It will also be useful for computations of higher point correlation functions in the microscopic theory. 
\subsection{One loop contribution to $F^{(2,0)}$}\label{aauebb}
The diagrams in figs.~\ref{directandexchange2pt} contribute in the following manner to the density two point functions,
\begin{equation}\label{rhorhoabc}
\begin{split}
\langle \rho_p \rho_{-p} \rangle^{\rm a} &= 2i\lambda\int_{k,k'} {\rm Tr} \left[\gamma^0 S_{p+k}S_k \right]{\rm Tr} \left[ S_{p+k'}\gamma^0 S_{k'} \right],\\
\langle \rho_p \rho_{-p} \rangle^{\rm b}  &= -2i\lambda\int_{k,k'} {\rm Tr} \left[\gamma^0 S_{p+k}S_{p+k'}\gamma^0 S_{k'}S_k\right]\\
\langle\rho_p \rho_{-p} \rangle^{\rm c} &=
2i\lambda\int_{k,k'}\left\{ {\rm Tr} \left[\gamma^0 S_{p+k} S_{p+k} \gamma^0 S_k\right]{\rm Tr}\left[S_{k'}\right]+ {\rm Tr} \left[\gamma^0 S_{p+k} \gamma^0 S_k S_k\right]{\rm Tr}\left[S_{k'}\right]\right\}\\
\langle\rho_p \rho_{-p} \rangle^{\rm d} &=-2i\lambda\int_{k,k'} \left\{{\rm Tr} \left[\gamma^0 S_{p+k} S_{k'}S_{p+k} \gamma^0 S_k\right]+ {\rm Tr} \left[\gamma^0 S_{p+k} \gamma^0 S_k S_{k'} S_k\right]\right\}\\
\end{split}
\end{equation}
where  $S_{k}\equiv S(\omega_k, \kv)$. The two contributions $ \rho \rho^{\rm a}$ and $\rho \rho^{\rm b}$ arise due to different trace structures in the left diagram of fig.~\ref{directandexchange2pt} while $\rho \rho^{\rm c}, \rho \rho^{\rm d}$ are the two different trace structures in the self-energy contribution.
Let us first evaluate $\rho\rho^{\rm a}$ and $\rho\rho^{\rm b}$. Since we are at weak coupling, we can re-express these traces in terms of two point functions of the conserved currents and a non-conserved operator $\mathcal{O}= \bar{\psi} \psi$ of the free fermion theory. The currents for free theory were defined in Eq.~\eqref{microconsu1} and after some algebra we get,  	
\begin{equation}\label{rrmicroeftdecomp}
\begin{split}
\langle \rho_p \rho_{-p}  \rangle^{\rm a}  &= i2\lambda \langle \rho_p \mathcal{O}_{-p}\rangle\langle \mathcal{O}_p \rho_{-p}\rangle,\\
\langle \rho_p \rho_{-p} \rangle^{\rm b} &= -i\lambda\left[\langle \rho_p \mathcal{O}_{-p}\rangle\langle \mathcal{O}_p \rho_{-p}\rangle+\langle \rho_p \rho_{-p}\rangle\langle \rho_p \rho_{-p}\rangle-\sum_{i=x,y}\langle \rho_p j^i_{-p}\rangle\langle j^i_p \rho_{-p}\rangle\right].\\
\end{split}
\end{equation}	
Now, instead of evaluating these correlators microscopically, we can use the identification of these operators with their EFT counterparts to enumerate the two point functions using the free gaussian EFT. Since we are interested in the leading free two point function, we only need our free EFT operators to linear order in the fluctuation $\phi$.	
\begin{eqnarray}
\rho(t,\x)&=&\frac{p_F}{(2\pi)^2}\int d\theta\, \nabla_n \phi+ O(\phi^2),\qquad j^i(t,\x)=v_F\int d\theta\, n^i \rho_{\rm EFT}+ O(\phi^2)\nonumber\\
\mathcal{O}(t,\x)&=&p_F\int \frac{ d\theta}{(2\pi)^2} \gamma(p_F)\nabla_n \phi + O(\phi^2),\nonumber\\
\end{eqnarray}
where in Sec.~ \ref{freefermionseft}, we found $\gamma(|\p|) = \frac{m}{\sqrt{p^2+m^2}}$ for a theory of free Dirac fermions. The two point functions in Eq. \eqref{rrmicroeftdecomp} can be computed using the free Gaussian EFT in Eq.~\eqref{eq_action_phi}
\begin{equation}\label{crossreft2pt}
\begin{split}
\langle \rho_p \mathcal{O}_{-p}\rangle &= \frac{i p_F m}{\epsilon_{F}}\int_\theta\,\frac{p_n}{(\omega- v_F p_n)},\qquad
\langle \rho_p j^i_{-p}\rangle = i v_F p_F\int_\theta\,\frac{n^i p_n}{(\omega- v_F p_n)},\\
\langle \rho_p\rho_{-p}\rangle &= i p_F \int_\theta\,\frac{p_n}{(\omega- v_F p_n)},
\end{split}
\end{equation}	
where, $\epsilon_F, v_F$ take free fermion values. Putting everything together, we obtain 	
\begin{equation}\label{rhoab}
\begin{split}
\langle \rho_p \rho_{-p} \rangle^{\rm a+b} &=\int_{\theta,\theta'}\left[\frac{i \lambda p^4_F}{\e_{F}^2}\left(  1- \cos\left(\theta-\theta'\right)\right)\right] \frac{ p_n p_{n'} }{ \left(\omega-p_n v_F\right)\left(\omega-p_{n'}v_F\right)}.
\end{split}
\end{equation}	
We have essentially computed the result without explicitly doing a microscopic fermion loop integral at finite density! We have checked that using explicit finite density fermion propagator as we have done in App.~\ref{doOm}, we recover the same result. 

It is now instructive to compare this result with the EFT prediction in Eq. \eqref{eq_rr_EFT}. In principle, the EFT parameters should be extracted only after performing the angular integrals. However, the three distinct angular structures in the integrals give rise to distinct analytic expressions in terms of $s=\frac{\omega}{v_F q}$. As a result, we can extract the Landau function $F^{(2,0)}$ directly at the integrand level. 
\begin{equation}
F^{(2,0)}(\theta, \theta') = \frac{\lambda p^3_F}{\e_{F}^2v_F}\left[  1- \cos\left(\theta-\theta'\right)\right].\\
\end{equation}
In the next subsection, we show that $\rho \rho^{\rm c+d}$ do not give rise to the same angular structure or similar analytic functions of $s$. 

\subsection{Self-energy}\label{self-energy}\label{self_energy}

In this subsection we evaluate the $O(\lambda)$ contribution to $\delta v_F , \delta p_F$. The self-energy diagrams in fig.~\ref{directandexchange2pt} has this information in principle but first we proceed to extract this information by deriving the one loop corrected quasi-particle propagator which is easier to evaluate. The advantage of this method will be clear through the process, we will also have access to $\delta \epsilon''_{F}$ which the density two point function cannot access. Using the explicit expressions for the fermion propagators, the one loop exact correction to the fermion propagator becomes, 
\begin{equation}\label{oneloopexact}
\begin{split}
\langle \Theta| T(\psi_\alpha (\omega, \vec{p}) \bar{\psi}_\beta (\omega', \vec{p}')) |\Theta\rangle &= \left[\frac{-i}{\cancel p + m - \Sigma(\omega, \p)} \right](2\pi)^3 \delta(\omega-\omega')\delta^2(\p-\p'), 
\end{split}
\end{equation}
where,
\begin{equation}
\begin{split}
i \Sigma(\omega, \p)_{\alpha \beta} = 2i \lambda\int_k \left( S_{\alpha \beta}(k)-\delta_{\alpha\beta} {\rm Tr}\left[S(k)\right]\right).
\end{split}
\end{equation}
Using the fact that in  $2+1$ space-time dimensions, two by two matrices are spanned by $(\delta_{\alpha\beta}, \gamma_{\alpha \beta}^\mu)$, we can decompose $\Sigma_{\alpha\beta}$ as follows, 
\begin{equation}
\begin{split}
\Sigma(\omega, \p)_{\alpha \beta}&= \Sigma^I \delta_{\alpha\beta} + \Sigma^\mu \gamma_{\alpha\beta}^\mu,\qquad
\Sigma^I=\frac{1}{2} {\rm Tr}\left[\Sigma(\omega,\p) \delta\right]= -\lambda\int_k {\rm Tr}\left[S(k)\right],\\
\Sigma^0&=\frac{1}{2} {\rm Tr}\left[\Sigma(\omega,\p) \gamma^0\right]= \lambda\int_k {\rm Tr}\left[\gamma^0 S(k)\right],\\
\Sigma^i&=-\frac{1}{2} {\rm Tr}\left[\Sigma(\omega,\p) \gamma^i\right]= -\lambda\int_k {\rm Tr}\left[\gamma^i S(k)\right],\qquad i=1,2
\end{split}
\end{equation}
Using the finite density propagator in Eq.~\eqref{finitedenprop} and dimensional regularization we obtain 
\begin{equation}\label{one_loop}
\begin{split}
\Sigma^I&= \frac{m \e_F \lambda}{2\pi },\qquad	\Sigma^0=\frac{-\lambda p_F^2}{4 \pi}.
\end{split}
\end{equation}

The other components $\Sigma^1, \Sigma^2$ evaluate to zero because of rotational symmetry. To extract $\delta \epsilon''_{F}$, $\delta v_F$, and $\delta p_F$ from the one-loop correction, we obtain the quasi particle propagator by expanding around the Fermi surface $\tilde{p}_F$ of the interacting theory. The expansion is typically of the form
\begin{equation}\label{quasipartprop}
G_{\alpha \beta}^{\rm Quasi-particle}=\frac{i Z_{\alpha \beta}}{\omega- (v_F+ \delta v_F) \hat{n}\cdot \delta \p-\frac12(\epsilon''_F+ \delta \epsilon''_F)\delta \p^2},\qquad \delta \p= \p-\tilde{p}_F \hat{n}.
\end{equation}
We will extract the required data by transforming our one-loop corrected finite-density propagator in Eq. \eqref{oneloopexact} into this form. We begin by rewriting the propagator as follows 
\begin{equation}\label{pole1}
\begin{split}
&\frac{-i}{\cancel p + m - \Sigma(\omega, \p)} = \frac{i (\left(\omega+\tilde{\mu}+\Sigma^0\right)\gamma^0-\p\cdot \vec{\gamma}+\left(m-\Sigma^{I}\right))}{\left(\omega-\left(\tilde{\epsilon}_p-\tilde{\mu}\right)\right)\left(\omega+\left(\tilde{\epsilon}_p+\tilde{\mu}\right)\right)},\\
\text{where},&\qquad \tilde{\epsilon}_p =\sqrt{\p^2+(m-\Sigma^I)^2},\qquad\tilde{\mu}=\mu+\Sigma^0
\end{split}
\end{equation}
Note that to $O(\lambda)$ all the calculated one loop corrections are independent of external momenta. In particular, this ensures there is no contribution of the one loop corrections to wave function renormalization. The correction to the Fermi momenta, the Fermi velocity and derivative of Fermi velocity can be solved by looking at the poles of the relativistic fermion propagator at small frequency and near the (shifted) Fermi surface $\tilde{p}_F$. Since we are interested in corrections up to $O(\lambda)$, the expansion entails   
\begin{equation}\label{qplimit}
\begin{split}
\omega \rightarrow 0,\qquad \p\rightarrow (p_F + \lambda \delta p_F) \hat{n} + \delta \p.\\
\end{split}
\end{equation}
The pole arises due to the first denominator of one-loop corrected the finite density propagator Eq. \eqref{pole1}. It is easy to convince oneself that the numerator cannot cancel this pole and the second denominator cannot give rise to such a singularity. In the limit of Eq.~\eqref{qplimit}, massaging the relevant denominator into the form of Eq. \eqref{quasipartprop}, we obtain the requisite data.
\begin{equation}\label{deltadepsinfM2}
\begin{split} 
\delta p_F &= \frac{1}{v_F}\left[\Sigma^0+\frac{m \Sigma^I}{\e_F}\right],\qquad
\delta v_F = \frac{\epsilon''_{F}}{v_F}\left[\Sigma^0+\frac{m \Sigma^I}{\e_F} \right] + \frac{v_F}{\e_F^2} m \Sigma^I\\
\delta \epsilon''_{F}&= \frac{1}{\epsilon _ {F}^3}\left[-2 m \Sigma ^I v_F{}^2+\frac{\epsilon _ {F}^2 \epsilon'''_{F} \left(\Sigma^0\epsilon _{F}+m \Sigma ^I\right)}{v_F}+m \Sigma ^I \epsilon _ {F} \epsilon _ {F}''\right]
\end{split} 
\end{equation}
While we explicitly do not record the wave function renormalization coefficient, it can similarly be obtained. For our microscopic theory, the corrections take the values, 	
\begin{equation}\label{deltadepsinfM}
\begin{split} 
\delta p_F &=-\frac{\lambda\left(p_F^2-2 m^2\right) \e_F}{4 \pi  p_F },\qquad  \delta v_F =\frac{\lambda \left(p_F^2 m^2+2 m^4\right)}{4 \pi  p_F \epsilon^2_{F}}\\
\delta \e_F''&=\frac{\lambda \left(-p_F^2 m^2-4 m^4\right)}{4 \pi \e_F^4}.
\end{split} 
\end{equation}

As a consistency check, we derive the Wilson coefficients $\delta p_F$ and $\delta v_F$ the self energy diagrams in fig.~\ref{directandexchange2pt}. We provide a brief outline of the computation of $\rho\rho^{\rm c}$ and $\rho\rho^{\rm d}$. The correlators involve a bubble integral, which can be computed independently of the remaining structure. After some manipulations, the trace structure of the integrand can be expressed as the partial static limit of free three-point functions.
\begin{equation}
\langle\rho_p\rho_{-p}\rangle^{\rm c+ d} \sim \lim_{\q \rightarrow 0}\lim_{\omega \rightarrow 0} \lambda \sum_i \alpha_i \langle \rho_p O^i_q \rho_{-p-q}\rangle
\end{equation}
where, $O^i_q=\{\rho_q, \mathcal{O}_q\}$ and $\alpha_i$ are the coefficients one gets from the one-loop computation. The contribution from $j^i$ evaluates to zero. 

From the partial static limit of free $\rho\rho\rho$ in obtained in Eq.~\eqref{rrr_static}, the partial static limit of the free $\mathcal{O}\rho\rho$ evaluated in \eqref{rtrrfinal} and using the explicit one-loop results evaluated in this subsection, we carry out this exercise and obtain agreement with $\delta v_F$ and $\delta p_F$ obtained from the corrections to the quasi-particle propagator. This computation confirms that the analytic structure of self-energy diagrams are different from the ones contributing to $F^{(2,0)}$ and together with $\rho^{\rm a+b}$ we obtain complete analytic agreement with Eq.~\eqref{eq_rr_EFT}. This agreement also presents a non-trivial check of our $\mathcal{O}$ operator in the free theory that we obtained in Eq.~\eqref{rhotilde2}. Verifying $\delta \epsilon''_{F}$ analogously would involve similar structures arising in three point function of density and it would entail the static limit of a four point function which we will not pursue here.

\section{Three point density correlation function }
In this section we detail the computation of three point function of density to leading order in the Landau parameters in the EFT. We also provide details of the analogous computation in the microscopic theory.
\subsection{From effective field theory}\label{3ptcalceft}

We use the cubic action in Eq.~\eqref{eq_action_phi} and the density operator to quadratic order in $\phi$,

\begin{equation}
\rho(t,\x)= {p_F}\int \frac{d\theta}{(2\pi)^2} \left[
	 \nabla_n \phi +  \frac{1}{2p_F}\nabla_{s^i} \left(\partial_{\theta_i} \phi \nabla_n \phi\right)
\right] + O(\phi^3)\, .
\end{equation}

The three point function of the density receives several contributions which we systematically evaluate below. These contributions can be understood as the weak coupling limit of our exact three point function eqn. \eqref{rrrexactf20f30}. 

\subsubsection*{$F^{(2,0)}$ in one of the arms of free correlation function}
We begin by enumerating the weak coupling limit of Eq.~\eqref{rrrexact1}. In this limit the diagrams contributing to the three point function are given by fig.~\ref{figfreeeft2} but instead of the exact propagator, we consider the perturbative expansion of the propagator in $F^{(2,0)}$. Operationally, this constitutes insertion of  the quadratic interaction term involving $F^{(2,0)}$ into one of the legs of the free three point function in fig.~\ref{figfreeeft}. We will express the result in terms of the free fermion angular integrand Eq.~\eqref{freepartsrrr} and some new angular structures.

\subsubsection*{WZW contribution}

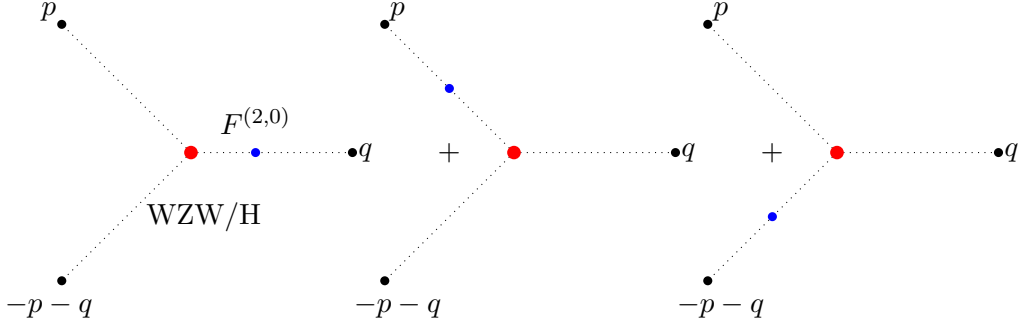
\begin{figure}[h!]
\begin{adjustwidth}{1cm}{}
\begin{tikzpicture}[scale=0.85]
\draw[dotted] (0,0) -- (2.5,0); 
\fill[black] (2.5,0) circle (2pt);  
\node at (2.7,0) {$q$};
\draw[dotted] (0,0) -- (-2,2);  
\fill[black] (-2.0,2) circle (2pt);  
\node at (-2.2,2.2) {$p$};
\draw[dotted] (0,0) -- (-2,-2); 
\fill[black] (-2.0,-2) circle (2pt);  
\node at (-2.2,-2.4) {$-p-q$};
\fill[red] (0,0) circle (3pt);  
\node at (0.2,-1) {WZW/H};

\fill[blue] (1,0) circle (2pt);
\node at (1,0.5) {$F^{(2,0)}$};

\node at (4,0) {$+$};

\draw[dotted] (5,0) -- (7.5,0); 
\fill[black] (7.5,0) circle (2pt);  
\node at (7.7,0) {$q$};
\draw[dotted] (5,0) -- (3,2);  
\fill[black] (3,2) circle (2pt);  
\node at (3.2,2.2) {$p$};
\draw[dotted] (5,0) -- (3,-2); 
\fill[black] (3,-2) circle (2pt);  
\node at (3.2,-2.4) {$-p-q$};
\fill[red] (5,0) circle (3pt);  

\fill[blue] (4,1) circle (2pt);

\node at (9,0) {$+$};

\draw[dotted] (10,0) -- (12.5,0); 
\fill[black] (12.5,0) circle (2pt);  
\node at (12.7,0) {$q$};
\draw[dotted] (10,0) -- (8,2);  
\fill[black] (8,2) circle (2pt);  
\node at (8.2,2.2) {$p$};
\draw[dotted] (10,0) -- (8,-2); 
\fill[black] (8,-2) circle (2pt);  
\node at (8.2,-2.4) {$-p-q$};
\fill[red] (10,0) circle (3pt);  

\fill[blue] (9,-1) circle (2pt);
\end{tikzpicture}
\end{adjustwidth}
\caption{$F^{(2,0)}$ insertion with the WZW/H vertex.}
	\label{f20wzw}
\end{figure}

We detail the calculation when the central vertex is the WZW term and $F^{(2,0)}$ is inserted in one of the arms of the three point function, represented in fig.~\ref{f20wzw}. The contribution of these diagrams to the three point function becomes, 
\begin{align}
\langle\rho\rho\rho\rangle_{\rm (1)}^{\rm WZW}&= v_F\int_{\theta,\theta'} \left[F^{(2,0)}(\theta, \theta') \frac{p_{n'}}{(\omega_p- v_F p_{n'})}\langle\rho_p\rho_q\rho_{-p-q}\rangle^{\theta}_\text{WZW}\right.\nonumber\\
&\left.+\frac{1}{3!} \partial_{\theta}F^{(2,0)}(\theta', \theta) \frac{p_{n'}g^\theta(p|q,-p-q)}{(\omega_p - v_F p_{n'})}+ \text{Cyclic} \right],\nonumber\\
g^\theta(p|q,-p-q)&=\frac{(p+q)_n q_s(\omega_p-\omega_q)+ q_n(p+q)_s(2\omega_p+\omega_q)}{(\omega_p-v_F p_n)(\omega_q-v_F q_n)(\omega_p+\omega_q-v_F (p+q)_n)},\label{gdefinition} 
\end{align}
where ``$\text{Cyclic}$" denotes the cyclic permutations, i.e., the $Z_3$ permutation of the external momenta $\{p,q,-p-q\}$.

\subsubsection*{H contribution}

The contribution to the three point function when the central vertex in fig.~\ref{f20wzw} is given by the cubic expansion of the free Hamiltonian ( second term in the second line of Eq. \eqref{eq_action_phi}) can be evaluated similarly to give, 

\begin{equation}\label{f20Hrrr}
\begin{split}
\langle\rho\rho\rho\rangle^{\rm H}_{(1)}
&= v_F\int_{\theta, \theta'} \left[F^{(2,0)}(\theta, \theta') \frac{p_{n'}}{(\omega_p- v_F p_{n'})}\langle\rho_p\rho_q\rho_{-p-q}\rangle^{\theta}_\text{H}+ \text{Cyclic} \right].\\
\end{split}
\end{equation}

\subsubsection*{$\rho^{2}$ contribution}
\begin{figure}[h!]
\begin{adjustwidth}{1.5cm}{}
	\begin{tikzpicture}[scale=0.85]
\fill[red] (8,2) circle (3pt);
\node at (8,2.3) {$q$};  
\node at (8,1.3) {$\rho^{(2)}$}; 
\draw[dotted] (6,-1) -- (8,2);   
\fill[black] (6,-1) circle (2pt);
\node at (6,-1.3) {$p$};
\draw[dotted] (10,-1) -- (8,2);   
\fill[black] (10,-1) circle (2pt);
\node at (10,-1.3) {$-p-q$};
\fill[blue] (7,0.5) circle (2pt);
\node at (7.2,0) {$F^{(2,0)}$};

\node at (12,0) {$+$};

\fill[red] (15,2) circle (3pt);
\node at (15,2.3) {$-p-q$};  
\node at (15,1.3) {$\rho^{(2)}$}; 
\draw[dotted] (13,-1) -- (15,2);   
\fill[black] (13,-1) circle (2pt);
\node at (13,-1.3) {$p$};
\draw[dotted] (17,-1) -- (15,2);   
\fill[black] (17,-1) circle (2pt);
\node at (17,-1.3) {$q$};
\fill[blue] (14,0.5) circle (2pt);
\node at (14.2,0) {$F^{(2,0)}$};

\node at (18,0) {$+$ Cyclic.};

\end{tikzpicture}
\end{adjustwidth}
	\caption{$F^{(2,0)}$ insertion in $\rho^{(2)}\rho\rho$.}
	\label{f20rho2}
\end{figure}
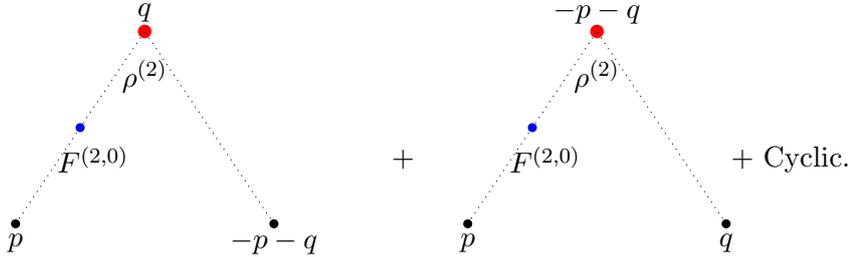

We now look at the contribution due to insertion of the quadratic interaction term in the three point function involving one nonlinear density operator (i.e., $\rho^{(2)}$) and two linear density operators. The diagrams are given by fig.~\ref{f20rho2}.
The right diagram can be obtained by $(q\leftrightarrow -p-q)$ of the left diagram, labelled as $\langle\rho^{(2)}_q\rho_p\rho_{-p-q}\rangle^{\rho^{(2)}}_{(1)}$ below. 

\begin{equation}
\begin{split}
&\langle\rho^{(2)}_q\rho_p\rho_{-p-q}\rangle^{\rho^{(2)}}_{(1)}+\langle\rho^{(2)}_{-q-p}\rho_p\rho_{q}\rangle^{\rho^{(2)}}_{(1)}
=\frac{v_F}{2} \int_{\theta', \theta} \frac{p_{n'}F^{(2,0)}(\theta,\theta')}{(\omega_p- v_F p_{n'})}\nonumber\\
&\times\left[\frac{q_{s}(p+q)_{n}}{(\omega_{p+q}-v_F(p+q)_{n})}\partial_{\theta}\frac{1}{(\omega_p- v_F p_{n})} + \frac{-q_{s}p_{n}}{(\omega_{p}-v_F(p)_{n})}\partial_{\theta}\frac{1}{(\omega_{p+q}- v_F (p+q)_{n})}\right.\label{f20rho2,1}\\
&\left.+(q \leftrightarrow -p-q)\right]+\frac{v_F}{2} \int_{\theta',\theta} \frac{p_{n'}\partial_{\theta}F^{(2,0)}(\theta',\theta)}{(\omega_p- v_F p_{n'})}h^{\theta}(p|q,-p-q),\nonumber\\
\end{split}
\end{equation}
where the function $h$ is defined as
\begin{equation}
h^{\theta}(p|q,-p-q)=\frac{q_{s}(p+q)_{n}}{(\omega_{p+q}-v_F (p+q)_{n})(\omega_{p}-v_F p_{n})} + \frac{(-p-q)_{s}q_{n}}{(\omega_{q}-v_F q_{n})(\omega_{p}-v_F p_{n})}.\label{hdefinition}
\end{equation}
The total $\rho^{(2)}$ contribution  becomes, 
\begin{equation}\label{rho2total}
\langle\rho_p\rho_q\rho_{-p-q}\rangle^{F^{(2,0)}}_{\rho^{(2)}}=\left( \langle\rho^{(2)}_q\rho_p\rho_{-p-q}\rangle^{F^{(2,0)}}_1+\langle\rho^{(2)}_{-q-p}\rho_p\rho_{q}\rangle^{F^{(2,0)}}_1 + \rm{Cyclic}\right).
\end{equation}

\subsubsection*{Contribution of $F^{(20)}$ class of vertices}
We now turn to the weak coupling expansion of the exact result involving cubic vertices recorded in Eq.~\eqref{rrrexact2}. The weak coupling limit for this class is relatively simple. Let us first consider the interaction term involving the $F^{(20)}$ parameter and its generalizations, given by, 
\begin{equation}
\begin{split}
L^{F^{(2,0)}} &=- \frac{v_F}{2}\int_{t\x\theta\theta'}\left(F^{(2,0)}(\theta,\theta') \Big[ \nabla_s (\nabla_n\phi \partial_{\theta}\phi) (\nabla_n\phi)' \Big]+  \partial_{\theta} F^{(2,0)}(\theta,\theta') \Big[ (\nabla_n\phi \nabla_s\phi) (\nabla_n\phi)' \Big] \right.\\
&\left.\qquad+  F_1^{(2,0)}(\theta,\theta') \left[ (\nabla_n\phi)^2(\nabla_n\phi)'\right]\right).
\end{split}
\end{equation}
The relevant diagrams continue to be given by the left most diagrams of \ref{figfreeeft2} with the central vertex as the $F^{(2,0)}$ class of cubic interactions but with free propagators now. Together with Eq. \eqref{rho2total}, the contribution from this term in the Lagrangian can be compactly expressed as, 
\begin{equation}
\begin{split}
&\langle\rho_p\rho_q\rho_{-p-q}\rangle^{F^{(2,0)}}_{(1)}+ \langle\rho_p\rho_q\rho_{-p-q}\rangle^{\rho^{(2)}}_{(1)}\\
&=  v_F\int_{\theta,\theta'} \left[F^{(2,0)}(\theta, \theta') \frac{p_{n'}}{(\omega_p- v_F p_{n'})}\langle\rho_p\rho_q\rho_{-p-q}\rangle^{\theta}_{\rho^{(2)}}+\frac{1}{2} \frac{p_{n'}\partial_{\theta}F^{(2,0)}(\theta',\theta)}{(\omega_p- v_F p_{n'})}h^{\theta}(p|q,-p-q)\right.\\
&\left.+ \text{Cyclic} \right]\\
&-\frac{v_F}{2 }\int_{\theta,\theta'}\partial_\theta F^{(2,0)}(\theta,\theta')\frac{p_{n'}}{(\omega_{p}-v_F p_{n'})} j^{\theta}(p|q,-q-p)+ \text{Cyclic}\\
& - \frac{v_F}2\int_{\theta,\theta'} \frac{F_1^{(2,0)}(\theta, \theta')p_nq_n (p+q)_{n'}}{(\omega_{p+q}-v_F (p+q)_{n'})(\omega_q-v_F q_n)(\omega_p-v_F p_n)}+\text{Perm}.\\
\end{split}
\end{equation}
where the function $h$ is defined in Eq. \eqref{hdefinition}, ``$\text{Perm}$" denotes the $S_3$ permutations of the external momenta $\{p,q,-p-q\}$ and 
\begin{equation}
j^{\theta}(p|q,-q-p)=\frac{q_s (p+q)_n+q_n (p+q)_s}{(\omega_{p+q}-v_F (p+q)_n)(\omega_q-v_F q_n)}.\label{jdefinition}
\end{equation}
\subsubsection*{Contribution of $F^{(3,0)}$ vertex}
Finally we evaluate the contribution from the last line of Eq. \eqref{eq_action_phi} or the $F^{(3,0)}$ vertex. We obtain, 
\begin{equation}\label{line4eft}
\begin{split}
\langle\rho_p\rho_q\rho_{-p-q}\rangle^{F^{(3,0)}}_{(1)} &=
-6 \int_{\theta,\theta',\theta''} \frac{  F^{(3,0)}(\theta,\theta',\theta'')p_{n}q_{n'}(p+q)_{n''}}{(\omega_p-v_F p_{n})(\omega_q-v_F q_{n'})(\omega_{p+q}-v_F (p+q)_{n''})}.
\end{split}
\end{equation}

We can now record the full interacting contribution, 
\begin{equation}\label{rrrFweak}
\begin{split}
&\langle\rho_p\rho_q\rho_{-p-q}\rangle_{(1)} =\\
& v_F\int_{\theta,\theta'} \left[F^{(2,0)}(\theta, \theta') \frac{p_{n'}}{(\omega_p- v_F p_{n'})}\langle\rho_p\rho_q\rho_{-p-q}\rangle^{\theta}_\text{WZW}+\frac{\partial_{\theta}F^{(2,0)}(\theta', \theta)}{3!}  \frac{p_{n'}g^\theta(p|q,-p-q)}{(\omega_p - v_F p_{n'})}\right.\\
&\left.+ \text{Cyclic} \right]+ v_F\int_{\theta,\theta'} \left[F^{(2,0)}(\theta, \theta') \frac{p_{n'}}{(\omega_p- v_F p_{n'})}\langle\rho_p\rho_q\rho_{-p-q}\rangle^{\theta}_\text{H}+ \text{Cyclic} \right]\\
&+v_F\int_{\theta,\theta'} \left[F^{(2,0)}(\theta, \theta') \frac{p_{n'}}{(\omega_p- v_F p_{n'})}\langle\rho_p\rho_q\rho_{-p-q}\rangle^{\theta}_{\rho^{(2)}}+ \frac{p_{n'}\partial_{\theta}F^{(2,0)}(\theta',\theta)}{2(\omega_p- v_F p_{n'})}h^{\theta}(p|q,-p-q)\right.\\
&\left.-\frac12\int_{\theta,\theta'}\partial_\theta F^{(2,0)}(\theta,\theta')\frac{p_{n'}}{(\omega_{p}-v_F p_{n'})} j^{\theta}(p|q,-q-p)+ \text{Cyclic}\right]\\
&-\frac{v_F}2\int_{\theta,\theta'} \frac{F_1^{(2,0)}(\theta, \theta')p_nq_n (p+q)_{n'}}{(\omega_{p+q}-v_F (p+q)_{n'})(\omega_q-v_F q_n)(\omega_p-v_F p_n)}+\text{Perm}\\
&-6 \int_{\theta,\theta',\theta''}\frac{  F^{(3,0)}(\theta,\theta',\theta'')p_{n}q_{n'}(p+q)_{n''}}{(\omega_p-v_F p_{n})(\omega_q-v_F q_{n'})(\omega_{p+q}-v_F (p+q)_{n''})},
\end{split}
\end{equation}

Where the functions $g,h$ and $j$ are given in Eqs. \eqref{gdefinition}, \eqref{hdefinition} and \eqref{jdefinition} respectively. 

\subsubsection*{Partial static limit and response to change of reference state}\label{pslim}

As a non-trivial check of our results, we  take the partial static limit of our three-point density function analogous to static susceptibilities, as discussed in subsection \ref{ssasc}, leading to thermodynamic constraints between Wilson coefficients of the EFT. Recall the static limit of density correlation relation from Eq.~\eqref{eq_chi_ell} to perturbative order in $F_0$, relating the change in the Fermi momentum due to a change in the chemical potential, the Fermi velocity, and the Landau parameter $F_0$, 
\begin{equation}
\frac{\partial p_F}{ \partial \mu}
	\equiv \lim_{q\to 0}\lim_{\omega\to 0} G^R_{\rho \rho}(\omega,\vec q)
	= \frac{p_F}{v_F} \frac{1}{1+F_0}\simeq \frac{p_F}{v_F} (1- F_0)\, .
\end{equation}
As explained in App.~F of \cite{Delacretaz:2021ufg} for the free EFT, the partial static limit of the three point function relates to the change of the two point correlation due to chemical potential.
\begin{equation}\label{F13interacting}
\begin{split}
\lim_{q \rightarrow 0}\lim_{ \omega_q \rightarrow 0} \langle \rho_p \rho_q \rho_{-p-q} \rangle = \frac{-i \partial}{\partial \mu} \langle \rho_p  \rho_{-q} \rangle 
\end{split}
\end{equation}
Analogous to the static susceptibilities from the two-point correlation functions, we expect the partial static limit of the three-point density correlations in the interacting theory to yield new thermodynamic relations. While this can be obtained from Eq.~\eqref{F13interacting}, an independent way of obtaining these relations is to examine the change of the action itself under a change of the reference state, 
\begin{equation}
\begin{split}
f^0_p \rightarrow f^0_p + \Delta p_F \delta(p_F-|\p|)
\end{split}
\end{equation}
The co-adjoint orbit action on this new reference state is given by,
\begin{equation}
\begin{split}
f\rightarrow f + \Delta p_F U^{-1} \delta(p_F-|\p|) U
\end{split}
\end{equation}
We now record the change of the action under this change of the Fermi surface. Note that the WZW term does not change since it is independent of the reference state. The total change therefore is obtained from the Hamiltonian part of the action, 
\begin{equation}
\begin{split}
S_H&=-\int_{t,x,p} \left(\epsilon_p f + \int_{p'} F^{2,0}(\p,\p') \delta f_p \delta f_{p'}+\int_{p',p''} F^{2,0}(\p,\p',\p'') \delta f_p \delta f_{p'} \delta f_{p''}\right)
\end{split}
\end{equation}
where the change in the reference state induces,
\begin{equation}
\begin{split}
\delta f \rightarrow \delta f+ \Delta p_F \left(\delta_p-\left\{\phi,\delta_p\right\}+\frac12\left\{\phi,\left\{\phi,\delta_p\right\}\right\}+O(\phi^3)\right)
\end{split}
\end{equation}
where for notational convenience, $\delta_p\equiv\delta(p_F-|p|)$. We look for the change in action to linear order in $\Delta p_F$ and $F$s. After a bit of algebra, the total change in the gaussian action under a change of the Fermi surface or the reference state to linear order in $\Delta p_F, F^{(2,0)}$ and $ F^{(3,0)}$ is given by, 
\begin{equation}
\begin{split}
&S\rightarrow -\int_{t,\x,\theta} \frac12\left(p_F+\Delta p_F\right)\left(v_F+\Delta p_F\left[\epsilon_{p_F}''+\frac{v_F}{p_F}\int_{\theta'}F^{(2,0)}_1(\theta,\theta')\right]\right) (\nabla_n \phi)^2\\
&-\int_{t,\x,\theta,\theta'}\frac12\left(p_F+\Delta p_F\right)\left(v_F+\Delta p_F\epsilon_{p_F}''\right)\\
&\times\left[F^{(2,0)}(\theta,\theta')+\frac{\Delta p_F}{p_F}\left\{\left(1-\frac{\epsilon''_{p_F}}{v_F}\right)F^{(2,0)}(\theta,\theta')+2F^{(2,0)}_1(\theta,\theta')+\frac6{v_F}\int_{\theta''}F^{(3,0)}(\theta,\theta',\theta'')\right\}\right]\nabla_n \phi \nabla_{n'}\phi \\
&+ O(\phi^3)
\end{split}
\end{equation}
Where we are only interested in $O(\Delta p_F)$ terms and  we are not keeping track of higher order terms $O({\Delta p_F}^2)$. We can therefore determine the change of the wilson coefficients $v_F$ and $F^{(2,0)}$ due to a change in the reference state to linear order in $\Delta p_F$ and $F^{(3,0)}$, 
\begin{equation}\label{change_F_pF}
\begin{split}
\frac{\partial v_F}{\partial p_F} &=\epsilon''_{p_F}+ \frac{\tilde{F}_0 v_F}{p_F}\\
\frac{\partial F^{(2,0)}(\theta,\theta')}{\partial p_F} &= \frac{1}{p_F}\left\{\left(1-\frac{\epsilon''_{p_F}}{v_F}\right)F^{(2,0)}(\theta,\theta')+2F^{(2,0)}_1(\theta,\theta')+\frac6{v_F}\int_{\theta''}F^{(3,0)}(\theta,\theta',\theta'')\right\}
\end{split}
\end{equation}

We now obtain the same constraints from taking the partial static limit of Eq.~\eqref{F13interacting}. This serves as a non-trivial check of our three point function computation in the EFT. The partial static limit  $\lim_{q, \omega_q \rightarrow 0} \langle \rho_p \rho_q \rho_{-p-q} \rangle$ with the leading interacting contribution as given in \eqref{rho3eft}  takes the form, 
\begin{equation}\label{rrr_static}
\begin{split}
&\lim_{q, \omega_q \rightarrow 0} \left(\langle \rho_p \rho_q \rho_{-p-q} \rangle_{(0)} + \langle \rho_p \rho_q \rho_{-p-q} \rangle_{(1)}\right)\\
&=\int_\theta \frac{p_n}{(\omega-v_F p_n)}\left[-\frac{F_0}{v_F}+ \frac{1}{v_F}\right]\\
&+ \int_\theta  \frac{p^2_n}{(\omega-v_F p_n)^2}\left[\left(\frac{p_F \epsilon''_{F}}{v_F}(1-F_0) + \tilde{F}_0\right)+\int_{\theta'} \left(\frac{2p_F \epsilon''_{F} p_{n'}}{(\omega- v_F p_{n'})}\right)F^{(2,0)}(\theta,\theta')\right]\\
&+\int_{\theta,\theta'} \frac{p_n p_{n'}}{(\omega -v_F p_n)(\omega -v_F p_{n'})}\left[2 F^{(2,0)}(\theta, \theta')+2 F_1^{(2,0)}(\theta,\theta')+\frac{6}{v_F}\int_{\theta''} F^{(3,0)}(\theta,\theta',\theta'')\right].
\end{split}
\end{equation}
The two point density correlation to first order in the Landau parameter $F^{(2,0)}$ is given by, 
\begin{equation}\label{rrF20}
\begin{split}
\langle \rho_p\rho_{-p}\rangle& = ip_F 
\left[\int_\theta \frac{q_n}{\omega-v_F q_n} + 
\int_{\theta,\theta'} v_F F^{(2,0)}(\theta-\theta')\frac{q_{n}}{\omega-v_F q_{n}}\frac{q_{n'}}{\omega-v_F q_{n'}}\right],
\end{split}
\end{equation}
which results in the following change in response to a variation in the chemical potential,%
\begin{equation}
\begin{split}
&\frac{-i \partial}{\partial \mu} \langle \rho_p  \rho_{-q} \rangle\approx\int_{\theta} \left[\frac{p_n}{\omega- v_F p_n}\frac{\partial p_F}{\partial \mu} \right]+\int_{ \theta, \theta'} \frac{p_n p_{n'} \partial_\mu (v_F p_F F^{(2,0)}(\theta, \theta'))}{(\omega_p - v_F p_n) (\omega_p - v_F p_{n'})}\\
&+\int_{\theta}\left[  \frac{p_n^2}{(\omega- v_F p_n)^2} \left(p_F\frac{\partial v_F}{\partial \mu}+2 p_F \epsilon''_{F} \int_{\theta'} \frac{p_{n' }F^{(2,0)}(\theta,\theta')}{(\omega- v_F p_{n'})}\right)\right],\\
\end{split}
\end{equation}

where in the second term in the second line we have replaced $v_F \frac{\partial v_F}{\partial \mu}\sim \epsilon''_{F}$ since we are considering the correlators to linear order in the Landau parameters. We get the following thermodynamic constraints 
\begin{equation}\label{change_F20_mu}
\begin{split}
\frac{\partial p_F}{\partial \mu}&= \frac{1}{v_F}\left(1-F_0\right),\qquad
\frac{\partial v_F}{\partial \mu}=\frac{ \epsilon''_{F}}{v_F}(1-F_0) + \frac{\tilde{F}_0}{p_F}\\
\partial_\mu\left[v_F p_F F^{(2,0)}(\theta, \theta')\right]&=\left[2 F^{(2,0)}(\theta, \theta')+2 F_1^{(2,0)}(\theta,\theta')+\frac{6}{v_F} \int_{\theta''} F^{(3,0)}(\theta,\theta',\theta'')\right].
\end{split}
\end{equation} 
While the first relation is what we get from analysis of static susceptibilities, the other two constraints are new and involve the generalized Landau parameters. Using, $\partial_{\mu} \equiv \frac{\partial p_F}{\partial\mu}\partial_{p_F}$ we obtain perfect agreement with \eqref{change_F_pF}, to linear order in $F$s.

\subsection{From microscopic theory: one loop contribution}\label{fmt}

We outline the computation of the density three point function in the microscopic theory. The explicit contribution is given by diagrams in fig.~\ref{microscopic}, 
\begin{equation}\label{rrrmicroL}
\begin{split}
\langle \rho_p \rho_q \rho_{-p-q}\rangle_\lambda &= 2i \lambda \int_k \left(\textrm{Tr} \left[\gamma^0 S_{p+k} S_k\right]\textrm{Tr} \left[ S_{p+k'} \gamma^0 S_{p+q+k'} \gamma^0 S_{k'}\right]\right.\\
&\left.-\textrm{Tr} \left[\gamma^0 S_{p+k}S_{p+k'} \gamma^0 S_{p+q+k'} \gamma^0  S_{k'}S_{k}\right]+ (q \leftrightarrow -p-q)\right)\\
+&\text{Cyclic}.
\end{split}
\end{equation}
Enumerating the free fermion three point function at finite density from first principles using our microscopic propagator is challenging. Subtle cancellations in fermion loops \cite{metzner1997Fermisystemsstrongforward} indicates there are non-trivial contributions from sub-leading order in the analysis contrary to the density two point function calculation in App.~\ref{doOm}.  Instead, we follow the method outlined in App.~\ref{domtpcf}. We identify our free microscopic currents with free EFT operators which reduces the problem to evaluating these correlators using  EFT. Similar to the two point function, we can express the one loop contribution \eqref{rrrmicroL} as follows 
\begin{equation}\label{rho3micro}
\begin{split}
\langle \rho_p \rho_q \rho_{-p-q}\rangle_\lambda &= i \lambda \left(\langle \rho_p \mathcal{O}_{-p}\rangle \langle\mathcal{O}_{p} \rho_q \rho_{-p-q}\rangle-\langle \rho_p \rho_{-p}\rangle \langle \rho_{p} \rho_q \rho_{-p-q}\rangle\right.\\
&\left.+ \langle \rho_p j^x_{-p}\rangle \langle j^x_{p} \rho_q \rho_{-p-q}\rangle +\langle \rho_p j^y_{-p}\rangle \langle j^y_{p} \rho_q \rho_{-p-q}\rangle\right)+ \text{Cyclic}.
\end{split}
\end{equation}
Since we need the identification only for the free theory, we can directly read off our results obtained in Sec.~ \ref{EFTops} up to $O(\phi^2)$. 	 
\begin{eqnarray}
\rho(t,\x)&=&\frac{p_F}{(2\pi)^2}\int d\theta\, \nabla_n \phi +  \frac{1}{2(2\pi)^2}\int d\theta\,\nabla_s \left(\partial_\theta \phi \nabla_n \phi\right) + O(\phi^3),\nonumber\\
j^i(t,\x)&=&v_F\int d\theta\, n^i \rho+ \frac{1}{2(2\pi)^2}\int d\theta\, \left[p_F \epsilon_{F}'' n^i(\nabla_n \phi)^2+ v_F s^i \nabla_s \phi \nabla_n \phi\right]+ O(\phi^3),\nonumber\\
\mathcal{O}(t,\x)&=&p_F\int \frac{ d\theta}{(2\pi)^2}\left[\gamma(p_F) \left(\nabla_n \phi + \frac1{2p_F} \nabla_s (\nabla_n\phi\partial_\theta \phi)\right) + \frac12 \gamma'(p_F)  (\nabla_n \phi)^2\right] + O(\phi^3),\nonumber\\
\end{eqnarray}
where we found $\gamma(|\p|) = \frac{m}{\sqrt{p^2+m^2}}$ for free Dirac fermions. The two point functions in Eq. \eqref{rho3micro} can be computed as done in Sec.~ \ref{den_corr} and we record the results,
\begin{equation}
\begin{split}
\langle \rho_p \mathcal{O}_{-p}\rangle^{\rm free} &= \frac{i p_F m}{\epsilon_{F}}\int_\theta\frac{p_n}{(\omega- v_F p_n)},\qquad
\langle \rho_p j^i_{-p}\rangle^{\rm free} = i v_F p_F\int_\theta\frac{n^i p_n}{(\omega- v_F p_n)}\nonumber\\
\end{split}
\end{equation}
We now obtain the relevant three point correlation functions involving two density operators and a single current.	 
\subsubsection*{Evaluating $ \langle j^i_{p} \rho_q \rho_{-p-q}\rangle$} 

Similar to the density three point function, the relevant correlator receives contribution from three parts: the WZW term, the Hamiltonian piece of the action Eq. \eqref{eq_action_phi} and finally we consider the contribution of the nonlinear part of the operators.  
\begin{equation}
\begin{split}
&\langle j^i_{p} \rho_q \rho_{-p-q}\rangle_{WZW}
= v_F \int_\theta n^i \langle \rho_p \rho_q \rho_{-p-q} \rangle^\theta_{WZW} + \frac{v_F}{3!} \int_\theta\,s^i g^{\theta}(p|q,-p-q),
\end{split}
\end{equation}
where $g$ has been defined in Eq. \eqref{gdefinition}. Similarly, 

\begin{eqnarray}
&&\langle j^i_{p} \rho_q \rho_{-p-q}\rangle_{H}=v_F \int_\theta\, n^i \langle \rho_p \rho_q \rho_{-p-q} \rangle^\theta_{H}.     
\end{eqnarray}
We now compute the contribution from the nonlinear parts of the currents.
$$\langle j^i_{p} \rho_q \rho_{-p-q}\rangle_{\rho^{(2)}}
=\langle j^{i, (2)}_{p} \rho_q \rho_{-p-q}\rangle + \langle j^i_{p} \rho^{(2)}_q \rho_{-p-q}\rangle+\langle j^i_{p} \rho_q \rho^{(2)}_{-p-q}\rangle.$$
We obtain, 
\begin{equation}
\begin{split}
\langle j^{i, (2)}_{p} \rho_q \rho_{-p-q}\rangle &= v_F\int_\theta\, n^i \langle\rho^{(2)}_{p} \rho_q \rho_{-p-q}\rangle^\theta + \frac{p_F \epsilon_{F}''}{2}\int_\theta\, n^i \frac{-2(q_n (p+q)_n)}{(\omega_q-v_F q_n)(\omega_{p+q}-v_F (p+q)_n)}\\
&-\frac{v_F}{2}\int_\theta\, s^i j^\theta(p|q,-p-q).
\end{split}
\end{equation}
Note that the middle tensor structure appearing is similar to first line of Eq. \eqref{line4eft} and $j$ is defined in Eq.~\eqref{jdefinition}. Finally, performing similar manipulations, 
\begin{equation}
\begin{split}
\langle j^{i}_{p} \rho^{(2)}_q \rho_{-p-q}\rangle +\langle j^{i}_{p} \rho_q \rho^{(2)}_{-p-q}\rangle&= v_F\int_\theta\, n^i \left(\langle\rho_{p} \rho^{(2)}_q \rho_{-p-q}\rangle^\theta+\langle\rho_{p} \rho_q \rho^{(2)}_{-p-q}\rangle^\theta\right)\\ &+ \frac{v_F }{2}\int_\theta\, s^i h^\theta(p|q,-p-q),
\end{split}
\end{equation}
where $h$ is defined in Eq.~\eqref{hdefinition}. Finally putting everything together, we have, 
\begin{equation}
\begin{split}
\langle j^{i}_{p} \rho_q \rho_{-p-q}\rangle &= v_F \int_\theta\,  n^i \langle\rho_{p} \rho_q \rho_{-p-q}\rangle^\theta + \frac{v_F}{3!} \int_\theta\,s^i g^{\theta}(p|q,-p-q)\nonumber\\
&+ \frac{p_F \epsilon_{F}''}{2}\int_\theta\, n^i \frac{-2(q_n (p+q)_n)}{(\omega_q-v_F q_n)(\omega_{p+q}-v_F (p+q)_n)}-\frac{v_F}{2}\int_\theta\, s^i j^\theta(p|q,-p-q)\\
&+ \frac{v_F }{2}\int_\theta\, s^i h^\theta(p|q,-p-q).
\end{split}
\end{equation}
See Eqs. \eqref{hdefinition}, \eqref{jdefinition} and \eqref{gdefinition} for the functions $h,j$ and $g$ respectively. As a non trivial check of our computation we have numerically verified the Ward identity in the euclidean domain $\omega\gg v_F p_n$.
\begin{equation}
\begin{split}
\omega_p \langle \rho_p \rho_q \rho_{-p-q}\rangle&= p_i \langle j^i_p \rho_q \rho_{-p-q}\rangle.\\
\end{split}
\end{equation}
\subsubsection*{Evaluating $ \langle\mathcal{O}_{p} \rho_q \rho_{-p-q}\rangle$} \label{rtrr}
Using the linear and nonlinear contributions of the operator $\mathcal{O}$ listed in Eqs. \eqref{rhotilde2} and \eqref{rhotilde},
\begin{equation}\label{rtrrfinal}
\begin{split}
&\langle\mathcal{O}_{p} \rho_q \rho_{-p-q}\rangle=\frac{m}{\mu} \langle\rho_{p} \rho_q \rho_{-p-q}\rangle+\frac{m p_F v_F}{\epsilon_{F}^2} \int_\theta \frac{q_n (p+q)_n}{(\omega_q-v_F q_n)(\omega_{p+q}-v_F (p+q)_n)}.
\end{split}
\end{equation}
Putting everything together, the one loop microscopic contribution to the three point density function is given by
\begin{equation}\label{3ptfnmicroeftblc}
\begin{split}
&\langle \rho_p \rho_q \rho_{-p-q}\rangle_\lambda\\ 
&=i  \lambda\int_{\theta, \theta'}\left[v_F^2\left(-1+\cos(\theta-\theta')\right)\left[\langle \rho_p \rho_{-p}\rangle^{\theta} \langle \rho_{p} \rho_q \rho_{-p-q}\rangle^{\theta'}+ \text{Cyclic}\right]\right.\\
&\left.+\left(\frac{m^2p_F v_F}{\epsilon_{p_F}^3}-p_F \epsilon_{F}''v_F(\cos(\theta-\theta'))\right)\left[\langle \rho_p \rho_{-p}\rangle^{\theta}\left[\frac{(q_{n'} (p+q)_{n'})}{(\omega_q-v_F q_{n'})(\omega_{p+q}-v_F (p+q)_{n'})}\right]\right.\right.\\
&\left.\left.+ \text{Cyclic} \right]+ v_F^2 \sin(\theta-\theta')\left[\frac{1}{3!}\langle \rho_p \rho_{-p}\rangle^{\theta}g^{\theta'}(p|q,-p-q) -\frac{1}{2}\langle \rho_p \rho_{-p}\rangle^{\theta}j^{\theta'}(p|q,-p-q)\right.\right.\\
&\left.\left.+\frac{1}{2 }\langle \rho_p \rho_{-p}\rangle^{\theta}h^{\theta'}(p|q,-p-q)+ \text{Cyclic}\right]\right],
\end{split}
\end{equation}

where, 

\begin{equation}
\langle \rho_p \rho_{-p}\rangle^{\theta}=\frac{i p_F p_n }{\omega_p - v_F p_n}. 
\end{equation}

\bibliography{main_arxiv_v3}
\bibliographystyle{JHEP}
	
\end{document}